\documentclass[11pt,a4paper]{article}
\pdfoutput=1
\usepackage{jheppub}
\usepackage{amsmath,amsfonts,amssymb}
\usepackage{hyperref, csquotes}
\usepackage{graphicx, color}
\usepackage{pgf,tikz}
\tikzstyle{wv} = [circle, inner sep=0.1pt, draw=black, minimum size=2mm]
\tikzstyle{bv} = [circle, inner sep=0.1pt, fill=black, minimum size=2mm]

\newcommand{\cjt}[1]{#1}
\newcommand{\red}[1]{#1}
\newcommand{\blue}[1]{#1}
\newcommand{\green}[1]{#1}

\renewcommand\[{\begin{equation}}
\renewcommand\]{\end{equation}}
\newcommand{\boxd}[1]{\boxed{\phantom{\Biggl(}#1\phantom{\Biggl)}}}
\newcommand{\Tr}{\mathrm{Tr}}
\newcommand{\tr}{\mathrm{tr}}
\def\extd{\mathrm {d}}
\newcommand{\id}{\mathbb I}

\newcommand{\gd}{d_{\textsc{g}}}	
\newcommand{\rk}{r}         		
\renewcommand{\ell}{c}
\newcommand{\ks}{{2\zeta}}          
\newcommand{\cs}{a}             	
\newcommand{\nb}{{n_b}}			    
\renewcommand{\next}{n_\partial} 
\newcommand{\wfr}{Z_k}
\newcommand{\cb}{\lambda_{b}}   	
\newcommand{\cbk}{\lambda_{b,k}}

\newcommand{\m}{\mu_k}
\newcommand{\mr}{\tilde{\mu}}
\newcommand{\ms}{\bar{\mu}}
\newcommand{\mc}{\bar{\mu}_*}
\newcommand{\cn}[1]{\lambda_{#1}}   
\newcommand{\cnr}[1]{\tilde{\lambda}_{#1}}   
\newcommand{\cns}[1]{\bar{\lambda}_{#1}} 
\newcommand{\cnc}[1]{\bar{\lambda}_{#1 *}} 
\newcommand{\sdd}{\omega^{\textrm{s.d.}}}
\newcommand{\gdeg}{\Omega}
\newcommand{\reg}{{\mathcal R}_k}
\renewcommand{\and}{\eta_k}
\newcommand{\scd}[1]{d_{#1}}		
\newcommand{\efd}{{d_{\textrm{eff}}}}
\newcommand{\uvd}{d_{\rk}}
\newcommand{\crd}{d_{\textrm{crit}}}
\newcommand{\nosed}{d_{\bullet}}
\newcommand{\noser}{r_{\bullet}}
\newcommand{\crr}{\rk_{\textrm{crit}}}
\newcommand{\Sk}{S_0}			    
\newcommand{\Sia}{S_\textsc{ia}}	
\newcommand{\gfc}{W}                
\newcommand{\gfi}{\Gamma}           
\newcommand{\gfk}{\Gamma_k}         
\newcommand{\kin}{\mathcal{K}}                
\newcommand{\pr}{P_{\textsc{r}}}
\newcommand{\nmax}{n}               
\newcommand{\ak}{N_k}  

\newcommand{\cct}{\ccu-1}
\newcommand{\ccu}{{N_\phi}}
\newcommand{\gp}{}
\newcommand{\gf}{\Phi}       		
\newcommand{\gfb}{\bar\Phi}

\newcommand{\cgf}{\chi}       		
\newcommand{\cgfb}{\bar\chi}
\newcommand{\vg}{\pmb{g}}
\newcommand{\vh}{\pmb{h}}
\newcommand{\dg}[1]{\delta(g_{#1},h_{#1})}
\newcommand{\mop}{(\gfb\cdot_{\hat{\ell}}\gf)}
\newcommand{\nop}{(\gfb\cdot_{{\ell}}\gf)}
\newcommand{\rep}{j}

\newcommand{\cas}{C}
\newcommand{\vrep}{{\pmb{\rep}}}
\newcommand{\crep}{{\hat{\pmb{\rep}}_\ell}}
\newcommand{\set}[2]{\mathbf{S}^{#1}_{#2}}
\newcommand{\ball}[2]{\mathbf{B}^{#1}_{#2}}
\newcommand{\vol}[1]{v_{#1}}
\newcommand{\vks}[1]{v_{#1}^{(\zeta)}}
\newcommand{\kv}[1]{V^{(#1)}}
\newcommand{\ku}[1]{I_0^{(#1)}}
\newcommand{\kw}[1]{I_{\ks}^{(#1)}}
\newcommand{\kf}[1]{J_f^{(#1)}}
\newcommand{\gr}{G_\rk}
\newcommand{\pot}{U_k}
\newcommand{\self}[1]{M_k^{(#1)}}
\newcommand{\simp}[1]{\Delta_{\ak}^{#1}}

\newcommand{\poch}[1]{P^{(#1)}}
\newcommand{\bco}[1]{\beta^{#1}}
\newcommand{\bcb}[1]{\bar{\beta}^{#1}}

\newcommand{\bfuv}[1]{\beta_{\textsc{uv}}^{#1}}

\newcommand{\delj}[1]{\delta_{\rep_{#1},0}}
\renewcommand{\aa}{i}
\newcommand{\bb}{j}
\newcommand{\cc}{z}
\newcommand{\dd}{u}
\newcommand{\suma}[2]{\mathcal{S}_#2^{(#1)}}

\begin{document}

\title{Phase transitions in TGFT: Functional renormalization group in the cyclic-melonic potential approximation and equivalence to $\textrm{O}(N)$ models}

\author[a,b]{Andreas G. A. Pithis}
\author[c,d]{and Johannes Th\"urigen}

\emailAdd{pithis@thphys.uni-heidelberg.de}
\affiliation[a]{Scuola  Internazionale  Superiore  di  Studi  Avanzati  (SISSA),\\ via  Bonomea  265,  34136  Trieste,  Italy, EU}
\affiliation[b]{Institut f\"ur Theoretische Physik, Universit\"at Heidelberg,\\ Philosophenweg 16, 69120 Heidelberg, Germany, EU}

\emailAdd{johannes.thuerigen@uni-muenster.de}
\affiliation[c]{Mathematisches Institut der Westf\"alischen Wilhelms-Universit\"at M\"unster\\ Einsteinstr. 62, 48149 M\"unster, Germany, EU}
\affiliation[d]{Institut f\"ur Physik/Institut f\"ur Mathematik der Humboldt-Universit\"at zu Berlin\\
Unter den Linden 6, 10099 Berlin, Germany, EU
}

\begin{abstract}
{In the group field theory approach to quantum gravity, continuous spacetime geometry is expected to emerge via phase transition. 
However, understanding the phase diagram and finding fixed points under the renormalization group flow remains a major challenge.
In this work we tackle the issue for a tensorial group field theory using the functional renormalization group method.
We derive the flow equation for the effective potential at any order restricting to a subclass of tensorial interactions called cyclic melonic and projecting to a constant field in group space. 
For a tensor field of rank $\rk$ on $\textrm{U}(1)$ we explicitly calculate beta functions and 
find equivalence with those of $\textrm{O}(N)$ models but with an effective dimension flowing from $\rk-1$ to zero.
In the $\rk-1$ dimensional regime, the equivalence to $\textrm{O}(N)$ models is modified by a tensor specific flow of the anomalous dimension with the consequence that the Wilson-Fisher type fixed point solution has two branches.
However, due to the flow to dimension zero, fixed points describing a transition between a broken and unbroken phase do not persist and we find universal symmetry restoration.
To overcome this limitation, it is necessary to go beyond compact configuration space.
}
\end{abstract}

\maketitle


\section{Introduction}

In numerous avenues to a quantum theory of gravity based on random discrete geometries the recovery of continuum spacetime, its symmetries and dynamics is expected to arise through a phase transition. Prominent representatives of such approaches are quantum Regge calculus~\cite{Williams:2007up}, dynamical triangulations~\cite{CDT},  tensor models~\cite{GurauBook}, covariant loop quantum gravity~\cite{Rovelli:2011tk,Perez:2012wv} and group field theories~\cite{Freidel:2005jy,Oriti:2012wt,Krajewski:2012wm,Carrozza:2013uq}. 
The latter form a class of combinatorially non-local quantum field theories which generalize matrix models for two-dimensional gravity~\cite{MM} to higher dimensions. In these, the building blocks of geometry are $\rk-1$ simplices generated by fields living on $\rk$ copies of a group. 
The action encodes how to glue these building blocks together to construct $\rk$-dimensional discrete geometries through a combinatorially non-local pairing pattern in the interaction. 
Like in matrix models these converge to various continuum geometries at criticality. 
Hence, it is highly important to understand such theories' phase structure to determine the conditions under which macroscopic continuum geometries can actually emerge from them.

The Kadanoff-Wilson formulation of the renormalization group is a crucial tool for understanding how a physical theory evolves along scales. The main idea is to implement a coarse graining operation which leads to the step-wise elimination of short length scale fluctuations~\cite{Wilson}. In this way, one obtains an effective action after each elimination step which accounts for the contribution of the eliminated modes. 
A powerful realization of this concept is the functional renormalization group (FRG)~\cite{Berges:2002ga,Kopietz,Delamotte}. 
Technically, it smoothly implements the Kadanoff-Wilson idea in the path integral by introducing a regulator function which depends on a continuous scale parameter. With this one can construct the so-called effective average action the scale-derivative of which satisfies an exact flow equation. 
It allows to practically interpolate between the classical action describing the microscopic dynamics of the system and the effective action which encodes its macroscopic dynamics. 
In particular, this equation allows to search for distinguished points in the theory space of the system, so-called critical or fixed points of the renormalization group, where the scale-derivative of the effective average action vanishes. In this way, the FRG provides a convenient tool to study the critical properties and the phase structure of a system, investigate the potential occurrence of phase transitions and study the breaking as well as restoration of symmetries. As such, it has found wide application in statistical as well as high-energy physics~\cite{Dupuis:2020fhh}.

The application of the FRG method to random and quantum geometry models is not straightforward because of the combinatorial non-local interactions which they exhibit. 
In spite of these difficulties, progress has been accomplished for matrix and tensor models for quantum gravity \cite{Eichhorn:2013jx,Eichhorn:2014cu,Eichhorn:2017wh,Eichhorn:2019dt,Eichhorn:2019kz,Eichhorn:2019hsa,Castro:2020dzt,Eichhorn:2020sla} and has been successfully transferred to tensorial group field theory (TGFT) which is a group field theory with a specific class of so-called tensorial interactions~\cite{Benedetti:2015et,TGFT1,TGFT2,BenGeloun:2015ej,BenGeloun:2016kw,BenGeloun:2016wq,TGFT3,BenGeloun:2018ekd}. 
An important insight has been that the FRG equation in TGFT is generically non-autonomous which is a consequence of an external scale $\cs$ in such theories. 
The phase structure has been studied in low-order truncations of the theory space~\cite{Benedetti:2015et,TGFT1,TGFT2,BenGeloun:2015ej,BenGeloun:2016kw} and in an autonomous limit~\cite{TGFT3,BenGeloun:2018ekd}. 
Typically, non-Gaussian fixed points are found in these works; however these results need further confirmation beyond low-order truncation. 
Furthermore, the full characterization of the IR properties of such theories has so far been out of reach. In  particular, understanding whether and under which conditions phase transitions and different phases can truly exist for such models has remained an open issue. Settling this point is also highly relevant for the group field theory condensate cosmology program~\cite{Gielen:2013cr,Gielen:2014gv,Gielen_2016,Oriti:2016qtz,deCesare:2016rsf,Gielen:2017eco,Oriti_2017,Pithis_2019} which assumes the existence of condensate phases with a tentative continuum geometric interpretation as a working hypothesis.

In this work we attack these points by establishing a local-potential approximation (LPA$'$)%
\footnote{According to standard FRG jargon, our approximation is not just an LPA but an LPA$'$ as we also consider here the flow of the anomalous dimension. However, it is not the full LPA$'$ of tensorial group field theory since we restrict our ensuing analysis to a specific class of infinitely many interactions.}
for tensorial group field theory without gauge constraint.
We achieve this by exploiting two approximations: A restriction to a specific subclass of interactions at any order called cyclic-melonic and a projection to uniform field configurations.
In this setting we are able to calculate the inverse of the full Hessian of the effective average action. 
As a side result, this allows us to identify a \enquote{Goldstone} and a \enquote{radial} mode in the resulting flow equations for the first time in the context of such theories. 
We explicitly derive the non-autonomous FRG equation at arbitrary rank for the case of the group $\textrm{U}(1)$. This result is valid at all renormalization group scales $k$.
In particular, we find that upon proper rescaling this scale occurs always as $\ak = \cs k$, i.e.~the scale $k$ is only meaningful compared to the external scale $\cs$. 
This clarifies also the relation of the FRG flow in TGFT with that in tensor models where the only available scale is the tensor size and this agrees with $\ak$ here.

As a result, our flow equations turn out to be equivalent to those known for $\textrm{O}(N)$ scalar field theories on $d$-dimensional Euclidean space~\cite{Berges:2002ga} but with a scale-dependent effective dimension $\efd$ flowing from a large-$\ak$ dimension $\uvd=\rk-1$ to $\efd=0$ at arbitrary small $\ak$.
More precisely, at large $\ak$ only the large-$N$ part of the $\textrm{O}(N)$ equations contributes and there is an additional relative factor of $\rk$ between the quadratic \enquote{mass} term and the other couplings.
Along the flow these modifications are continuously removed and at small $\ak$ the equivalence to the  $\textrm{O}(N)$ model equation with $d=0$ dimension is exact,
with $N=1$ for a $\mathbb{Z}_2$-symmetric real scalar field theory and $N=2$ for the $\text{U}(1)\simeq\textrm{O}(2)$-symmetric complex scalar field theory.
This effective zero-dimensionality
agrees with results on scalar field theory on compact spaces~\cite{Benedetti,Serreau:2011fu,Guilleux:2017ig}
and with our previous work on Landau-Ginzburg mean field theory in group field theory~\cite{Pithis:2018bw}.
It has the immediate consequence that there is only the symmetric phase at small $\ak$ and there can thus be no phase transition to a broken phase.
Indeed, we discover that the global $\mathbb{Z}_2$ or $\textrm{U}(1)$ symmetry is universally restored in this regime. 
The essential reason for this are the isolated zero modes in the spectrum of the theory due to the compactness of the field domain. 
Thus, we conjecture that this result applies not only in the cyclic-melonic potential approximation but also to the full theory space including arbitrary tensor-invariant interactions on any compact domain.

The phase diagram at large-$\ak$ has a much richer structure.
Since the tensorial theory has the same scaling dimensions as a local theory in $\uvd=\rk-1$ dimensions there is a critical rank $\crr=5$ above which the Gaussian fixed point describes phase transitions with mean-field exponents.
On the grounds of the equivalence with large-$N$ $\textrm{O}(N)$ models, there is a Wilson-Fisher type fixed point below $\crr$, in particular for $\rk=4$ in the LPA. 
However, we find that taking into account also the flow of the anomalous dimension, properties of this fixed point solution of the flow equations are more drastically modified in the LPA$'$. 
At finite-order truncation the solution in the tensorial case has two branches such that there can be two non-Gaussian fixed points but their exact domain of convergence is difficult to exactly determine due to computational limitations at truncations of order $n>12$. 
For the non-Gaussian fixed point smoothly connected to the Gaussian fixed point we find divergence already for $\rk>4$ in the LPA$'$. 
Our results indicate that the non-Gaussian fixed point on the other branch exists at the critical rank $\rk=5$.
Though all of these fixed points do not persist to smaller scales $k$ for compact groups of fixed volume size $\cs$ due to the dimensional flow to zero, these large-$\ak$ results can be seen as valid on all scales $k$ in the large volume limit. 
In particular, in this limit our equations are a generalization of the quartic truncation  
for TGFT on $\mathbb{R}^\rk$~\cite{BenGeloun:2016kw, BenGeloun:2015ej}.

The paper is structured in the following way: 
In Section~\ref{Section:methodandmodel} we introduce the relevant theory space of TGFT and the FRG methodology, discuss the concrete model with cyclic-melonic interactions and compute the non-autonomous FRG equation for this model on $\text{U}(1)^r$. 
In Section~\ref{sec:results} we present the separate analysis of these FRG equations in the large-$\ak$ and small-$\ak$ regimes where they become autonomous as well as the analytic and numerical arguments for symmetry restoration of the full non-autonomous case.
In the final Section~\ref{Section:conclusion} we summarize and contextualize our results, discuss limitations of our analysis and propose further studies. 
The Appendix~\ref{sec:Traces}--\ref{sec:Dimensions} supplements the main Sections of this work with more detailed calculations needed for the computation of the FRG equations and clarifies differences between the concepts of canonical and scaling dimensions in field theories with tensorial interaction.

\section{FRG equation for TGFT in the cyclic-melonic approximation}\label{Section:methodandmodel}

The FRG approach is based on a functional differential equation which determines the flow of the effective action under the renormalization group \cite{Wetterich:1993im,Morris:1994kg}.
Formally, this can be applied to TGFT in a straightforward way \cite{Benedetti:2015et} but to gain any information from it some truncation of the effective action is necessary.
Here, we decide not to truncate at a finite order of the polynomial potential; instead we follow the idea of the local potential approximation (LPA) and consider the potential at any order but restrict to a specific class of dominant interactions.
Projecting to constant field configurations $\rho$, we then find an explicit FRG equation for the effective potential which we can expand in $\rho$ to derive flow equations for the couplings of cyclic-melonic interactions at any order.

\subsection{Functional renormalization group for TGFT}
\label{Subsection:method}

To set the framework and fix notation we start with the definition of group field theory and tensorial invariance and in a second step introduce the functional renormalization group for such theory.

\subsubsection{Group field theory and tensorial invariance}\label{Subsubsection:TGFT}

Group field theories are field theories distinguished by two properties: 
a direct group product as configuration space
and combinatorial non-locality.
Specific models might be amended by further structure such as closure constraints or Plebanski constraints to capture gravitational degrees of freedom in some way already in the microscopic theory~\cite{Freidel:2005jy,Oriti:2012wt,Krajewski:2012wm}.
Here we will not consider any such constraint and leave the generalization of our results to such models to future work.

For the first property, the fields $\gf$ are real- or complex-valued%
\footnote{In this work, explicit formulae are almost always for the complex case but we will consider the real case along side and provide explicit results when necessary. 
}
functions of $\rk$ arguments $\vg = (g_1 , g_2 , ..., g_\rk)$ each in a Lie group $G$, that is  on the configuration space $G^{\times\rk}$.
More specifically, for $G$ compact
one considers square-integrable functions $\gf,\gf' \in L^2(G^{\times\rk})$ with respect to the inner product
\[
(\gf,\gf') = \int \extd\vg\, \gfb(\vg) \gf'(\vg)
\]
defined in terms of the Haar measure which we choose here to be dimensionful,
\[
\int_G \extd g = \cs \,.
\]
The volume scale $\cs$ will be crucial for physically meaningful rescaling of couplings.
It furthermore allows to take a large-volume limit $\cs\rightarrow\infty$ related to the definition of 
the theory on $\mathbb{R}^\rk$ where this limit is used to define an IR regularization \cite{BenGeloun:2015ej,BenGeloun:2016kw}.
As ``momentum" transform we consider the expansion in the matrix coefficients of unitary irreducible representations labelled by a multi-index $\vrep=(\rep_1,...,\rep_\rk)$
\[\label{eq:transform}
\gf(\vg) = \sum_{\rep_1,...,\rep_\rk} \left(\prod_{\ell=1}^\rk d_{j_\ell}\right) \tr_\vrep \left[ \gf_\vrep\bigotimes_{\ell =1}^\rk D^{\rep_\ell}(g_\ell) \right],
\]
where $D^{\rep}(g)$ are the representation matrices on $d_j$-dimensional representation space, the coefficients of which form a countable complete orthogonal basis of $L^2(G)$ according to the Peter-Weyl theorem. Thus, also the field transform $\gf_\vrep$ is matrix-valued with respect to each representation $\rep_\ell $ and $\tr_\vrep$ is the trace over all the representation spaces~\cite{jacques2008analysis}.

Second, combinatorial non-locality is the property that the interaction part $\Sia$ in the action $S=\Sk+\Sia$ can be expanded in field monomials with a specific convolution pattern:
each argument $g_i$ of an occurrence of a field is paired with exactly one argument $g_j$ of another field occurrence.
As a consequence, the Feynman diagrams labelling the perturbative expansion are stranded diagrams instead of ordinary graphs. 
With some additional structure such diagrams are bijective to $\rk$-dimensional combinatorial (pseudo) manifolds thus describing quantum geometry as random geometry~\cite{Gurau:2010iu,GurauBook} generalizing matrix models~\cite{DiFrancesco:1992cn} to higher dimensions.

One example of combinatorial non-local interactions leading to $\rk$-dimensional (pseudo) manifolds are tensor invariants.
This invariance refers to a symmetry under transformations of the field $\gf$ as a rank-$\rk$ covariant complex tensor, i.e.~transforming under unitary transformations $U^\ell:L^2(G)\rightarrow L^2(G)$ in each argument individually,
\[
\gf(\vg) \mapsto \left(\bigotimes_{\ell=1}^\rk U^\ell \,\gf \right)(\vg) 
=  \int \extd h_1 ... \extd h_\rk \prod_{\ell=1}^\rk U^\ell(g_c,h_c)  \gf(h_1,...,h_\rk)
\]
which is also called \emph{tensorial} symmetry.
Consequently, as known from tensor models~\cite{GurauBook}, there is an infinite class of invariant interactions labelled by bipartite%
\footnote{For real tensor fields the symmetry is orthogonal instead of unitary, leading to $\rk$-coloured graphs without the property of bipartiteness~\cite{Carrozza:2016ff}.}
$\rk$-coloured graphs $b\in B$ where ``$\rk$ colouring'' means that each vertex is $\rk$-valent and adjacent to an edge of each colour $\ell=1,...,\rk$.
Thus, one has a theory space given by
\[\label{eq:SIA}
\Sia[\gf,\gfb] = \sum_{b\in B} \cb \Tr_b[\gf,\gfb] 
\]
where $\Tr_b$ denotes the convolution of fields as described by the $\rk$-coloured graph $b$. 
We refer to a GFT with tensor-invariant interactions as \emph{tensorial group field theory} (TGFT) in this sense.


\subsubsection{FRG equation in TGFT}

For a non-perturbative analysis of a field theory it is useful to work with the effective average action.
The definitions are completely the same as for standard field theories \cite{Kopietz}: 
Starting with the generating function of connected Green functions
\[
e^{\gfc[\bar{J},J]} = \int D \gfb D \gf \, e^{-S[\gf,\gfb] + (J,\gf) + (\gf,J)}
\]
one obtains the effective (one-particle-irreducible) action via 
Legendre transform
\renewcommand{\gf}{\varphi}       
\renewcommand{\gfb}{\bar\varphi}
\[
\gfi[\gf,\gfb] = \sup_{\bar{J},J}\{(\gf,J) + (J,\gf) - \gfc[\bar{J},J]\}
\]
where the sources $\bar{J},J$ are functions of $\gfb,\gf$ obtained from inverting the expectation value of the field
\[
\gf(\vg) :=  \langle\Phi(\vg)\rangle  =  \frac{\delta W[\bar{J},J]}{\delta \bar{J}(\vg)}
\quad,\quad 
\gfb(\vg) :=  \langle\bar{\Phi}(\vg)\rangle  =  \frac{\delta W[\bar{J},J]}{\delta J(\vg)}.
\]

To implement the renormalization group, one modifies propagation in a scale-dependent way.
That is, one modifies the kinetic part $\Sk$ of the action by an IR regulator $\reg$ depending on the cutoff scale $k$,
\[\label{eq:regularizedkinetic}
S_{0,k}[\gf,\gfb] = (\gf,\kin\gf) - (\gf, \reg \gf)
\]
where the common kinetic operator is
$
\kin = (-1)^{\gd} \Delta + \mu 
$
with $\mu$ the coupling at quadratic order and $\Delta = \sum_\ell \Delta_\ell$ the Laplace-Beltrami operator on $G^{\times \rk}$ which is a sum over Laplace-Beltrami operators $\Delta_\ell$ on each copy of the group $G$.
Transforming to representation space along Eq.~\eqref{eq:transform}, the Laplacian becomes the diagonal Casimir operator $\cas_\vrep = \sum_\ell \cas_{\rep_\ell}$ together with a factor $\cs^{-2}$ for dimensional reasons.

As natural from the field theoretic perspective, we associate the scale $k$ with the modes of the group field labelled by representations $\rep_\ell$. 
Thus, in representation space the regulator should also be diagonal and reduces to a function $\reg=\reg(\cs^{-2}\cas_\vrep/k^2)$.
For a meaningful regulator this function has to satisfy the following properties~\cite{Berges:2002ga,Kopietz,Benedetti:2015et}: 
positivity 
$\reg(z)\geq 0$, 
monotonicity
$\frac{d}{dz}\reg(z)\leq 0$, 
and as a third condition
$\reg(0)>0$ together with $\lim_{z\to \infty}\reg(z)=0$. 
This third condition guarantees that the cutoff is removed at $k\to 0$. When a UV cutoff $\Lambda$ is present, typically the condition $\lim_{k\to\Lambda}\reg=\infty$ is supplemented.
Note that, in the usual field theory jargon, we call large $k$ UV and small $k$ IR, even though there is no direct relation to a notion of energy in TGFT.

As a consequence, the generating function $\gfc_k$ and the average field $\gf$ become dependent on $k$ and the effective average action \`{a} la Wetterich is
\[
\gfk[\gf,\gfb] = \sup_{\bar{J},J}\{(\gf,J) + (J,\gf) - \gfc_k[\bar{J},J]\} - (\gf, \reg \gf)
\]
which for the theory space of tensor-invariant interactions is of the general form
\[\label{eq:tensorialWetterichaction}
\gfk [\gfb , \gf] = (\gf,\kin_k\gf)  
+  \sum_{b\in B} \cbk \Tr_b[\gf,\gfb] 
\quad , \quad \kin_k = (-1)^{\gd} \wfr \Delta + \m
\]
where the dependence on the scale $k$ is captured by effective couplings $\m,\cbk$ as well as the wave-function renormalization parameter $\wfr$.
The effective average action $\gfk$ smoothly interpolates between the microscopic action $S$ at $k=\Lambda$ and an effective action $\Gamma_{k=0}$.
Note that, even if in the interaction-part of the action \eqref{eq:SIA} only couplings $\cb = \lambda_{b,k=\Lambda}$ corresponding to connected graphs $b\in B_{\textrm{con}}$ are non-vanishing, interactions labelled by graphs with arbitrary number of connected components $b\in B$ generically appear in $\gfk$.
This is because $\gfk$ captures all interactions generated in the flow, that is any possible boundary graphs according to the action $S$.

The renormalization group flow of the effective average action is determined by the functional equation \cite{Wetterich:1993im,Morris:1994kg,Benedetti:2015et}
\[\label{eq:Wetterich} \boxd{
k \partial_k \gfk[\gf,\gfb] = 
\frac{1}{2} \overline{\Tr} \left[ \cjt{\gp} \left( \Gamma_k^{(2)}[\gf,\gfb] + \reg \id_2 \right)^{-1} k \partial_k \reg \id_2 \right] }
\]
with initial condition $\Gamma_{k=\Lambda}[\gf,\gfb]=S[\gf,\gfb]$ at the UV scale $\Lambda$.
Here 
the trace $\overline{\Tr}$ denotes summation over all field excitations; for a complex field, this is a sum over the $G^{\times\rk}$ degrees of freedom of both $\gf$ and its complex conjugate $\gfb$.
In particular, 
the Hessian of the effective average action $\gfk^{(2)}$ 
%
%
is a quadratic form in configuration space and a $2\times2$ matrix with respect to $\gf$ and $\gfb$,
\begin{equation}\label{Hessian2x2}
\gfk^{(2)} = \begin{pmatrix}
    \kin_k + F & F_{12} \\
    F_{21} & \kin_k + F 
\end{pmatrix}
\end{equation}
with $k$-dependent interaction derivatives
\[
F[\gf,\gfb](\vg,\vh) :=
\frac{\delta^2\Gamma_k[\gf,\gfb]}{\delta\gf(\vg)\delta\gfb(\vh)}
= \frac{\delta^2\Gamma_k[\gf,\gfb]}{\delta\gfb(\vg)\delta\gf(\vh)}
\]
and
\[
F_{12}[\gf,\gfb](\vg,\vh) :=
\frac{\delta^2\Gamma_k[\gf,\gfb]}{\delta\gf(\vg)\delta\gf(\vh)}
\quad , \quad
F_{21}[\gf,\gfb](\vg,\vh) :=
\frac{\delta^2\Gamma_k[\gf,\gfb]}{\delta\gfb(\vg)\delta\gfb(\vh)}.
\]
This aspect of the complex field in TGFT seems to have been overlooked so far in the literature~\cite{BenGeloun:2015ej,BenGeloun:2016kw,BenGeloun:2016wq, TGFT3,Carrozza:2017dl,Lahoche:2018fx,Lahoche:2018uv,Lahoche:2019bb,Lahoche:2019et,Lahoche:2019fq,Lahoche:2019wm,Lahoche:2020bi,Lahoche:2020df,Lahoche:2020ib,Lahoche:2020kt,Baloitcha:2020vh}.

Under some mild conditions it is straightforward to obtain the $2\times 2$ trace of the inverse of the operator 
\[
\left( \Gamma_k^{(2)} + \reg \id_2 \right) 
= \begin{pmatrix}
    \pr + F & F_{12} \\
    F_{21} & \pr + F 
\end{pmatrix}
\]
where we use the (inverse of the) regulated propagator
\[
\pr = \kin_k + \reg = (-1)^{\gd} \wfr\Delta + \m + \reg \, .
\]
One has to be careful with inversion of that operator in two ways.
First, the inverse with respect to $\gf,\gfb$ is the usual matrix inverse only if the quadratic forms $\pr+F$, $F_{12}$ and $F_{21}$ commute with each other, $[F_{12},F_{21}]\equiv F_{12} F_{21}-F_{21} F_{12} = 0$ etc.,
since
\[
\begin{pmatrix}
    \pr + F & F_{12} \\
    F_{21} & \pr + F 
\end{pmatrix}
\begin{pmatrix}
    \pr+F & -F_{12} \\
    -F_{21} & \pr+F 
\end{pmatrix}
= \begin{pmatrix}
    (\pr + F)^2 - F_{12} F_{21} & [(\pr + F),F_{12}] \\
    [F_{21}, (\pr + F)] & (\pr + F)^2 - F_{21} F_{12}
\end{pmatrix}.
\]
Furthermore, for the inverse to exist there must exist an inverse of the $2\times2$ determinant $(\pr+F)^2 + F_{12}F_{21}$ with respect to the group.
If the operators obey these two conditions, 
the full trace $\overline{\Tr}$ in the FRG equation Eq.~\eqref{eq:Wetterich} simplifies to the trace $\Tr_{G^{\times\rk}}$ over group space yielding
\begin{eqnarray}
k \partial_k \gfk 
&=& \frac{1}{2} \Tr_{G^{\times\rk}} \left[ \cjt{\gp} \left((\pr+F)^2 - F_{12}F_{21}\right)^{-1} 2(\pr + F)\, k \partial_k \reg \right] \, .
\label{eq:2x2Wetterich}
\end{eqnarray}
Since there are no differential operators in the interaction differentials $F$, $F_{12}$ and $F_{21}$, one has to mainly check the conditions on the regularized propagator $\pr$. 




\subsection{Cyclic-melonic potential and projection}\label{Subsection:model}

In this work we aim to calculate the renormalization group flow of a tensorial group field theory with a full potential, that is a class of interactions of arbitrary order $(\gfb\gf)^n$.
We achieve this by restricting to cyclic-melonic interactions and projecting onto a constant group field on configuration space.

\subsubsection{Restriction to cyclic-melonic interactions}

\begin{figure}
\centering
\begin{tikzpicture}
\begin{scope}[xshift=-1cm]
\foreach \i in {1,2}{
	\begin{scope}[rotate=\i *180]
	\node [wv]		(w\i)	at (-.6,.6)	{};
	\node [bv]		(b\i)	at (.6,.6)	{};
	\end{scope}
	}
\foreach \i in {1,2}{
	\path	(w\i) edge 				node 	{}	(b\i)
		(w\i)	edge [bend left=30]	node 	{}	(b\i)
		(w\i)	edge [bend right=30]	node 	{}	(b\i);
	}
\foreach  \i/\j in {1/2,2/1}{
	\draw (w\i) -- (b\j);
	}
\node (c) at (-.4,0) {\scriptsize{$c$}};
\node (l) at (-1,0) {$\cn{2}^\ell$};
\end{scope}
\begin{scope}[xshift=1.7cm]
\foreach \i in {1,2,3}{
	\begin{scope}[rotate=\i *120]
	\node [wv]		(w\i)	at (-.5,.8)	{};
	\node [bv]		(b\i)	at (.5,.8)	{};
	\end{scope}
	}
\foreach \i in {1,2,3}{
	\path	(w\i) edge 				node 	{}	(b\i)
		(w\i)	edge [bend left=30]	node 	{}	(b\i)
		(w\i)	edge [bend right=30]	node 	{}	(b\i);
	}
\foreach  \i/\j in {1/2,2/3,3/1}{
	\draw (w\i) -- (b\j);
	}
\node (c) at (-.6,.4) {\scriptsize{$c$}};
\node (l) at (-1.5,0) {, $\cn{3}^\ell$};
\end{scope}
\begin{scope}[xshift=5cm]
\foreach \i in {1,2,3,4}{
	\begin{scope}[rotate=\i *90]
	\node [wv]		(w\i)	at (-.4,.9)	{};
	\node [bv]		(b\i)	at (.4,.9)	{};
	\end{scope}
	}
\foreach \i in {1,2,3,4}{
	\path	(w\i) edge 				node 	{}	(b\i)
		(w\i)	edge [bend left=30]	node 	{}	(b\i)
		(w\i)	edge [bend right=30]	node 	{}	(b\i);
	}
\foreach  \i/\j in {1/2,2/3,3/4,4/1}{
	\draw (w\i) -- (b\j);
	}
\node (c) at (-.9,.9) {\scriptsize{$c$}};
\node (l) at (-1.6,0) {, $\cn{4}^\ell$};
\end{scope}
\begin{scope}[xshift=8.8cm]
\foreach \i in {1,2,3,4,5}{
	\begin{scope}[rotate=\i *63]
	\node [wv]		(w\i)	at (1.1,.3)	{};
	\node [bv]		(b\i)	at (1.1,-.3)	{};
	\end{scope}
	}
\foreach \i in {1,2,3,4,5}{
	\path	(w\i) edge 				node 	{}	(b\i)
		(w\i)	edge [bend left=30]	node 	{}	(b\i)
		(w\i)	edge [bend right=30]	node 	{}	(b\i);
	}
\foreach  \i/\j in {1/2,2/3,3/4,4/5}{
	\draw (w\i) -- (b\j);
	}
\path	(w5) edge [dashed, bend right=30]	node {} (b1);
\node (c) at (-1.2,.6) {\scriptsize{$c$}};
\node (s) at (-2,0)	{, $\dots \cn{n}^\ell$};
\end{scope}
\end{tikzpicture}
\caption{Cyclic-melonic interaction vertices diagrammatically described by coloured graphs. }\label{fig:cyclicmelonic}
\end{figure}
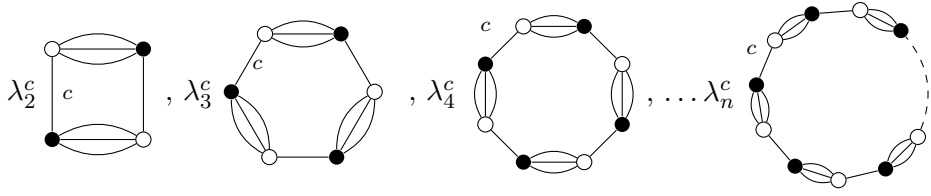

Cyclic melonic interactions are simply closed chains of open melons (see Fig.~\ref{fig:cyclicmelonic} for the corresponding coloured graphs).
More precisely,
an open melon of colour $\ell$ is defined by an operator $\mop$ on $G$ with kernel
\[
\mop(g_\ell,h_\ell) 
:= \int \Bigl(\prod_{b\ne\ell}\extd g_b\Bigr)\, \gfb(g_1,...,g_\ell,...,g_\rk)\gf(g_1,...,h_\ell,...,g_\rk ) 
\]
and a cyclic melonic operator of order $n$ is then $\mop^n$ where, as usual, multiplication of operators $\mop$ is explicitly given by convolution of their kernels
\[
\mop^2(g_c,g'_c) = \int \extd h_c \mop(g_c,h_c)\mop(h_c,g'_c) .
\]
The theory space of cyclic melonic interactions, as described by the effective average action, is then
\[
\Gamma_k [\gfb , \gf] = \int \extd \vg \, \gfb( \vg ) \kin_k 
\gf( \vg )
+ \sum_{\ell=1}^\rk \Tr_G V^\ell_k\mop 
\]
where 
the potential is determined by a single function
\renewcommand{\cbk}{\lambda^\ell_{n,k}}
\[\label{eq:potentialfunction}
V^\ell_k ( z ) = \sum_{n = 2}^{\infty} \frac{1}{n!}\cbk z^n , 
\]
with real scale-dependent coefficients $\cbk$.
This theory space is approximately stable under the FRG in the large $k$ regime as shown in the $\rk=3$ case with an additional closure constraint \cite{TGFT3}. There, the interactions are called ``non-branching" melons since they correspond to the subset of non-branching rooted trees according to the classification of melonic interactions in terms of a bijection to rooted trees~\cite{Bonzom:2011cs}. 
We prefer to call them cyclic melons since it is their cyclicity which will reduce all calculations to the simple potential function $V_k$ upon projection onto a constant field.
Furthermore, this potential is very similar to the one in matrix theories, precisely because it is the generalization of complex-matrix interactions to melonic tensor interactions which preserves cyclicity.
Still, it covers essential tensorial structure because of the different colours which mix in a non-trivial way at higher loop orders. We further comment on our choice of theory space and its limitations in particular when projecting onto uniform field configurations in the following subsection.

\subsubsection{The Hessian projected on constant fields}

The Wetterich equation for this theory space nicely simplifies when projecting on a constant field.
To show this, we have to derive the second derivatives on the right-hand side of the equation.
The diagonal entries 
are
\[
F[\gf,\gfb](\vg,\vh) :=
\frac{\delta^2\Gamma_k[\gf,\gfb]}{\delta\gf(\vg)\delta\gfb(\vh)}
= \sum_{c=1}^\rk \sum_{n=2}^\infty  \frac{n}{n!}\cbk \sum_{p=0}^{n-1}  \mop^{p}(g_c,h_c)
\nop^{n-p-1}(\hat\vg_c,\hat\vh_c)
\]
where $\nop$ 
on $G^{\rk-1}$ is the operator with kernel
\[
\nop(\hat\vg_c,\hat\vh_c) := \int \extd g_\ell\, \gf(g_1,...,g_\ell,...,g_\rk)
\gf(h_1,...,g_\ell,...,h_\rk)
\]
using the notation $\hat\vg_c = (g_1,...,g_{c-1},g_{c+1},...,g_\rk)$. 
Multiplication of such operators is thus convolution on $G^{\rk-1}$.
Furthermore, we have used the standard definition that a zero exponent yields unity, that is
\[
\mop^{0}(g_\ell,h_\ell) \equiv \dg{c} \quad , \quad
\nop^{0} (\hat\vg_c,\hat\vh_c)\equiv \prod_{b\ne c}\dg{b} .
\]
Both Dirac delta terms occur when differentiating with respect to neighbouring fields in the interaction.
Spelling them out explicitly, we have 
\begin{align}
F[\gf,\gfb](\vg,\vh) 
= \sum_{c=1}^\rk \sum_{n=2}^\infty  \frac{n}{n!} \cn{n}^\ell \biggl[& 
\prod_{b\ne c}\dg{b} \,\mop^{n-1}(g_c,h_c)
+ \dg{c}\nop^{n-1}(\hat\vg_c,\hat\vh_c) \nonumber\\
&+  \sum_{p=1}^{n-2} \mop^{p}(g^c,h^c)\nop^{n-p-1}(\hat\vg_c,\hat\vh_c)
\biggr].  
\end{align}
Both $\mop^p$ and $\nop^p$ are monomials of order $p$ in $\gfb\gf$,
furthermore respectively containing $\rk p-1$ and $\rk p-(\rk-1)$ group integrals.
Thus, upon projection onto a field
\[
\gf(\vg) = \cgf
\]
which is constant on group configuration space we can simply express $F[\cgfb,\cgf](\vg,\vh)$ again in terms of the potential functions $V_k^\ell$,
\renewcommand{\dg}[1]{\cs{\delta(g_{#1},h_{#1})}}
\begin{eqnarray}\label{eq:Hessian}
F[\cgfb,\cgf](\vg,\vh) &=& \sum_{c=1}^\rk \sum_{n=2}^\infty  \frac{n}{n!} \cn{n}^\ell \cs^{(n-2)\rk} \left(\prod_{b\ne c}\dg{b} + \dg{c} + n-2\right)(\cgfb\cgf)^{n-1} \nonumber\\
&=& \cs^{\rk} \sum_{c=1}^\rk \left[\left(\prod_{b\ne c}\dg{b} + \dg{c} - 1\right){V_k^\ell}'(\rho)  + \rho {V_k^\ell}''(\rho)\right] ,
\end{eqnarray}
where in the second line the constant field is expressed by its norm
\[
\rho := (\cgf,\cgf) = \cs^{\rk}\,\cgfb\cgf \,.
\]
All the combinatorially non-local information of the cyclic melonic interactions after projection on $\cgf$ is now captured by the operator
\[\label{eq:deltaoperatorg}
\mathcal{O}^\ell(\vg,\vg') := \prod_{b\neq \ell }\dg{b} + \dg{c} - 1 \, .
\]

In the off-diagonal terms of the Hessian $F_{12}[\gf,\gfb] = F_{21}[\gfb,\gf]$ no Dirac deltas occur since there are no derivatives with respect to neighbouring fields,
\begin{align}
 F_{12}[\gf,\gfb](\vg,\vh) = \sum_{c=1}^\rk \sum_{n=2}^\infty  \frac{n}{n!} \cn{n}^\ell \sum_{p=0}^{n-2}
&\int \extd g'\, \mop^{p}(h_\ell,g')\gfb(g_1,...,g',...,g_\rk) \nonumber\\
\times&\int \extd h'\, \mop^{n-p-2}(g_\ell,h')\gfb(h^1,...,h',...h_\rk) \, .   
\end{align}
Thus, projection on constant $\gf(\vg) = \cgf$ simply leads to
\[
F_{12}[\cgfb,\cgf](\vg,\vh) = \cs^{\rk}\,\cgfb^2\,\sum_{c=1}^\rk {V^\ell}''(\rho) \quad , \quad 
F_{21}[\cgfb,\cgf](\vg,\vh) = \cs^{\rk}\,\cgf^2\, \sum_{c=1}^\rk {V^\ell}''(\rho) \,.
\]

{The case of a real field easily follows from these calculations.
A real rank-$\rk$ tensorial group field is invariant under orthogonal instead of unitary transformations.
As a consequence, diagrams are still edge-coloured graphs, but not bipartite anymore.
Still, in the cyclic-melonic approximation the interactions are the same.
But for the quadratic derivative we find both kind of terms at the same time, that is for a real field $\gf$
\begin{align}
F_{\mathbb{R}}[\gf](\vg,\vh) = \frac{\delta^2\Gamma_k[\gf,\gf]}{\delta\gf(\vg)\delta\gf(\vh)} &= F[\gf,\gf](\vg,\vh) + F_{12}[\gf,\gf](\vg,\vh)
\end{align}
which upon projection becomes
\[
F_{\mathbb{R}}[\cgf](\vg,\vh) 
= \cs^{\rk} \sum_{c=1}^\rk \left( 2\rho {V_k^\ell}''(\rho) + \mathcal{O}^\ell(\vg,\vg'){V_k^\ell}'(\rho)  \right)
\]
Thus, the real field result just yields the diagonal entry of the complex case with an additional factor of two for the $V''$ term.
}

Notice that, in spite of restricting ourselves to cyclic-melonic interactions, the renormalization group flow generates also any other tensor-invariant interactions \cite{PerezSanchez:2018fn}, both of melonic and non-melonic type. 
In particular, also disconnected interactions are generated the interactions of which factorize into a product of traces. 
In FRG studies of TGFT, non-melonic interactions are commonly omitted since they are suppressed at large tensor size~\cite{GurauBook} (first attempts beyond are \cite{Carrozza:2017dl,BenGeloun:2018ekd}).
The physical relevance of disconnected interactions, akin to multi-trace operators in matrix models for $2d$ gravity~\cite{Das:1990gp,Korchemsky:1992uj,AlvarezGaume:1992np}, is so far not clear in TGFT (except for first results in the quartic truncation ~\cite{BenGeloun:2018ekd}).

The projection onto uniform field configurations as applied here washes these combinatorial subtleties out. It only retains essential and general non-local information in the flow equations to the effect that contributions stemming from all interactions are considered as if they behaved like cyclic-melonic ones. 
While this is arguably a limitation of our method, it facilitates for the first time the computation of flow equations with a full potential of arbitrary order for a TGFT. 
As we demonstrate below, it allows in particular to compare TGFT to $\textrm{O}(N)$ models on Euclidean space. 
In addition, the phenomenon of symmetry restoration at small scales $k$ is not sensitive to the combinatorial details of the interactions but is rather universal as we show below. Hence, focusing on cylic-melonic interactions in the effective average action is less restrictive than it may seem on first sight.

\subsubsection{FRG equation for projected fields}

{In group configuration space $G^{\times\rk}$ the functional derivatives $F,F_{12},F_{21}$ are not diagonal.}
Therefore, it is hard to find the inverse operator $( \Gamma_k^{(2)} + \reg )^{-1}$.
This task simplifies when transforming to group momentum space $\hat{G}^{\times\rk}$ given by representations $\rep_\ell$ according to Eq.~\eqref{eq:transform}.
All operators are diagonal there. The diagonal of the Laplacian is the Casimir operator
\[
(-1)^{\gd} (\Delta)_\vrep = \frac{1}{\cs^2} C_{\vrep} \equiv \frac{1}{\cs^2} \sum_{\ell=1}^\rk C_{\rep_\ell}
\]
with volume factor $\cs$.
The diagonal of the non-local operator is
\begin{equation}\label{eq:deltaoperator}
\mathcal{O}^\ell_{\vrep} :=  \delta_{0\rep_\ell} + (1-\delta_{0\rep_\ell}) \prod_{b\neq \ell }\delta_{0\rep_b} .
\end{equation}
Thus, in this basis the conditions for the inversion of $( \Gamma_k^{(2)} + \reg )$ according to Eq.~\eqref{eq:2x2Wetterich} are fulfilled.
The trace over the $2\times2$ part of the operators in the FRG equation for the complex field yields
\begin{equation}\label{eq:Wetterichprojected}
k \partial_k \pot(\rho) = \frac{1}{2}\Tr_{\hat{G}^{\rk}}\left[
\frac{\gp k\partial_k \reg}{\pr+\sum_\ell \mathcal{O}_\vrep^\ell {V^\ell_k}'(\rho)}
+ \frac{\gp k\partial_k \reg}{\pr+\sum_\ell \mathcal{O}_\vrep^\ell {V^\ell_k}'(\rho)+ 2 \rho \left(\prod_c \delta_{0\rep_c}\right) \sum_\ell  {V^\ell_k}''(\rho)}
\right]
\end{equation}
where the effective average action $\gfk$ on the left-hand side has been reduced upon projection to the effective potential
\[\label{eq:effectivepotential}
\pot(\rho):= \m \rho + \sum_{\ell=1}^{\rk} V_k^\ell (\rho).
\]
For the real field theory one directly finds
\begin{equation}\label{eq:Wetterichprojectedreal}
\frac{1}{2}\Tr_{\hat{G}^{\rk}}\left[ \frac{\gp k\partial_k \reg}{\pr+F_\mathbb{R}(\rho)}
\right]
= \frac{1}{2}\Tr_{\hat{G}^{\rk}}\left[\frac{\gp k\partial_k \reg}{\pr+\sum_\ell \mathcal{O}_\vrep^\ell {V^\ell_k}'(\rho)+ 2 \rho \left(\prod_c \delta_{0\rep_c}\right) \sum_\ell  {V^\ell_k}''(\rho)}
\right]
\end{equation}
which is simply the second of the two parts of the complex case. 

As a result we have an effective propagator with a characteristic modification by means of Kronecker symbols in $\mathcal{O}^\ell$.
This is the effect of the combinatorially non-local interactions after projection onto uniform field configurations.
It is the same kind of non-local operator as found for the connected two-point correlation function using Landau-Ginzburg mean field theory~\cite{Pithis:2018bw}.

The FRG equation for TGFT in the cyclic-melonic approximation has striking similarities to the FRG equation of $\textrm{O}(N)$ model in the local-potential approximation.
The FRG equation of the dimensionful effective potential for the $\textrm{O}(N)$ model on $d$-dimensional Euclidean space 
\cite{Wetterich:1991be,Berges:2002ga,Delamotte} reads
\begin{equation}\label{FRGON}
    k\partial_k U_k(\rho)=\frac{1}{2}\int\frac{\text{d}^dq}{(2\pi)^d} \left(\frac{\left(N-1\right)k\partial_k\mathcal{R}_k(q)}{Z_k q^2+\mathcal{R}_k(q)+U'_k(\rho)}+\frac{k\partial_k\mathcal{R}_k(q)}{Z_k q^2+\mathcal{R}_k(q)+U'_k(\rho)+2\rho U''_k(\rho)}\right)
\end{equation}
where $q$ is the $d$-dimensional momentum. 
This flow equation implies the existence of Goldstone bosons in the $\textrm{O}(N)$ model (with $N \geq 2$)~\cite{Berges:2002ga,Kopietz}. As well known, for negative mass term $\m$ 
there is a spontaneous breaking of the global $\textrm{O}(N)$ symmetry of the theory down to its $\textrm{O}(N-1)$ subgroup~\cite{Nair:2005iw}. Correspondingly, the term proportional to $N-1$ accounts for the physics of the Goldstone bosons dominating in the low temperature phase of the theory while the second term represents the contribution of the massive radial mode~\cite{Wetterich:1991be,Wetterich:1993im,Berges:2002ga}. 
For $N=2$ and with $\textrm{O}(2)\cong \textrm{U}(1)$, the comparison allows us to identify the first term in Eq.~\eqref{eq:Wetterichprojected} as the contribution stemming from a GFT \enquote{Goldstone} mode and the second from a \enquote{radial} mode. This attribution of modes is only possible when the $2 \times 2$ structure of the Hessian in Eq.~\eqref{Hessian2x2} is considered and thus has so far been overseen in the literature. Their physical interpretation deserves to be better understood and is left to future investigations. 

To herald the next subsection, we would like to note that the expression for the FRG equation~\eqref{FRGON} for the $\text{O}(1)$ and $\text{O}(2)$ models could be retained by replacing sums with integrals in Eqs.~\eqref{eq:Wetterichprojectedreal} and~\eqref{eq:Wetterichprojected}, respectively, and by setting the Kronecker deltas therein to unity. In fact, the latter would eliminate all footprints of the non-locality of the GFT interactions. In the following, we will explore the relation of our model with the $\mathrm{O}(N)$ model on Euclidean space by explicitly computing the FRG equations for the concrete setting where $G=\mathrm{U}(1)$ and by studying their behaviour in the small- and large-scale limits, in particular.

\subsection{FRG equation for Abelian group with generalized propagator}\label{Subsection:beta-functions}

For an explicit summation of the trace in the functional RG equations one has to specify a group $G$.
All compact Lie groups have discrete spectrum but they differ in the multiplicities of the eigenvalues, that is the dimension $d_\rep$ of the representation labelled by $\rep$.
For the rest of this work we consider $G = \textrm{U}(1)$ for which all representations $\rep \in \mathbb{Z}$ have $d_\rep=1$. 
As the spectrum given by the Casimir $\cas_\rep = j^2$ only depends on the absolute value, one can alternatively understand it as a spectrum $\rep\in\mathbb{N}$ with $d_\rep=2$ except for the single zero mode $\rep=0$.
\renewcommand{\cas}{C^{(\zeta)}}
For a more systematic understanding of the renormalization group flow it will be useful to generalize the $\textrm{U}(1)$ Casimir to
\[\label{eq:generalizedkinetic}
\frac{1}{\cs^2}C_\vrep  \mapsto
\frac{1}{\cs^\ks}C^{(\zeta)}_\vrep  := \frac{1}{\cs^\ks} \sum_{\ell = 1}^\rk |\rep_\ell|^\ks
\]
for $0 < \zeta < 1$. 
Accordingly, the mass term has canonical (``mass'') dimension $[\m]=\ks$.
The renormalization and scaling properties of field theories on $\textrm{U}(1)$ with tensorial interactions are already well understood~\cite{BenGeloun:2014gp}.
Besides the actual Casimir, $\zeta = 1$, we will in particular also consider $\zeta = 1/2$.
These choices are also known as short-range and long-range theories~\cite{Fisher:1972}.

As a regulator function \eqref{eq:regularizedkinetic} which satisfies the usual properties 
we choose the optimized regulator \cite{Litim:2001ek}
\begin{align}
\reg &= \wfr \left(k^\ks - \cs^{-\ks}\cas_\vrep \right) \theta\left(k^\ks - \cs^{-\ks}\cas_\vrep \right)
\end{align}
where $\theta$ is the Heaviside function. 
The scale derivative of this regulator is
\begin{align}
k\partial_k\reg &= \left(\ks k^\ks \wfr + \partial_k\wfr\left(k^\ks - \cs^{-\ks}\cas_\vrep \right)\right) \theta\left({(ak)^\ks} - \cas_\vrep \right)\nonumber
\\
&=  k^\ks \wfr \left(\ks - {\eta_k} \left(1- {(ak)^{-\ks}}{\cas_\vrep}\right) \right) \theta\left(1- {(ak)^{-\ks}}{\cas_\vrep} \right)
\end{align}
where the second line spells it out in terms of the anomalous dimension 
\[
\eta_k = - k\partial_k \log Z_k \, .
\]
Note that this regulator is only consistent for 
$\eta_k < \ks$ 
due to the regulator condition $\lim_{k\to\Lambda}\reg=\infty$ \cite{Meibohm:2016gt, Eichhorn:2019dt}.
With this regulator the FRG equation~\eqref{eq:Wetterichprojected} becomes
\[\label{eq:WetterichLitim}
k \partial_k \pot(\rho) =  \wfr k^\ks  \left(\prod_{c=1}^\rk \sum_{\rep_c\in\mathbb{Z}} \right) \sum_{\epsilon=0,1}
\frac{\zeta - \frac{\eta_k}{2} \left(1 - \frac{\sum_{c} |\rep_c|^\ks}{(ak)^\ks}\right)
\theta\left({(ak)^\ks} - \sum_{c} |\rep_c|^\ks \right)}
{\wfr k^\ks+ \m +   \sum_\ell \mathcal{O}_\vrep^\ell {V^\ell_k}'(\rho)+ \epsilon \left(\prod_c \delta_{0\rep_c}\right) 2\rho\sum_\ell {V^\ell_k}''(\rho) }
\]
{where the $\epsilon=0$ term is absent for a real field.}
Thus, the trace is a sum over integers $\rep_\ell$ up to a dimensionless cutoff 
\[
\ak := a k \,.
\] 
{In this sense there is a precise relation between the FRG of TGFT with a dimensionful scale $k$ and the FRG of tensor models where there are no a priori scales and the dimensionless tensor size $\ak$ is usually used to parametrize the renormalization group flow~\cite{Eichhorn:2013jx,Eichhorn:2014cu,Eichhorn:2017wh,Eichhorn:2019kz,Eichhorn:2019dt,Eichhorn:2019hsa}. 
We see here that such a dimensionless cutoff $\ak$ is a natural consequence of the compactness of the Lie group.
In particular, we note that in the field theoretic context it is not necessary to fix the FRG setting by consistency conditions; in TGFT, details like the $(\cs k)^\ks$ scaling of the regulator naturally follow from the definition of the field theory. 
} 

\subsubsection{Full non-autonomous equations}\label{subsubsection:fullnonautonomousequations}

At this point we add one more simplification:
We identify couplings of different colour $\ell = 1,...,\rk$ at each order $\cbk = \lambda_{n,k}/\rk$, and consequently potential functions $V_k^\ell = V_k/\rk$ such that the effective potential \eqref{eq:effectivepotential} is 
\[\label{eq:isopotential}
\pot = \m\rho+V_k(\rho) = \m\rho + \sum_{n=2}^\infty \frac{1}{n!}\lambda_{n,k} \rho^n  \,.
\]
As shown in detail in appendix App.~\ref{sec:Traces}, we then find for the momentum sum over representations
\begin{align}
{k \partial_k \pot(\rho)}
= \left(\zeta-\frac{\eta_k}{2}\right) {k^\ks \wfr} &\bigg(
\frac{1}{k^\ks \wfr+ \pot'(\rho) + 2\rho\,\pot''(\rho)}
\label{eq:Wetterichresult} \\
&+\frac{\cct + \rk\ccu \ku{1}(\ak)}{k^\ks \wfr+ \pot'(\rho)}
+\ccu\sum_{s=2}^\rk  \binom{\rk}{s}\frac{\ku{s}(\ak)}{k^\ks \wfr+ \self{s}(\rho)}\bigg)
\nonumber\\
+\frac{\eta_k}{2} {k^\ks \wfr} &\frac1 {\ak^\ks}\bigg(
\frac{\rk  \ccu \kw{1}(\ak)}{k^\ks \wfr + \pot'(\rho)} +
{ \ccu} \sum_{s=2}^{\rk} \binom{\rk}{s}\frac{s \kw{s}(\ak)}{k^\ks \wfr + \self{s}(\rho)}\bigg)
\nonumber
\end{align}
where $\ccu=1$ for the case of a real field and $\ccu=2$ for complex field.
At each order $s$ in the scale $\ak=a k$ there is an effective mass 
\[
\self{s}(\rho) := \m + \frac{\rk-s}{\rk} V_k'(\rho) 
\]
given by the derivative of the effective potential $\pot$ modified by a factor in all couplings but $\m$.
In lowest order this simply is $\self{0}=\pot'$.
The cutoff dependence is captured by threshold functions 
\[\label{eq:definitionthresholdfunctions}
I^{(s)}_\gamma ({\ak}) = \left(\prod_{c=1}^\rk \sum_{\rep_c\in\mathbb{Z}\setminus\{0\}} \right) 
|\rep_\ell|^\gamma \, \theta\left({\ak^\ks} - \sum_{c} |\rep_c|^\ks \right) \, .
\]
These functions have exact closed results only in specific cases. 
In particular, the standard case $\zeta=1$, that is the sum over lattice points in an $s$-dimensional ball without zeros, can only be approximated.
We use three summation schemes and find that they yield the same qualitative results.
The three schemes are
\begin{itemize}
    \item ``box'' scheme: approximation by a sum over the hypercube%
    \footnote{The box cutoff would follow from a regulator factorizing in the directions $\ell$, that is with $\prod_{c}\theta\left({\ak^\ks} -  |\rep_c|^\ks \right)$ instead of $\theta\left({\ak^\ks} - \sum_{c} |\rep_c|^\ks \right)$. But such a regulator would violate the positivity condition $\reg\ge0$. This might be cured by a regulator factorizing not only in the Heaviside function but completely,
\[
\reg = \wfr \cs^{-\ks} \prod_{\ell=1}^\rk \left(\ak^\ks - |\rep_\ell|^\ks \right) \theta\left(\ak^\ks - |\rep_\ell|^\ks \right) \, \nonumber .
\]
However, this propagator does not lead to the simplification of the inverse propagator 
$|\rep_\ell|^\ks \mapsto k^\ks$ 
in the trace which is crucial to evaluate the sum explicitly.
},
    that is $\ks\rightarrow\infty$,
    \item ``integral'' scheme: approximation of discrete sums by their integral counterparts which is sufficient in the $\ak\rightarrow\infty$ limit,
    \item ``simplex'' scheme: for $\zeta=1/2$ we can evaluate the trace exactly. We also use the resulting simplex sum as a lower bound for the other cases. This is important to control the effect of contributions of lower order in $\ak$.
\end{itemize}
For these cases we have (details are elaborated on in App.~\ref{sec:Traces})
\[\label{eq:I0versions}
\ku{s}(\ak) = \Bigg\{\begin{array}{ll}
        (2\ak)^s & \quad\text{   box} \\
        \vks{s}\ak^s & \quad\text{   integral, } \vks{s}:= \frac{\left(2\Gamma\left(1+1/{\ks}\right)\right)^s}{\Gamma\left(1+{s}/{\ks}\right)}\\
        \frac{2^s}{s!}\frac{\Gamma(\ak+1)}{\Gamma(\ak+1-s)} & \quad\text{   simplex}
            \end{array}
\]
which is simply the $s$-dimensional volume, in particular $\vks{s}$ is the volume of the unit ball in $L^{\ks}$ norm.
For the trace over the Laplacian we have
\[\label{eq:I2versions}
\kw{s}(\ak) = \Bigg\{\begin{array}{ll}
        2 H_{\ak}^{(-\ks)} (2\ak)^{s-1} & \text{ box} \\
        \frac{\vks{s}}{s+\ks}\ak^{s+\ks} & \text{ integral}\\
        \frac{2^s
        }{(s+1)!}\prod_{i=0}^s(\ak+i) & \text{ simplex, } \zeta = 1/2 \\ 
        \frac{2^s
        }{(s+2)!}(2\ak+s)\prod_{i=0}^s(\ak+i) & \text{ simplex, } \zeta = 1
        \end{array}
\]
where $H_n^{(m)}$ is the $n$'th generalized harmonic number of order $m$. 
For the simplex cutoff, the case of arbitrary $\zeta$ is more involved but for the analysis below we only need the cases shown.


As the threshold functions exemplify in detail, the full FRG equation is a non-auto\-nomous ordinary differential equation in the scale $k$.
This is a well known feature of TGFT on compact Lie groups~\cite{Benedetti:2015et, BenGeloun:2015ej,BenGeloun:2016kw,TGFT1,TGFT2,TGFT3}.
From our derivation one can see that the specific non-autonomous form is a result of the interplay of compact domain and non-local combinatorics. 
The specific non-locality of interactions amounts to various combinations of delta functions in the Hessian encoded in the operator $\mathcal{O}^\ell_\vrep$, Eq.~\eqref{eq:deltaoperator}; as a consequence, the trace amounts to summation over sub-spaces of the $\rk$-dimensional space of various dimension $0\le s \le \rk$.
In particular, terms in $\mathcal{O}^\ell_\vrep$ with a single $\delta_{j_\ell,0}$ are the only ones contributing beyond linear order in $\ak$ (as can be seen in detail in Eq.~\eqref{eq:trace}).
From a perturbative perspective (e.g.~the $\pr^{-1}F$-expansion), these terms result from diagrams with a maximal number of faces.
In this way, the usual dominance of melonic diagrams in tensorial theories \cite{GurauBook} occurs in the FRG equation.

\subsubsection{Flow equations at all orders}\label{subsubsection:flowequationsall}

We will now derive flow equations for the couplings at any order in terms of an expansion in the effective average field $\rho$.
Due to the lack of autonomy as well as the effective potential appearing with various multiplicities $(\rk-s)/\rk$, 
directly solving the full FRG equation Eq.~\eqref{eq:Wetterichresult} for the effective potential is a hard problem. 
Instead, we expand left- and right-hand side of the equation around $\rho=0$ and compare at each order in $\rho^n$ which yields flow equations for each individual coupling $\lambda_{n,k}$.

\renewcommand{\m}{\mu}
To start with, the left-hand side of the equation is a formal power series in the projected average field $\rho$ by definition \eqref{eq:isopotential},
\[\label{eq:lhsflowequation}
k \partial_k  \pot(\rho) 
= k\partial_k \m \,\rho + \sum_{n=2}^\infty \frac{1}{n!} k \partial_k\cn{n}\,\rho^n \, 
\]
where from now on we drop the subscript $k$ for the sake of readability, i.e.~all couplings  $\m=\mu_k$ and $\cn{n}=\lambda_{n,k}$ are always understood as evaluated at $k$.
For the right-hand side we need the Taylor series around zero for a function $f$ of its fraction
\begin{align}
\frac1{f(\rho)} = \sum_{n=0}^\infty \frac{1}{n!}{c^{n}\left( 
\{f^{(i)}(0)\}_{0\le i\le n} \right)} \rho^n
=\frac{1}{f(0)}
-\frac{f'(0)}{f(0)^2}  \rho
+\frac{2 f'(0)^2-f(0) f''(0)}{2 f(0)^3}\rho ^2
+..., 
\end{align}
where the expansion coefficients $c^{n}$ at order $n$ depend on the $i$'th derivatives $f_i=f^{(i)}(0)$ up to $i=n$.
These coefficients can be expanded in $f_0$ as
\[
c^{n} = \sum_{l=1}^n (-1)^l \frac{l!}{f_0^{l+1}} c^n_l
\]
where coefficients $c^n_l$ are sums over products $\prod_{j=1}^n f_j^{s_j}$ with $\sum s_j = l$.

In the full FRG equation Eq.~\eqref{eq:Wetterichresult} there is a sum over two types of fractions with
\[
f_1(\rho) = \wfr k^\ks + 
\m+\frac{\rk-s}{\rk}{V}_k'(\rho)
\quad
\textrm{ and }
\quad
f_2(\rho) = \wfr k^\ks + 
\m+{V}_k'(\rho)+ 2{\rho}{V}_k''(\rho) \, .
\]
The $i$'th derivative of the derivative of the potential at $\rho=0$ is $V^{(i+1)}(0)\equiv f_1^{(i)}(0)=\cn{i+1}$ 
such that
\begin{align}
\frac1{f_1(\rho)} &= \sum_{n=0}^\infty \frac{1}{n!}c^{n}\left(
\wfr k^{\ks} +\m,\frac{\rk-s}{\rk}\cn2,...,\frac{\rk-s}{\rk}\cn{n+1} 
\right) \rho^n \\
&\equiv \sum_{n=0}^\infty \frac{1}{n!}\bco{n}\left(\m,\frac{\rk-s}{\rk}\cn{i}\right) \rho^n \, ,
\end{align}
where we denote the order-$n$ coefficients $\bco{n}$.
These coefficients expand in the couplings
\[\label{eq:betaexpansion}
\bco{n}(\m,\frac{\rk-s}{\rk}\cn{i}) = \sum_{l=1}^n \left(\frac{\rk-s}{\rk}\right)^l \bco{n}_l(\m,\cn{i})
\quad , \quad \bco{n}_l(\cn{i}) = \frac{(-1)^l l!}{(\wfr k^\ks+\m)^{l+1}} c^n_l(\cn{2},...,\cn{n+1}) \, .
\]
For the expansion of $f_2$, we have $(\rho V'')^{(i)}(0) = i\cn{i+1}$ with the consequence that couplings $\cn{i}$ always appear with a factor $1+2(i-1)=2i-1$ such that
\begin{align}
\frac1{f_2(\rho)} &= \sum_{n=0}^\infty \frac{1}{n!}c^{(n)}\left(
\wfr k^{\ks}+\m,3\cn2,5\cn{3}...,(2n+1)\cn{n+1} 
\right) \rho^n\\
&\equiv \sum_{n=0}^\infty \frac{1}{n!}\bcb{n}(\m,\cn{i}) \rho^n
\end{align}
denoting 
these coefficient functions $\bcb{n}(\cn{i})$ to include the multiplicities $2i-1$ in each coupling argument. 
We note that for each term $\prod_i \cn{i}^{s_i}$ in the numerator of the coefficients $\bco{n}$ and $\bcb{n}$ one has 
\[\label{eq:betagrading}
\sum_{i=1}^{n+1} s_i = n \quad \textrm{and} \quad \sum_{i=1}^{n+1} i\cdot s_i = 2 n .
\]


Comparing now left- and right-hand side of the $\eta_k$-independent part of the full FRG equations, we find
\begin{align}
{k\partial_k\cn{n}}  = {\zeta\wfr k^\ks} \bigg[ \bcb{n}(\m,\cn{i})
&+ ({\cct + 2 \ccu \rk \ak})\bco{n}(\m,\cn{i}) \\
&+\ccu\sum_{s=2}^\rk  \binom{\rk}{s}\ku{s}(\ak) \bco{n}\left(\m,\frac{\rk-s}{\rk}\cn{i}\right) \bigg] \, . \nonumber
\end{align}
Because of the $s$ dependence inside the $\bco{n}$ coefficients, they do not factor from the dependence on the scale $\ak$.
Only when expanding these coefficients in the power of couplings using \eqref{eq:betaexpansion},
\[ \boxd{
k\partial_k\cn{n} = \zeta\wfr k^\ks  \bigg[\bcb{n}(\m,\cn{i})  
+\ccu \sum_{l=1}^n F_\rk^l(\ak) \bco{n}_l(\m,\cn{i}) \bigg]
}\,,
\]
the non-autonomous part at order $l$
\[
F_\rk^l(\ak) := {\frac{\cct}{\ccu} + 2\rk \ak} + \frac{1}{\rk^l}\sum_{s=2}^\rk \binom{\rk}{s}\left(\rk-s\right)^l \ku{s}(\ak)
\]
factorizes from the coefficients $\bco{n}_l$.
For example, the flow equation at quadratic order are only linear in the couplings such that
\[
{k\partial_k \m}  = \zeta \wfr k^\ks \left(3+\ccu F_\rk^1(\ak)\right)\frac{-\cn{2}}{(\wfr k^\ks + \m)^2} \, .
\]
These are the flow equations order by order in the cyclic-melonic LPA.

\subsubsection{Anomalous dimension}\label{subsubsection:anomalousdimension}

Adding the second part in the full FRG equation~\eqref{eq:Wetterichresult} depending on the anomalous dimension $\eta_k$, we get from the series expansion in $\rho$ 
\begin{align}
{k\partial_k\cn{n}}  = \frac{1}{2}\wfr {k^\ks} &\bigg[
\left(\ks-\eta_k\right) 
\left(\bcb{n}(\m,\cn{i}) + (\cct)\bco{n}(\m,\cn{i}) \right) \\
& + \ccu \rk \left( \left(\ks-\eta_k\right)  2\ak  
+\eta_k \ak^{-\ks} \kw{1}(\ak) \right) \bco{n}(\m,\cn{i}) \nonumber\\
&+\ccu\sum_{s=2}^\rk  \binom{\rk}{s}\left( \left(\ks-{\eta_k}\right) \ku{s}(\ak) + s \eta_k \ak^{-\ks}{\kw{s}(\ak)} \right)
\bco{n}\left(\m,\frac{\rk-s}{\rk}\cn{i}\right) \bigg] \nonumber
\end{align}
which we can again expand in powers of couplings $\cn{i}$ as
\[\label{eq:betaequations}\boxd{
k\partial_k\cn{n} = \wfr k^\ks  \bigg[\left(\zeta-\frac{\eta_k}{2}\right) \bcb{n}(\m,\cn{i})  
+\ccu \sum_{l=1}^n \left(\zeta F_\rk^l(\ak) - \frac{\eta_k}{2}\gr^l(\ak) \right) \bco{n}_l(\m,\cn{i}) \bigg]
}
\]
where the non-autonomous functions for the $\eta_k$ term are 
\begin{align}
\gr^l(\ak) &:= F_\rk^l(\ak) - \rk\ak^{-\ks}\kw{1}(\ak) - \frac{1}{\rk^l}\sum_{s=2}^{\rk} \binom{\rk}{s}(\rk-s)^l s\,\ak^{-\ks} \kw{s}(\ak) \, .
\end{align}
For example, at order $n=1$ we have
\[\label{eq:massflow}
{k\partial_k \m}  = \wfr k^\ks
\left( \zeta(3+\ccu F_\rk^1(\ak)) -\frac{\eta_k}{2}(3+\ccu \gr^1(\ak)) \right)
\frac{- \cn{2}}{(\wfr k^\ks + \m)^2} \, .
\]
In this way, we have included the anomalous dimension in the flow equations of the effective potential and its coupling coefficients.

\

Though we have used a constant field projection it is nevertheless of interest to consider the flow of the wave function renormalization $\wfr$, and thus of the anomalous dimension $\eta_k$, too.
In particular, for TGFT it is known~\cite{Benedetti:2015et} that the anomalous dimension can be relatively large due to pro\-pa\-gating internal momenta already at one loop, that is at quadratic order in the field in the FRG equation.
Thus, as the wave function renormalization $\wfr$ is the parameter associated with the Laplacian in the effective average action $\gfk$, Eq.~\eqref{eq:tensorialWetterichaction}, it is necessary to go beyond the local-potential approximation (LPA) to obtain a flow equation for $\eta_k = - k\partial_k \log\wfr$.
As standard, we refer to this extended local-potential approximation as LPA$'$.
For this, we simply complement our flow equations for the effective potential of the projected average field $\rho$ by equations with the full average field $\gf$ at quadratic order $\gfb\gf$ and up to order $\ks$ in derivatives. 
This yields the exact flow equations for the mass and anomalous dimension.

For the derivative expansion it is essential to evaluate the trace over regulated functions properly.
Since the dependence of the result on momenta enters via the regulator, the expansion is only meaningful when the trace over representation space $\hat{G}^{\times\rk}$ is summed over a ball in $l_{\ks}$ norm for a kinetic term with $|\rep_\ell|^\ks$, Eq.~\eqref{eq:generalizedkinetic}.
We find for the anomalous dimension
\[\label{eq:etaflow} \boxd{
\eta_k
= -\cn{2}\frac{2(\rk-1)\ak + {\sum\limits_{s=1}^{\rk-1} \binom{\rk-1}{s} s\,\vks{s}\ak^{s}}}
{\frac{2\rk}{\ccu}(\wfr k^{\ks}+ \m)^2 
- \cn{2} \left( \rk +
2(\rk-1)\ak  + {\sum\limits_{s=1}^{\rk-2} \binom{\rk-1}{s} \frac{s + {\ks} }{\ks}\vks{s}\ak^{s}}
+\vks{\rk-1}\ak^{\rk-1}
\right)} }
\]
for which all details can be found in App. \ref{sec:etaequation}.
As a by-product we find for the  flow of $\m$
\[
k\partial_k \m = -\wfr k^{\ks} \cn{2} \ccu \bigg(\frac{\ks-\eta_k}{2}
\frac{2 + 2\ak + \sum
\binom{\rk-1}{s} \vks{s} \ak^{s}
}{(\wfr k^{\ks}+ \m)^2} 
+ \frac{\eta_k}{2} \frac{
{\frac{2}{1+\ks}\ak}
{+\frac{(\rk-1)\vks{\rk-1}}{\rk-1+\ks}\ak^{\rk-1}}
}{(\wfr k^{\ks}+ \m)^2}  
\bigg)
\]
which is in surprisingly good agreement with the result of the constant field projection, Eq.~\eqref{eq:massflow}.
The first term is exactly the same for $\ccu=2$ and differs only by a constant term 2 versus 3 in the numerator for $\ccu=1$. 
The second term is exactly the same both at leading order $\ak^{\rk-1}$ and at lowest order $\ak$; the difference is only that the projection results in additional terms at intermediate orders.
We take this surprisingly good agreement as further evidence that the constant field projection is a meaningful approximation.

With the set of equations \eqref{eq:betaequations}, \eqref{eq:massflow} and \eqref{eq:etaflow} we have derived the full content of the FRG equation for a real or complex tensorial group field on $\text{U}(1)^{\times\rk}$ in the cyclic melonic LPA$'$ at any scale $k$.
In the following, we will explore the resulting phase structure.

\section{Results: Phase structure of the cyclic-melonic theory space}\label{sec:results}

Due to non-autonomy of the FRG equation, a standard analysis of the phase structure of TGFT in the cyclic-melonic potential approximation is only feasible in specific regimes, in particular in the large-$\ak$ and the small-$\ak$ regime.
In these regimes we find precise relations of the theory to $\textrm{O}(N)$ models in $\uvd=\rk-1$ dimensions (at large $\ak$) and effectively zero dimension (at small $\ak$).
At large $\ak$, however, the tensor-specific flow of the anomalous dimension $\eta_k$ modifies the phase structure in a significant way, in particular changing the details of the Wilson-Fisher type non-Gaussian fixed point.
For intermediate regimes we find that the equations keep the resemblance to $\textrm{O}(N)$ models, but with an effective scale-dependent dimension $\efd(k)$ flowing between the asymptotic values.

In the following, we first discuss the asymptotic large-$\ak$ regime. Then we address the intermediate regimes and the issue of symmetry restoration in the full theory.

\subsection{The large-$\ak$ regime}

From a physical point of view, the large-$\ak$ limit can be seen in two ways.
On a compact group $G$ with fixed volume scale $\cs$ it is equivalent to the large-$k$ limit since $\ak = \cs k$.
From this perspective, it is still necessary to understand the flow under the full non-autonomous FRG equations.
Complementary, the large-$\ak$ limit is also the large-volume limit 
corresponding to the TGFT on $\mathbb{R}^\rk$ with a thermodynamic limit removing the IR regularization~\cite{BenGeloun:2015ej,BenGeloun:2016kw}. 
Indeed, our equations will agree with those results~\cite{BenGeloun:2015ej,BenGeloun:2016kw} in quartic truncation but generalize them to arbitrary order.
In this way one can also view the results of this section as a full description of the phase space of the  theory in a non-compact limit.

To find fixed points of the renormalization group flow in phase space it is necessary to rescale the couplings.
In the following we will first show how the rescaling which is natural from the point of view of renormalization leads to autonomous FRG equations at large $\ak$.
We will then calculate and discuss the fixed point structure, first with zero anomalous dimension (LPA) and then taking into account the flow of the anomalous dimension (LPA$'$) as well.

\subsubsection{Rescaled flow equations}

The scaling behaviour of couplings is special in field theories with tensorial interaction \cite{BenGeloun:2014gp,BenGeloun:2016kw}. We provide a systematic discussion of scaling dimensions from the renormalization perspective in Appendix~\ref{sec:Dimensions}.
The important lesson from that discussion is that the scaling dimension differs from the canonical dimension  which is $\ks n$ for a melonic $(\gf\gfb)^n$ interaction.
It only depends on the scaling power $\ks$ of the kinetic part $\kin \sim k^{\ks}$, very much in contrast to standard, combinatorially local field theories.
On the other hand, very similar to combinatorially local field theories, the \emph{scaling} dimension of such interaction is
\[\label{eq:scalingdimension}
\scd{n} = \ks n - \gd(\rk-1)(n-1)
\equiv \uvd - (\uvd-\ks)n   
\, . 
\]
That is, the scaling dimension in TGFT has the usual form but with a special dimension
\[
\uvd := \gd(\rk-1)
\]
depending on the rank $\rk$ of the tensor field and the dimension $\gd$ of the Lie group $G$ (that is $\gd=1$ here for $G=\textrm{U}(1)$).
We will call $\uvd$ the \emph{UV dimension} of a tensorial group field theory since the scaling and power counting is the same as for a standard QFT with that dimension.
 
The difference between canonical and scaling dimension is another sign for the necessity of the second scale $\cs$ in the theory.
To extract scale-free information in the renormalization group flow it is necessary to rescale with  $k$ in powers of the scaling dimension
\[
\cn{n} = \wfr^{n} k^{\scd{n}} \hat{\cn{}}_n
\]
but the rescaled coupling $\hat{\cn{}}_n$ still has dimension due to the difference of $\scd{n}$ to the canonical dimension.
This is fixed by rescaling also in $\cs$
\[
\hat{\cn{}}_n 
= \cs^{(1-n)\uvd} \cnr{n}
\]
such that
\[\label{eq:UVrescaling}
\cn{n} = \wfr^{n} k^{\scd{n}} \cs^{(1-n)\uvd} \cnr{n} 
\]
and in particular $\m =  {\wfr k^\ks} \mr$.
With this rescaling the flow equations \eqref{eq:betaequations} become
\begin{align}\label{eq:betaUV}
k\partial_k \cnr{n} +& \uvd\cnr{n} - n(\uvd -\ks + \eta_k)  \cnr{n}  \\
&=  (\cs k)^{-\uvd} \sum_{l=1}^n \left[ 
\left(\zeta-\frac{\eta_k}{2}\right)   \bcb{n}_l(\mr,\cnr{i})
+  \ccu\left( \zeta F_\rk^l(ak) - \frac{\eta_k}{2}\gr^l(ak) \right) \bco{n}_l(\mr,\cnr{i}) \right] \nonumber
\end{align}
after dividing both sides by the rescaling $\wfr^n k^{\scd{n}} \cs^{(1-n) \uvd}$. 
This is because according to Eq.~\eqref{eq:betagrading} the rescaled coefficients $\bcb{n}$ and $\bco{n}$ scale in $k$ with power
\begin{align}
-\ks(n+1) + \sum_i s_i \scd{i} 
&=-\ks(n+1) + n\uvd - 2n(\uvd - \ks) \\
&= \ks(n-1) - n\uvd
= \scd{n}  - \uvd - \ks
\end{align}
and there is the additional factor $\wfr k^\ks$ on the right-hand side due to $k\partial_k\reg$.
The scaling in the external scale $\cs$ is simply
\[
\sum_i s_i (1-i)\uvd
=  -n \uvd.
\]
In this way, the flow equations depend now only on the combination $\ak=\cs k$. Accordingly, it is equivalent to take the UV limit $k\to\infty$ or the large-volume limit $\cs\to\infty$.

In this limit, only the leading-order contributions of the $\ak$-dependent functions $F_\rk^l$ and $\gr^l$ survive. 
As a result we have the flow equations
\begin{eqnarray}
\eta_k &=& c_{\uvd}\frac{\uvd-\eta_k}{\ks\rk}\frac{-\cnr{2}}{(1+ \mr)^2} \\
\left[k\partial_k + \ks -  \eta_k \right] \mr
&=& c_{\uvd} \left(1 -\frac{\eta_k}{\uvd+\ks} \right) \frac{-\cnr{2}}{(1+\mr)^2}\\
\left[k\partial_k + \uvd - (\uvd - \ks  + \eta_k )n\right] \cnr{n}
&=& \rk c_{\uvd}  \left(1 -\frac{\eta_k}{\uvd+\ks} \right) \bco{n}(\mr,\cnr{i}/\rk)
\end{eqnarray}
with a constant $c_{\uvd} = \zeta\vks{\uvd}\ccu$. 
This constant $c_{\uvd}$ is actually not important and can be removed from the equations by a rescaling, removing at the same time any distinction between the real- ($\ccu=1$) and complex field case ($\ccu=2$). Furthermore, it is convenient for the large-$\ak$ equations to also rescale the factor $1/\rk$.
Thus, we define for $n\ge2$
\[\label{eq:UVrescaling2}
\cnr{n} \mapsto \cns{n} := c_{\uvd}^{1-n}\frac{\cnr{n}}{\rk}
 = \wfr^{-n} k^{-\scd{n}} \left(\frac{c_{\uvd}}{\cs}\right)^{1-n} \frac{\cnr{n}}{\rk}
\]
while $\ms:=\mr$.
Note that from the perspective of this rescaling, the momentum space volume factor $c_{\uvd}$ can just be seen as modification of the configuration space volume $\cs$.
This rescaling simplifies the large-$\ak$ flow equations to
\begin{eqnarray}
\eta_k &=& \frac{\uvd-\eta_k}{\ks}\frac{-\cns{2}}{(1+ \ms)^2}  \label{eq:etaUV}\\
k\partial_k\ms = \bfuv{1}(\ms,\cns{i})
&:=& \quad\quad \left(- \ks + \eta_k \right) \ms \quad+\quad \rk \left(1 -\frac{\eta_k}{\uvd+\ks} \right) \frac{-\cns{2}}{(1+\ms)^2} \label{eq:massflowUV}\\
k\partial_k \cns{n} = \bfuv{n}(\ms,\cns{2})
&:=& \left(- \uvd + (\uvd - \ks  + \eta_k )n\right) \cns{n}
+ \left(1 -\frac{\eta_k}{\uvd+\ks} \right) \bco{n}(\ms,\cns{i}) \label{eq:UVequations}
\end{eqnarray}
for $n\ge2$. 
Alternatively, the flow equation of the anomalous dimension solved for $\eta_k$ is
\[
\eta_k = - \frac{\uvd \cns{2}}{\ks(1+\ms)^2 - \cns{2} } \, .
\]
These large-$\ak$ equations are consistent with earlier results in quartic ($\nmax=2$) truncation%
\footnote{In Ref.~\cite{Benedetti:2015et}, Eq.~(4.5--4.7), the same equations are found for rank $\rk=3$, real field $\ccu=1$ and linear kinetic term $\zeta=1/2$ setting $\cs=3$.
In \cite{TGFT1}, Eqs.~(95--97), flow equations are derived for arbitrary rank $\rk$ and quadratic kinetic term $\zeta=1$ with closure constraint which results in an effective dimension $\uvd=\rk-2$ (see App.~\ref{sec:Dimensions}). Their calculations lead to a different constant $c_{\uvd}=4\vol{\rk}/(\rk\sqrt{\rk-1})$ which is again slightly different in the flow equation for $\eta_k$ being $c_{\uvd} = 4\vol{\rk}/(\rk-1)^{3/2}$.
Furthermore, in Ref.~\cite{BenGeloun:2016kw}, Eq.~(56), a complex theory of arbitrary rank $\rk$ with quadratic kinetic term $\zeta=1$ is considered ignoring the $2\times2$ structure of $\gfk^{(2)}$, thus effectively with $\ccu=1$ in the equations. 
There, the trace sums are evaluated as integrals in a thermodynamic limit $\cs\rightarrow\infty$ which results in non-autonomous functions with only $(ak)^{\uvd}$ and linear $ak$ contribution. 
On the grounds of the $\cs\rightarrow\infty$ limit, the authors consider then only the $(ak)^{\uvd}$ part.
Thus, our framework explains in particular why these results at large $\cs$  \cite{BenGeloun:2015ej,BenGeloun:2016kw} agree with the UV results in \cite{TGFT1}.
}.
The agreement with this literature allows us to improve on the claims on the fixed point structure made there from the perspective of our cyclic-melonic approximation.

\

Our large-$\ak$ flow equations \eqref{eq:UVequations} are exactly the same as those of $\uvd$-dimensional $\textrm{O}(N)$ models in the large-$N$ limit. 
Thereby, the scale $\ak = \cs k$ has nothing to do with the number of field components $N$ of the $\textrm{O}(N)$-symmetric scalar field. The tensor field here still has $\ccu=1$ or $\ccu=2$ components.
The equivalence is due to the fact that in the FRG equation \eqref{eq:Wetterichresult} the leading-order term in $\ak$ does not depend on a second derivative $V_k''$ of the potential $V_k$.
Thus, both the parts in Eq.~\eqref{eq:Wetterichprojected} related to the \enquote{radial} mode and the \enquote{Goldstone} mode have leading order contributions of the form of the Goldstone term.
Importantly, the reduction of the dimension $\rk$ of the generated combinatorial pseudo manifolds to the dimension $\uvd=\rk-1$ is due to the fact in the melonic diagrams there is a maximal number of $\rk-1$ faces per melon which accordingly contributes $\rk-1$ propagating degrees of freedom.

This equivalence is modified by an additional factor $\rk$ in the flow equation for $\m$ \eqref{eq:massflowUV}.
That is, there is a relative factor $\rk$ in front of $\m$ between the left- and right-hand side.
The difference to $\textrm{O}(N)$ models becomes more explicit when transforming back from flow equations at any order to the full FRG equation.
The flow equations \eqref{eq:massflowUV} and \eqref{eq:UVequations} are the Taylor expansion of the equation
\[\label{eq:UVequation} \boxd{
\left(k\partial_k + \uvd  - \left(\uvd-\ks+\eta_k\right)\rho\partial_\rho \right) 
\left(\frac{1}{\rk}\ms\rho+\bar{V}(\rho)\right)
= \frac{1-\frac{\eta_k}{\uvd+\ks}}{1 + \ms+ \bar{V}_k'(\rho)}
}
\]
where $
\bar{V}_k(\rho)$ is the rescaled effective potential, that is the the power series \eqref{eq:potentialfunction} with rescaled coefficients $\cns{n}$. 
These equations are exactly the large-$N$ limit of $\uvd$-dimensional $\textrm{O}(N)$ models up to a factor $1/\rk$ in front of $\ms$ on the left-hand side. 
The reason for this relative factor is that in the leading order effective mass $\self{\rk-1}$ in Eq.~\eqref{eq:Wetterichresult} there is a factor $1/\rk$ in front of all the couplings $\cn{i}$ except for $\m$.

Another difference is that in TGFT in the LPA$'$ the flow of the anomalous dimension is special.
The flow of the potential, Eq.~\eqref{eq:UVequation}, has the same dependence on $\eta_k$ as in $\textrm{O}(N)$ models~\cite{Codello_2015}. 
Only, the flow equation of the wave function renormalization is substantially different.
In particular, the minus sign in Eq.~\eqref{eq:etaUV} is absent in standard (combinatorially local) field theories. 
The reason is that there are substantial contributions to the flow of the anomalous dimension of propagating internal momenta $\rep_\ell$ in the tensorial theory.
These are coupled to the external momenta only via the regulator depending on $\ak^\ks-\sum_\ell |\rep_\ell|^\ks$.
Consequently, at  order $\ks$ in the derivative expansion in $\rep_\ell$ there is always a minus sign.

In the following, we firstly show that the $\rk$-factor qualitatively leads to the same results known from $\textrm{O}(N)$ models in the large-$N$ limit in the LPA. 
Then we show how the tensorial anomalous dimension modifies the results in the LPA$'$.

\subsubsection{Phase structure in the LPA} 

In this section we analyse the phase diagram in the LPA, that is neglecting the anomalous dimension. 
The difference to $\textrm{O}(N)$ models consists in a relative factor $\rk$ between $\ms$ and the other couplings $\cns{n}$ in the flow equations.

In general, to explore the phase diagram of the theory, one calculates the fixed points of the renormalization group flow as well as their critical exponents.
Fixed points $(\mc,{\cn{2}}{}_{*},...)$ are those points in the phase diagram where $k\partial_k {\cns{n}}{}_{*}=\bfuv{n}({\cns{i}}{}_{*})=0$ for all couplings (including $\mc$ as $n=1$).
Their stability is determined by the critical or scaling exponents $\theta_i$ which are the eigenvalues of the stability matrix 
$\left(-\partial_{\cns{i}}\bfuv{j}\right)_{ij}$.
Positive eigenvalues are related to IR repulsive, or respectively, UV attractive directions. 
They are UV relevant and correspond to renormalizable couplings.
Explicit calculations are mostly possible only  for a finite set of couplings $\cns{i}$, $i=1,2,...,\nmax$, that is truncating the theory at order $\nmax$.
An exception is the origin of the phase diagram where the path integral is merely a Gaussian measure, thus called \emph{Gaussian} fixed point (GFP). 
There, the scaling exponents are directly given by the scaling dimension, Eq.~\eqref{eq:scalingdimension}, explicitly listed in Tab.~\ref{tab:GPexponents}.

\begin{table}
  \centering
  \caption{\label{tab:GPexponents}%
Scaling exponents at the Gaussian fixed point 
for $\zeta=1$ (left) and $\zeta=1/2$ (right).
}
\vspace*{0.5 cm}
\resizebox{0.95\textwidth}{!}{\begin{minipage}{1.25\textwidth}
\centering
\begin{tabular}{|r|c|c|c|c|c|}
\hline
$\uvd$ & $\theta_1$ & $\theta_2$ & $\theta_3$ & $\theta_4$ & $\theta_5$\\ \hline\hline
2 & 2 & 2 & 2 & 2 & 2\\
3 & 2 & 1 & 0 & -1& -2\\
4 & 2 & 0 & -2& -4& -6\\
5 & 2 &-1 &-4 &-7 &...\\
\hline
\end{tabular}
\hspace*{1.0 cm}
\begin{tabular}{|r|c|c|c|c|c|}
\hline
$\uvd$ & $\theta_1$ & $\theta_2$ & $\theta_3$ & $\theta_4$ & $\theta_5$ \\ \hline\hline
1 & 1 & 1 & 1 & 1 & 1 \\
2 & 1 & 0 &-1 &-2 &-3\\
3 & 1 &-1 &-3 &-5& ...\\
\hline
\end{tabular}
\end{minipage}}
\end{table}

There are two distinguished values of dimension.
The \emph{critical} dimension $\crd$ is the dimension above which the GFP only has a single non-negative scaling exponent.
From Eq.~\eqref{eq:scalingdimension} it is clear that the critical dimension is $\uvd 
= \crd = 4\zeta$.
Correspondingly, the critical rank of the $G=\text{U}(1)$ tensorial theory is $\crr=5$ for a quadratic kinetic term ($\zeta=1$) and $\crr=3$ for a linear one ($\zeta=1/2$).
Above $\crr$ the theory is trivial in the sense that only the non-interacting theory is renormalizable at the GFP.
From the IR perspective, above the critical rank there is a critical surface of IR-attractive directions around the GFP which has co-dimension one. This surface splits the phase space in two distinct regions and the phase transition is captured by the GFP, that is it can be described by mean-field theory.
As a direct consequence, in the large-size limit $\cs\rightarrow\infty$ the tensorial theory in the cyclic-melonic LPA has a phase transition described by mean-field exponents for $\rk>\crr=4\zeta+1$.

The second special case is the dimension below which all GFP scaling exponents are positive. 
Again, from Eq.~\eqref{eq:scalingdimension} we see that this is the case for $\uvd\equiv\rk-1\le\ks$.
In between this dimension and the critical dimension, i.e. $\ks < \uvd < 4\zeta$, 
there is a finite number of relevant couplings at the GFP.
Thus, the theory is asymptotically free in the UV.
From the IR perspective, it is known that $\textrm{O}(N)$ models with dimension $d$ in this range (with $\zeta=1$) have a phase transition described by a non-Gaussian fixed point (NGFP), see e.g. Refs.~\cite{Codello_2013,Codello_2015}, which is related  to the Wilson-Fisher fixed point%
\footnote{For $\textrm{O}(N)$ models in $d<\crd$, the Wilson-Fisher fixed point starts to branch off from the Gaussian fixed point. At $d=3$ these provide a description of the universality classes of the Heisenberg, Ising and XY models, among others~\cite{Pelissetto_2002}. 
}
in the $\epsilon=4-d$ approximation \cite{Wilson:1971dc}.
That is, there is a continuous dependence in $\epsilon$ by which the NGFP connects to the GFP for $\epsilon=0$.
From analytic solutions of $\textrm{O}(N)$ models in the large-$N$ limit \cite{Tetradis_1996,D_Attanasio_1997} it is known that this NGFP has scaling exponents
$\theta_i = d - 2 i$. 
In particular, we find converging exponents also for $\zeta=1/2$ (tested for $d=2.1, 2, 1.9, 1.5, 1.1$)
with values%
\footnote{Solving up to truncation order $\nmax=12$ we find that convergence of scaling exponents is very fast close to the critical dimension (tested cases $d = 2.1, 2, 1.9$) while much slower away from it (tested cases $d=1.5,1.1$).
We calculate fixed point solutions and scaling exponents throughout this article as exact algebraic solutions using computer algebra (Mathematica).}
\[\label{eq:exponentsON}
\theta_i = d - \ks i \quad , \quad i = 1, 2,... \, .
\]
The question is now: How does the result change in the large-$\ak$ regime of the tensorial theory considered here; that is, what is the effect of the factor $1/\rk$ in Eq.~\eqref{eq:UVequation}? 

\begin{figure}
\includegraphics[width=7.5cm]{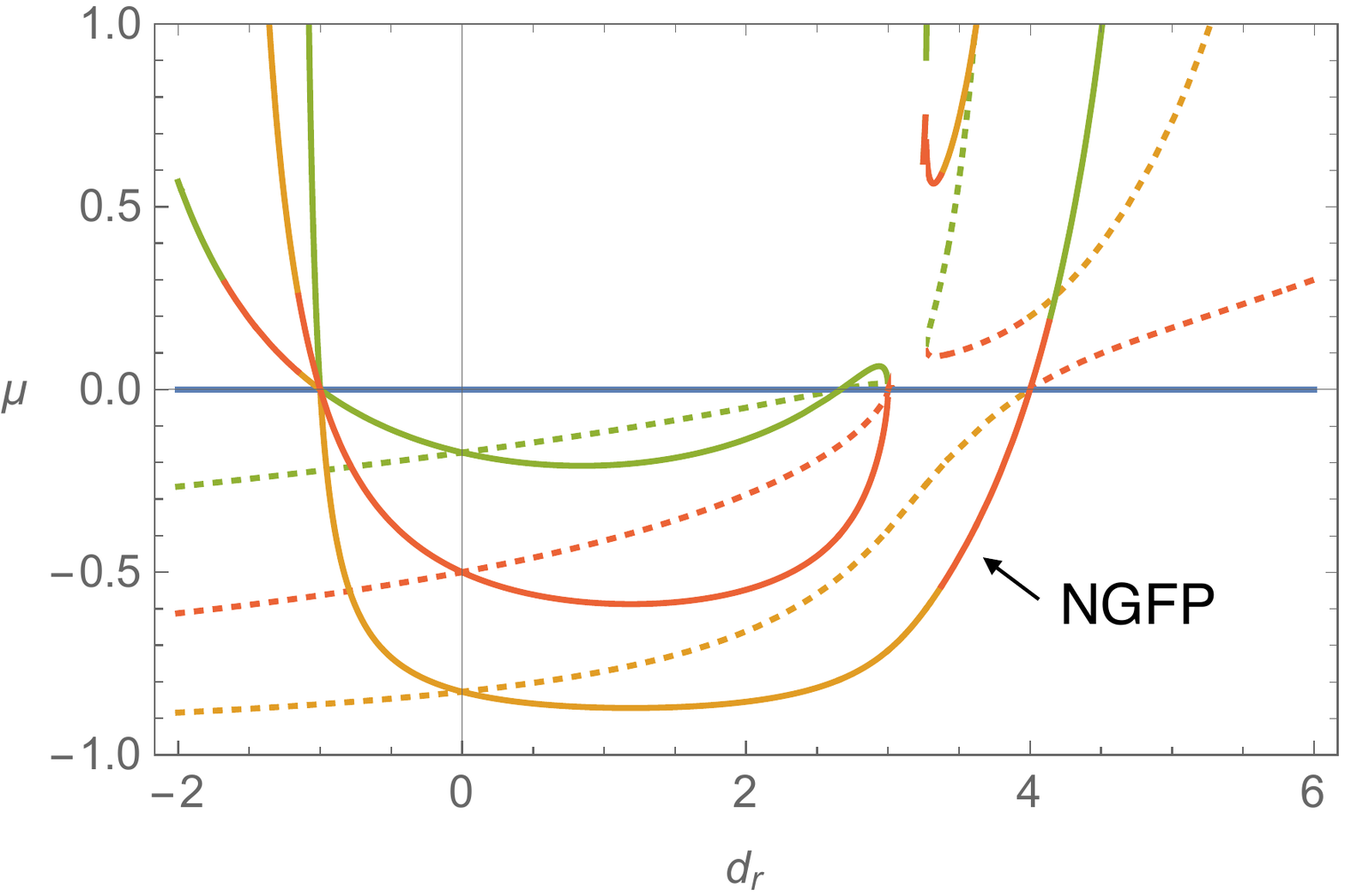}
\includegraphics[width=7.5cm]{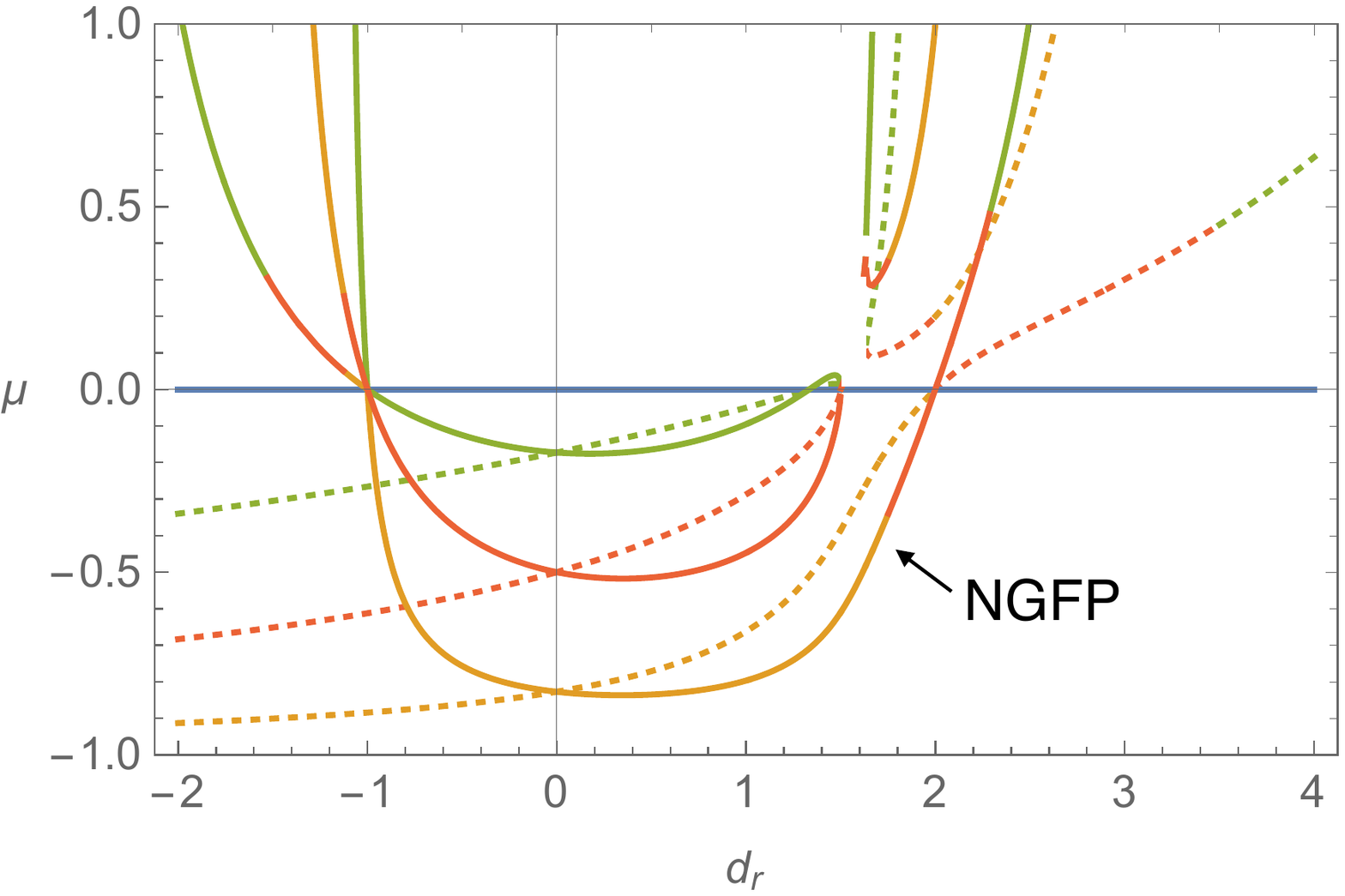}
\centering
\caption{Comparison of the fixed point solutions $\mc$ as a function of dimension $\uvd$ at truncation order $\nmax = 4$ without anomalous dimension, $\eta_k\equiv0$, for the FRG equation \eqref{eq:UVequation} (continuous curves) compared to the large-$N$ equation of $\textrm{O}(N)$ models (i.e.~without the factor $1/\rk$, dashed curves), for $\zeta=1$ (left) and $\zeta=1/2$ (right).
Different colours are due to different algebraic roots but only the continuous curves are relevant here.
The Wilson-Fisher type fixed point corresponds to the solution branch which has $\mc=0$ at the critical dimension $\uvd=\crd=4\zeta$.
In contrast to $\textrm{O}(N)$ models, we have another zero of $\mc$ at $\uvd=\rk-1 = -1$ which, however, is neither a physical dimension nor rank. 
}\label{fig:massbranches}
\end{figure} 

\

\begin{table}
  \centering
  \caption{\label{tab:exponentsr4}%
Values of the coupling constants and scaling exponents at the non-Gaussian fixed point in the large-$\ak$ regime for $\rk=\uvd+1=4$ without $\eta_k$ in $(\gfb\gf)^{\nmax}$ truncation for $\zeta=1$. 
}
\vspace*{0.5 cm}
\resizebox{0.95\textwidth}{!}{\begin{minipage}{1.25\textwidth}
\centering
\begin{tabular}{|r|c|c|c|c|c|c|c|c|c|c|}
\hline
$\nmax$ & $10\mr$& $10^2\cns2$& $10^3\cns3$ & $10^4\cns4$ & $10^5\cns5$ & $10^6\cns6$ & $10^7\cns7$ & $10^8\cns8$ & $10^9\cns9$ & $10^{10}\cns{10}$ \\ \hline\hline
 6 & -7.1817 & 2.8522 & 3.5074 & 3.7706 & 1.3424 & -6.2297 & \text{} & \text{} & \text{} & \text{} \\
 7 & -7.1720 & 2.8680 & 3.5233 & 3.7406 & 1.0193 & -8.3707 & -17.591 & \text{} & \text{} & \text{} \\
 8 & -7.1740 & 2.8647 & 3.5200 & 3.7469 & 1.0866 & -7.9239 & -13.910 & 41.128 & \text{} & \text{} \\
 9 & -7.1751 & 2.8630 & 3.5182 & 3.7503 & 1.1232 & -7.6812 & -11.912 & 63.425 & 304.07 & \text{} \\
 10 & -7.1750 & 2.8631 & 3.5184 & 3.7501 & 1.1205 & -7.6994 & -12.062 & 61.750 & 281.24 & -358.82 \\
 11 & -7.1749 & 2.8633 & 3.5186 & 3.7497 & 1.1167 & -7.7245 & -12.268 & 59.449 & 249.87 & -851.88 \\ 
 12 & -7.1749 & 2.8633 & 3.5186 & 3.7497 & 1.1166 & -7.7252 & -12.274 & 59.384 & 248.98 & -865.87 \\ 
\hline
\end{tabular}
\newline
\vspace*{0.5 cm}
\newline
\begin{tabular}{|r|c|c|c|c|c|c|c|c|c|c|c|c|}
\hline
$\nmax$ & $\theta_{1}$& $\theta_2$& $\theta_3$ & $\theta_4$ & $\theta_5$ & $\theta_6$ & $\theta_7$ & $\theta_8$ & $\theta_9$ & $\theta_{10}$ 
\\ \hline\hline
 6 & 0.44448 & -1.9006 & -6.1670 & -11.553 & -16.454 & -28.527 & \text{} & \text{} & \text{} & \text{} \\
 7 & 0.45290 & -1.8256 & -4.7984 & -9.8777 & -13.603 & -21.312 & -34.652 & \text{} & \text{} & \text{} \\
 8 & 0.45314 & -1.8669 & -4.1832 & -8.2540 & -12.239 & -17.179 & -26.712 & -41.022 & \text{} & \text{} \\
 9 & 0.45218 & -1.8834 & -4.0306 & -7.0618 & -11.165 & -14.647 & -21.814 & -32.301 & -47.464 & \text{} \\
 10 & 0.45205 & -1.8787 & -4.0690 & -6.3878 & -10.063 & -13.168 & -18.442 & -26.782 & -38.014 & -53.954 \\
 11 & 0.45214 & -1.8757 & -4.1043 & -6.1630 & -9.0992 & -12.228 & -16.073 & -22.864 & -31.940 & -43.840 \\ 
 12 & 0.45217 & -1.8761 & -4.1011 & -6.1886 & -8.4649 & -11.474 & -14.452 & -19.951 & -27.551 & -37.247 \\ 
\hline
\end{tabular}
\end{minipage}}
\end{table}

We find that also for the large-$\ak$ tensorial theory in the cyclic-melonic LPA there is a Wilson-Fisher-type NGFP for $\ks<\uvd<4\zeta$ but with scaling exponents modified by a deviation $\delta\theta_i$.
As a first guiding line for the effect of the additional $\rk$ factor in $\bfuv{1}$, we consider the fixed point solutions as functions of the UV dimension $\uvd$ (Fig.~\ref{fig:massbranches}).
For order-$\nmax$ truncations the equations are algebraic of order $\nmax$, thus having $\nmax-1$ solutions additional to the GFP. 
The main difference is that, in the $\textrm{O}(N)$ model case, fixed point solutions have only one pole above the critical dimension while in the tensor case with factor $\rk$ there is an additional second pole at negative dimension.
These poles correspond to the $\mc=-1$ pole in the flow equations, thus solutions beyond are not physically relevant.
Since only positive integer values of dimension are meaningful, this difference is not of interest though.

Both with and without the $\rk$ factor, there is one solution curve $\mc(\uvd)$ which is real for any value of dimension and vanishes at the critical dimension, $\mc(\crd)=0$ (and all other couplings vanish as well). This is the curve of non-Gaussian fixed points related to the Wilson-Fisher fixed point.
The other $\nmax-2$ solutions are related to multi-criticality%
\footnote{
This multi-critical structure should be related to the one found for $\textrm{O}(N)$ models at large $N$, see the works~\cite{Yabunaka_2017,Yabunaka_2018,Katsis_2018,defenu2020fate} and references therein. We leave it to future work to thoroughly investigate the multi-critical behavior of our setting.
}
and we will not consider them here but focus on the Wilson-Fisher-type NGFP.

\begin{figure}
\includegraphics[width=8cm]{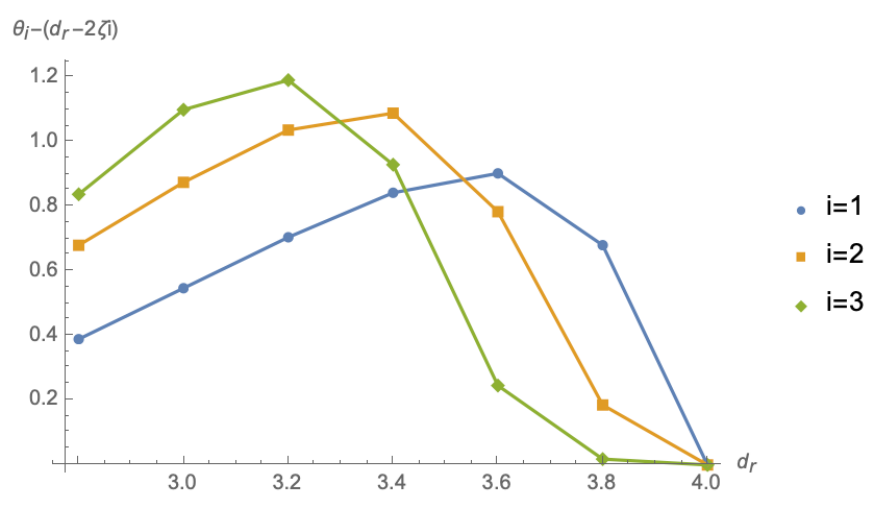}
\includegraphics[width=7cm]{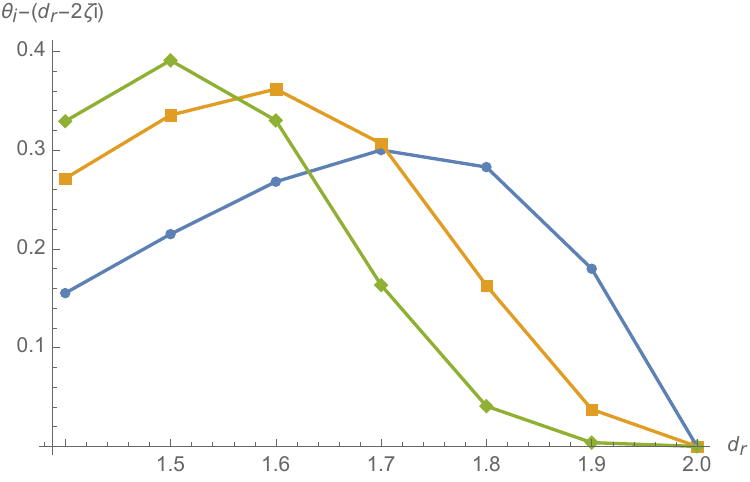}
\centering
\caption{Difference $\delta\theta_i(\rk,\zeta)=\theta_i-(\uvd-\ks i)$ of the first three scaling exponents $i=1,2,3$ for $\zeta=1$ (left) and $\zeta=1/2$ (right) as a function of $\uvd=\rk-1$ (without anomalous dimension, large-$\ak$ limit). All exponent values are converged up to two significant digits at least. 
}\label{figure:exponentsr}
\end{figure}

For quadratic kinetic term%
\footnote{This is a direct consequence of the condition $\ks<\uvd<4\zeta$. 
Thus at $\rk=4$ this NGFP exists for $3/4<\zeta\le1$ while it is there at $\rk=3$ for $1/2<\zeta<1$.
In particular, for a linear kinetic term $\zeta=1/2$ there is no integer $1<\uvd<2$, thus no NGFP for integer rank.} 
only the theory of rank $\rk=4$ has the Wilson-Fisher-type NGFP in the cyclic-melonic LPA.%
\footnote{Note that at the lower bound $\uvd=\ks$ we do neither see the fixed line of the two-dimensional $N=2$ vector model, related to the Berezinskii-Kosterlitz-Thouless transition~\cite{Berezinsky:1970fr,Berezinsky:1972rfj,Kosterlitz:1973xp,Grater:1994qx,VonGersdorff:2000kp,Jakubczyk:2016rvr}, since the correspondence here is with large-$N$ vector models.}
We present the converging values of the NGFP and its scaling exponents $\theta_i$ in truncation up to order $\nmax=12$ in Table \ref{tab:exponentsr4}.
They qualitatively agree with the NGFP of large-$N$ $\textrm{O}(N)$ models. In particular, they have negative $\mc$ but positive couplings.
Furthermore, there is one positive exponent and thus the NGFP describes a phase transition between a broken and unbroken phase of the global symmetry.
Quantitatively, the exponents are of the form
\[\label{eq:exponentsr}
\theta_i = \uvd - \ks i + \delta\theta_i(\rk,\zeta)
\]
with a deviation $\delta\theta_i(\rk,\zeta)$ compared to Eq.~\eqref{eq:exponentsON} depending both on the rank $\rk$ and the exponent in the kinetic term $\zeta$.
To better understand this deviation, we have calculated the NGFP exponents also for various fractional dimensions. 
Some of the results are shown in Fig.~\ref{figure:exponentsr}. 
Though one can clearly see a pattern indicating in particular a $(\uvd-4\zeta)^i$ dependence in this plot, the complete functional dependence of $\delta\theta_i$ on $i$ and $\rk$ seems to be too intricate to guess from these values.
As an idea for future work, one might use a strategy known from $\textrm{O}(N)$ models at large $N$ \cite{Tetradis_1996,D_Attanasio_1997}
to analytically solve the full equation \eqref{eq:UVequation} which could lead to an exact result for $\delta\theta_i$.


The NGFP solution curve also extends above the critical dimension, $\uvd>\crd = 4\zeta$, and we find evidence that scaling exponents still converge to the values given by Eq.~\eqref{eq:exponentsON} and~\eqref{eq:exponentsr} with truncations of larger and larger order in $n$.
Thus, this NGFP would have two positive exponents which could mean that the theory is asymptotically safe at this point.
There are similar findings in the context of $\textrm{O}(N)$ models with arbitrary $N$: 
In spite of the well-known result that these models have only the Gaussian fixed point for $d>4$, see e.g. Refs.~\cite{Codello_2013,Codello_2015}, the existence of non-trivial universality classes has been suggested for $d\geq 4$ in Refs.~\cite{Fei_2014,nakayama2014five}. However, a critical examination \cite{Percacci_2014} has shown that the effective potential at this NGFP is unbounded from below.
Indeed, we find also here that though $\mc$ is positive, all other couplings are negative.
We will leave the question of existence of an asymptotically safe NGFP above the critical rank for the tensorial theory in the cyclic-melonic LPA for future work.

\subsubsection{Results with anomalous dimension}

The anomalous dimension changes the picture drastically.
As discussed, due to the specific one-loop diagrams of tensorial interactions it has the opposite sign compared to $\textrm{O}(N)$ models, see Eq.~\eqref{eq:etaUV}.
As a consequence, we find that there are two candidate non-Gaussian fixed points 
which are 
continuously connected in $\uvd$ with the Gaussian fixed point at $\uvd=\crd=4\zeta$.

\begin{figure}
\includegraphics[width=7.5cm]{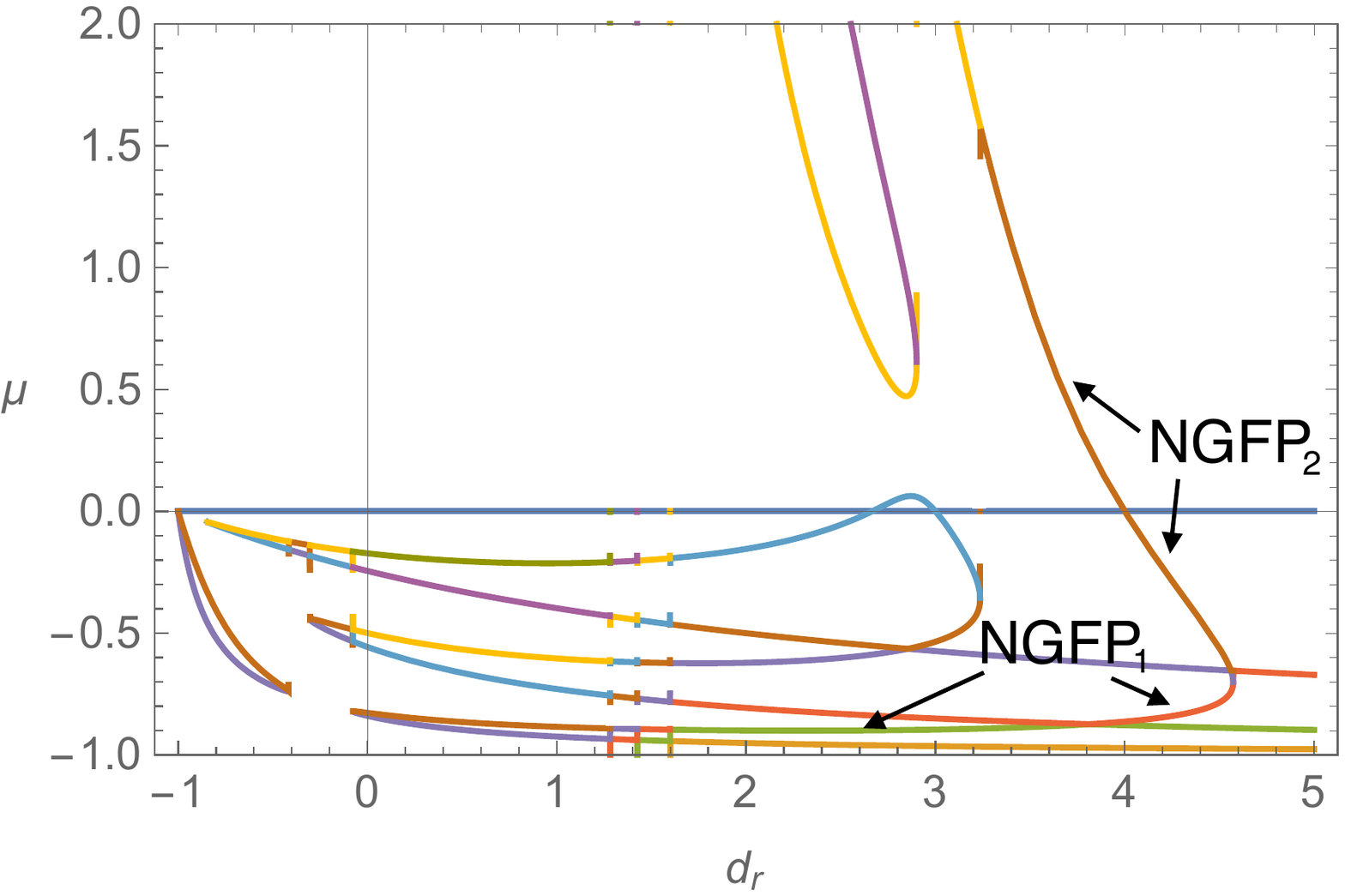}
\includegraphics[width=7.5cm]{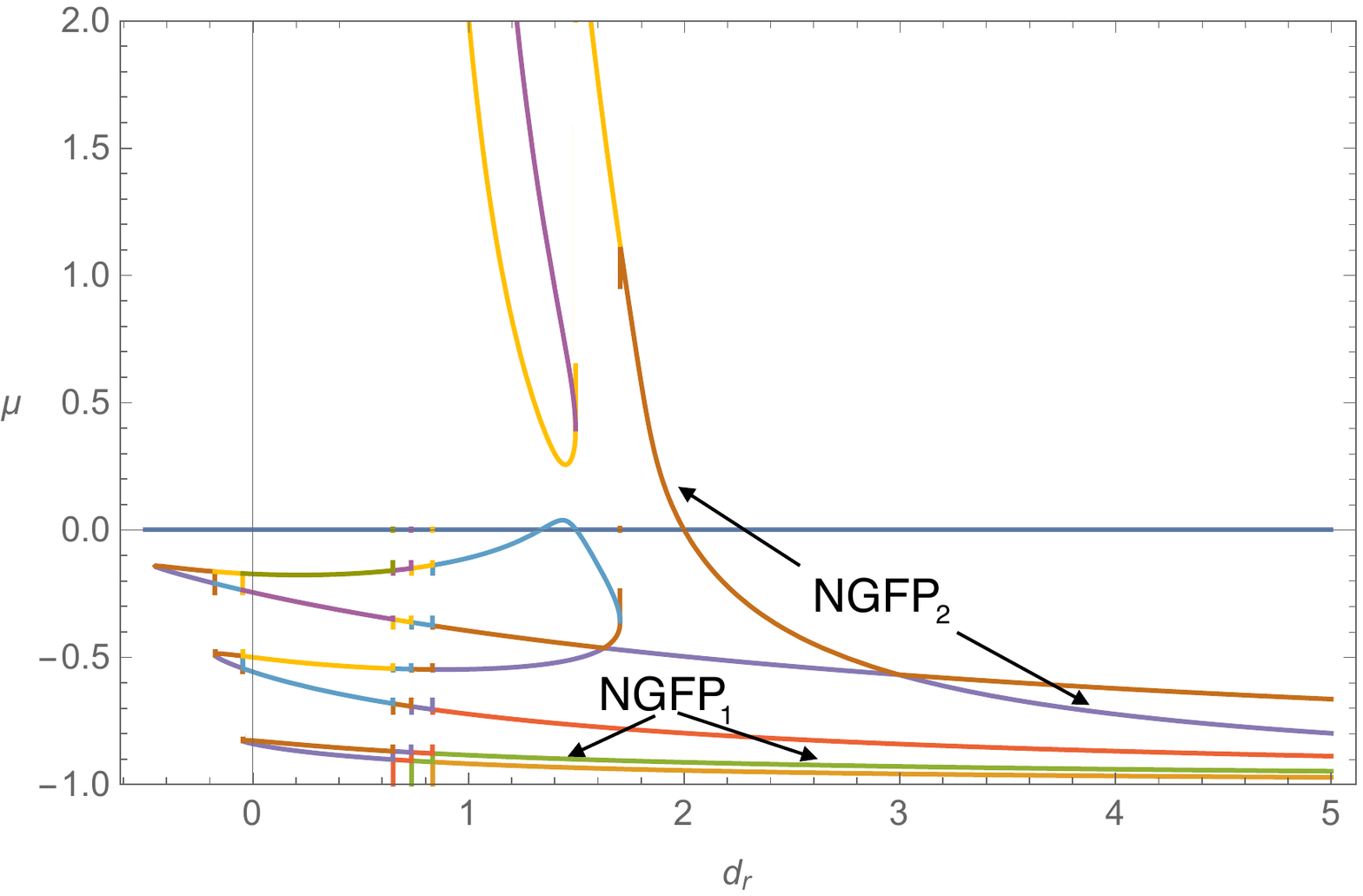}
\centering
\caption{Fixed point value $\mc$ as a function of the UV dimension $\uvd$ at truncation order $\nmax = 4$ for $\zeta=1$ (left) and $\zeta=1/2$ (right).
The NGFP candidates lie on the solution branch which has $\ms=0$ at the critical dimension $\uvd=\crd$. 
This curve turns (has a branching point) for $\zeta=1$ at $\nosed\approx4.5$ while for $\zeta=1/2$ we find this point at $\nosed\approx103$.
}\label{fig:massbrancheseta}
\end{figure}

The crucial effect of the anomalous dimension is that it deforms the NGFP solution as a function of the UV dimension $\uvd$ beyond $\uvd=4\zeta$.
As in the LPA case, this is best visualized plotting for example the curve $\mc(\uvd)$, see Fig.~\ref{fig:massbrancheseta}. 
We observe that the fixed point equations \eqref{eq:etaUV}--\eqref{eq:UVequations} taken together are algebraic of order $1+3(\nmax-1)$ in truncation of order $\nmax$. 
Thus, there are $3(\nmax-1)$ solutions additional to the GFP of which we find that up to $2(\nmax-1)$ solutions are real in the domain of interest.
Comparing Fig.~\ref{fig:massbrancheseta} to the case without $\eta_k$ (cf. Fig.~\ref{fig:massbranches}), this doubling of solutions is related to the fact that each of the $\nmax-1$ solutions does not extend to zero at $\uvd=-1$ anymore but has a branching point at small negative $\uvd$ with a second solution. 
Again, we can only suspect that these new partner solutions are related to an $\eta_k$-modified structure of multi-critical fixed points the further analysis of which we leave for future work.

More importantly, there is still a solution which coincides with the GFP at the critical dimension $\uvd=\crd=4\zeta$.
However, the slope of the curve $\mc(\uvd)$ at this zero is inverted compared to the LPA and it has a branching point with another solution at some value $\uvd=\nosed>\crd$. 
On these grounds we understand the curve consisting of both these solution branches as the generalization of the solution curve in the LPA. 
To distiguish them, we call the NGFP at the lower $\mc$ branch NGFP$_1$ and the one at the upper branch NGFP$_2$.
We find that, at given truncation, the NGFP$_1$ qualitatively agrees (with respect to signs of couplings and scaling exponents) with the Wilson-Fisher-type fixed point for $\uvd<\nosed$ in the LPA but its domain of convergence is not clear.
The NGFP$_2$ converges between some $3\zeta<\uvd<\crd$ and $\nosed$ and corresponds to the NGFP in the LPA 
with respect to exponents but its couplings have different signs. 

For the dimension $\nosed$ where both NGFPs coincide, we find convergence to $\nosed\approx4.55$ for the quadratic kinetic term while for the linear kinetic term $\nosed > 100$ up to $n=12$ truncation.
In the former case, $\zeta = 1$, this means that the main difference to the LPA is that instead of an NGFP only for integer $\rk = \uvd + 1 = 4$ there is now the candidate NGFP$_1$ also at critical rank $\rk=\crr=5$.
A similar result has been already discussed in Ref.~\cite{TGFT1} for a theory with closure constraint, thus for $\rk = \uvd + 2 = 6$ (cf.~App.~\ref{sec:Dimensions}) in quartic $\nmax = 2$ truncation%
\footnote{Also in \cite{BenGeloun:2015ej,BenGeloun:2016kw} both NGFP$_1$ and NGFP$_2$ are found in the quartic truncation ($\nmax=2$) and discussed for rank $\rk=3$, that is $\uvd=2$. 
As found here, the NGFP$_2$ clearly diverges with larger truncations for $\rk\le4$.
In \cite{TGFT3} for rank-3 TGFT on $\text{SU}(2)$ with closure constraint, thus $\uvd=3(3-2)=3$, there are also indications for a second NGFP in order $\nmax=5,6$ truncation; however, as that point does not exist for $n<5$ it is not related to the fixed points discussed here (both NGFP$_1$ and NGFP$_2$ are present for all $\nmax\ge 2$).
}.
However, we have to emphasize that even with our results up to truncation order $\nmax=11$ (see Tab.~\ref{tab:exponents1r5e}) the question of convergence remains inconclusive.
All further ranks $\rk\ge6$ are above the dimension $\nosed$ up to which both NGFPs exists.
In contrast, for a linear kinetic term ($\zeta=1/2$) there is not only the special case $\rk=\crr=3$ with the NGFP$_1$
but also a wide range of integer ranks $3<\rk<\nosed+1$ which have both the NGFP$_1$ and NGFP$_2$ in a given truncation.
While the NGFP$_2$ converges (see e.g.~the case $\rk=4$ in Tab.~\ref{tab:exponents2r4ez}), the results for the scaling exponents of the NGFP$_1$ for $\rk>\crr$ show rather clearly divergence.%
\footnote{However, we cannot say anything about ranks closer to $\rk=\nosed+1>100$ because for large ranks $\rk\gg1$ the couplings at the NGFP$_1$ become smaller than machine size such that it is not possible to obtain meaningful exponents anymore.
}

\begin{table}
  \centering
  \caption{\label{tab:exponents1r5e}%
Values of the coupling constants and scaling exponents of the NGFP$_1$ in the large-$\ak$ regime for $\crr=\crd+1=5$ (for $\zeta=1$) in $(\gfb\gf)^{\nmax}$ truncation of the LPA$'$. 
}
\vspace*{0.5 cm}
\resizebox{0.95\textwidth}{!}{
\begin{minipage}{1.25\textwidth}
\centering
\begin{tabular}{|r|c|c|c|c|c|c|c|c|c|c|}
\hline
$\nmax$ & $\eta$ & $10\mr$& $10^2\cns2$& $10^3\cns3$ & $10^4\cns4$ & $10^5\cns5$ & $10^6\cns6$ & $10^7\cns7$ & $10^8\cns8$ & $10^9\cns9$ 
\\ \hline\hline
 6 & -1.3361 & -9.1769 & 0.33927 & 0.22945 & 0.19608 & 0.18257 & 0.14841 & \text{} & \text{} & \text{}  \\
 7 & -1.3697 & -9.2980 & 0.25141 & 0.15244 & 0.12120 & 0.11195 & 0.10805 & 0.088978 & \text{} & \text{}  \\
 8 & -1.3925 & -9.3783 & 0.19963 & 0.11076 & 0.082324 & 0.073318 & 0.072410 & 0.072704 & 0.061212 & \text{}  \\
 9 & -1.4083 & -9.4333 & 0.16724 & 0.086464 & 0.060702 & 0.051958 & 0.050701 & 0.053224 & 0.055673 & 0.048156  \\
 10 & -1.4194 & -9.4714 & 0.14634 & 0.071649 & 0.048060 & 0.039722 & 0.037994 & 0.040106 & 0.044547 & 0.048517 \\
 11 & -1.4269 & -9.4974 & 0.13284 & 0.062486 & 0.040504 & 0.032564 & 0.030564 & 0.032083 & 0.036292 & 0.042429 \\ 
\hline
\end{tabular}
\newline
\vspace*{0.5 cm}
\newline
\begin{tabular}{|r|c|c|c|c|c|c|c|c|c|c|c|c|}
\hline
$\nmax$ & $\theta_1$ & $\theta_2$& $\theta_3$ & $\theta_4$ & $\theta_5$ & $\theta_6$ & $\theta_7$ & $\theta_8$ & $\theta_9$ & $\theta_{10}$ 
\\ \hline\hline
 6 & 0.71133 & -10.992 & -37.387 & -62.663 & -85.886 & -141.10 & \text{} & \text{} & \text{} & \text{} \\
 7 & 0.62816 & -10.250 & -35.566 & -67.696 & -86.591 & -130.58 & -201.46 & \text{} & \text{} & \text{} \\
 8 & 0.55378 & -9.4521 & -33.178 & -67.216 & -90.691 & -123.76 & -182.94 & -268.26 & \text{} & \text{} \\
 9 & 0.48821 & -8.6312 & -30.569 & -63.833 & -94.369 & -118.62 & -170.16 & -240.43 & -339.22 & \text{} \\
 10 & 0.43094 & -7.8041 & -27.876 & -59.373 & -94.466 & -115.74 & -159.03 & -220.89 & -300.97 & -411.98 \\
11 & 0.38153 & -6.9837 & -25.171 & -54.478 & -90.459 & -114.78 & -148.64 & -204.00 & -273.70 & -362.46 \\ 
\hline
\end{tabular}
\end{minipage}
}
\end{table}

\begin{figure}
\includegraphics[width=6.8cm]{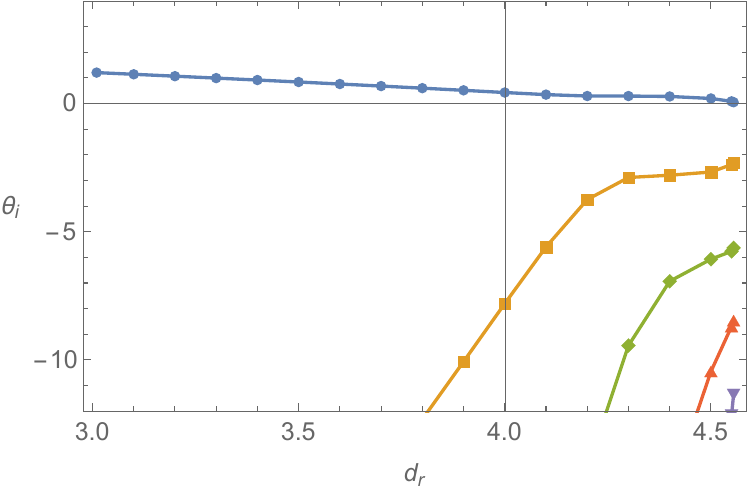}
\includegraphics[width=6.8cm]{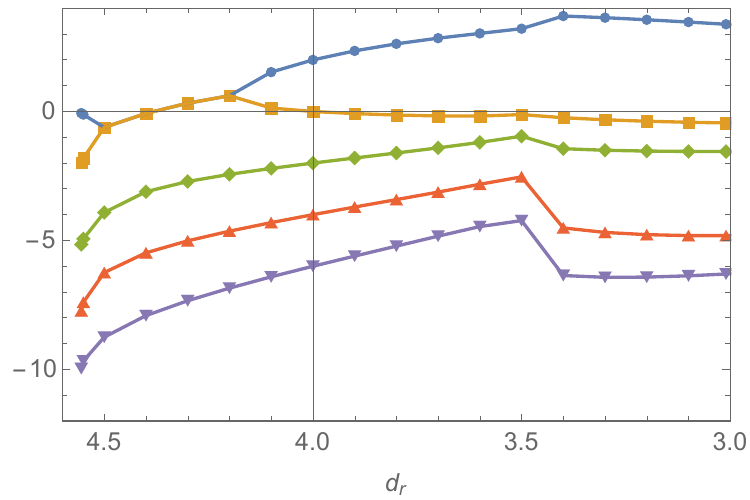}
\includegraphics[width=1cm]{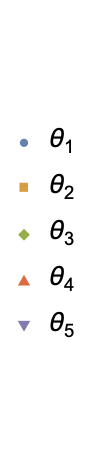}
\centering
\caption{Scaling exponents $\theta_i$, $i=1,2,3,4,5$, of the NGFP$_1$ (left) and NGFP$_2$  (right, $\uvd$ axis inverted to highlight continuity through the branching point)
as a function of dimension  in the large-$\ak$ limit for $\zeta = 1$ at order $\nmax=10$ truncation in the LPA$'$.
For the NGFP$_1$ only the first exponents for $\uvd$ close to $\crd\approx4.55$ are converged; For the NGFP$_2$ convergence breaks down around $\uvd=3.5$. In between $4.1<\uvd\le 4.5$ the first two exponents are complex $\theta_{1/2}=\theta_\pm = x\pm i y$, the real part is shown.
}\label{fig:exponents10e}
\end{figure}

In the LPA$'$, the scaling exponents seem to have a completely different behaviour quantitatively compared to the LPA.
Qualitatively, the NGFP$_1$, if it converges, is still of the Wilson-Fisher type:
The fixed point has negative $\mc$ but all other couplings are positive%
\footnote{Close to the upper dimension $\uvd-\nosed=\mathcal{O}(10^{-1})$ and in a given truncation of order $\nmax$ we find that the highest couplings $\cnc{\nmax}, \cnc{\nmax-1}...$ start to fluctuate around zero. With the techniques used it is not clear whether this is an artefact or not.}
and the first scaling exponent is positive while the others are all negative.
Thus we expect it to describe a phase transition between a broken and unbroken phase.
The exact values seem to be very different to Eq.~\eqref{eq:exponentsr} which is similar to the $\textrm{O}(N)$ models, Eq.~\eqref{eq:exponentsON}.
To understand their behaviour we have again considered also fractional dimensions, see Fig.~\ref{fig:exponents10e}. However, on a standard computer only truncations up to order $\nmax=10$ are feasible in finite time and in this truncation exponents of the NGFP$_1$ are converging very slowly (if they converge at all) except close to the maximal dimension $\nosed$. 
Other methods, probably beyond finite-order truncations, would be necessary to determine the domain of convergence and the values of the exponents exactly.

Scaling exponents of the NGFP$_2$ show some resemblance of the exponents in the LPA and of $\textrm{O}(N)$ models.
For $\zeta=1$ there is convergence of these exponents for $3.5\lessapprox\uvd<\nosed$ (see Fig.~\ref{fig:exponents10e}).
In particular, higher exponents seem to have a linear dependence on the dimension, though with the opposite sign $-\uvd$ as compared to Eq.~\eqref{eq:exponentsON}. This could be seen as being in accordance with the fact the fixed point curve around $\uvd=\crd$ is inverted (Fig.~\ref{fig:massbrancheseta}).
A clear exception is the second exponent $\theta_2$ which has a small negative value for $\uvd<\crd$.
Since there is thus only one positive exponent $\theta_1$ the NGFP$_2$ would be a candidate for describing a phase transition as well. 
However, the signs of the couplings are exactly opposite to the Wilson-Fisher fixed point which would allow for a symmetric ground state $\rho=0$ but leaves the issue of unbounded potential already discussed above for the LPA. Anyway, this result applies only to fractional $\uvd$ and thus to no integer rank $\rk$.

\begin{table}[ht!]
  \centering
  \caption{\label{tab:exponents2r4ez}%
Values of the coupling constants and scaling exponents at the NGFP$_2$ in the large-$\ak$ regime for $\rk=\uvd+1=4$ and $\zeta=1/2$ in the  $(\gfb\gf)^{\nmax}$ truncation of the LPA$'$. 
}
\vspace*{0.5 cm}
\resizebox{0.95\textwidth}{!}{
\begin{minipage}{1.25\textwidth}
\centering
\begin{tabular}{|r|c|c|c|c|c|c|c|c|c|c|c|}
\hline
$\nmax$ & $\eta$ & $10\mr$& $10^2\cns2$& $10^3\cns3$ & $10^4\cns4$ & $10^5\cns5$ & $10^6\cns6$ & $10^7\cns7$ & $10^8\cns8$ & $10^9\cns9$ & $10^{10}\cns{10}$ \\ \hline\hline
 6 & -0.82953 & -5.7181 & 3.9716 & 3.3926 & 1.0150 & -2.7624 & -1.2999 & \text{} & \text{} & \text{} & \text{} \\
 7 & -0.82922 & -5.7170 & 3.9724 & 3.3945 & 1.0297 & -2.6415 & -0.10387 & 15.822 & \text{} & \text{} & \text{} \\
 8 & -0.82923 & -5.7171 & 3.9723 & 3.3944 & 1.0291 & -2.6468 & -0.15644 & 15.126 & -11.587 & \text{} & \text{} \\
 9 & -0.82924 & -5.7171 & 3.9723 & 3.3943 & 1.0285 & -2.6513 & -0.20110 & 14.535 & -21.428 & -195.40 & \text{} \\
 10 & -0.82924 & -5.7171 & 3.9723 & 3.3943 & 1.0286 & -2.6507 & -0.19502 & 14.615 & -20.088 & -168.80 & 613.17 \\
 11 & -0.82924 & -5.7171 & 3.9723 & 3.3943 & 1.0286 & -2.6506 & -0.19369 & 14.633 & -19.796 & -162.99 & 746.98 \\ 
 12 & -0.82924 & -5.7171 & 3.9723 & 3.3943 & 1.0286 & -2.6506 & -0.19406 & 14.628 & -19.877 & -164.60 & 709.90 \\ 
\hline
\end{tabular}
\newline
\vspace*{0.5 cm}
\newline
\begin{tabular}{|r|c|c|c|c|c|c|c|c|c|c|c|c|}
\hline
$\nmax$ & $\theta_{1/2}$& $\theta_3$ & $\theta_4$ & $\theta_5$ & $\theta_6$ & $\theta_7$ & $\theta_8$ & $\theta_9$ & $\theta_{10}$ 
\\ \hline\hline
 6 & 0.24545$\pm1.33237 i$ & -2.3665 & -4.8104 & -9.1162 & -16.366 & \text{} & \text{} & \text{} & \text{} \\
 7 & 0.24662$\pm1.33276 i$ & -2.4041 & -4.3120 & -7.2845 & -12.247 & -20.146 & \text{} & \text{} & \text{} \\
 8 &  0.24671$\pm1.33264 i$ & -2.4115 & -4.2725 & -6.3724 & -9.9634 & -15.513 & -23.995 & \text{} & \text{} \\
 9 & 0.24663$\pm1.33264 i$ & -2.4074 & -4.3130 & -6.0972 & -8.6451 & -12.814 & -18.883 & -27.893 & \text{} \\
 10 &  0.24664$\pm1.33265 i$ & -2.4073 & -4.3116 & -6.1298 & -8.0245 & -11.124 & -15.803 & -22.338 & -31.831 \\
11 & 0.24664$\pm1.33265 i$ & -2.4076 & -4.3078 & -6.1566 & -7.9189 & -10.141 & -13.775 & -18.904 & -25.863 \\ 
 12 & 0.24664$\pm1.33265 i$ & -2.4076 & -4.3083 & -6.1486 & -7.9787 & -9.7579 & -12.459 & -16.564 & -22.099 \\ 
\hline
\end{tabular}
\end{minipage}
}
\end{table}

\subsection{Dimensional flow and symmetry restoration}\label{Section:IR}

For the question of phase transitions it is necessary to understand the full FRG equation describing the flow through all scales.
If we do not consider the large-$\ak$ limit as a large-$\cs$ but a large-$k$ limit, the results discussed in the last section describe only the flow and phase diagram for scales $k\approx\Lambda$ close to the UV scale $\Lambda$.
In particular, to determine the fate of the IR fixed points of the large-$\ak$ equation \eqref{eq:UVequation} under the full flow it is necessary to understand the content of the full equation \eqref{eq:Wetterichresult}.
Since we cannot solve it directly due to its intricate non-autonomy, we will analyse it in two complementary ways analytically and numerically.
First, we consider autonomous approximations for intermediate regimes which yields equivalence to the zero-dimensional $\text{O}(\ccu)$ model in the small-$k$ limit.
Then, generalizing the rescaling, we find that the full flow can be effectively understood as a flow of the dimension of the theory.
Finally, we will argue for a generic symmetry restoration and illustrate this with numerical solutions for specific initial conditions.

\subsubsection{The small-$k$ limit}

The UV rescaling can be generalized to any intermediate scale and in particular to an autonomous small-$k$ limit.
The functions $F_\rk^l$ and $\gr^l$ decribing the non-autonomous part of the FRG equations \eqref{eq:betaequations}, are polynomials in $\ak$ of order $\uvd=\rk-1$.
Autonomy in the large-$\ak$ stems from the fact that the UV scaling dimension Eq.~\eqref{eq:scalingdimension} of the tensorial theory has the same form as the scaling dimension of standard local QFT but with an effective dimension $\uvd$.
As a consequence, we can use this scaling dimension with an effective dimension $0 < \efd < \uvd$ to obtain approximately autonomous FRG equations at the intermediate regime where the dominating contribution in the full equations is of order $\efd$ in $\ak$.
That is, one generalizes the UV rescaling $\eqref{eq:UVrescaling}$ to
\[
\cn{n} = \wfr^{n} k^{\efd -(\efd-\ks)n} \cs^{(1-n)\efd} \cnr{n} .
\]
to obtain 
\begin{align}
k\partial_k \cnr{n} =  -\efd\cnr{n} + n(\efd - \ks  +& \eta_k )\cnr{n} 
+\left(\zeta-\frac{\eta_k}{2}\right){\ak^{-\efd}}  \bcb{n}(\mr,\cnr{i}) \\
&+\ccu \sum_{l=1}^n \left(\zeta {F}_\rk^l(\ak) - \frac{\eta_k}{2} \gr^l(\ak) \right) {\ak^{-\efd}} \bco{n}_l(\mr,\cnr{i}) \nonumber
\end{align}
in analogy to the UV case Eq.~\eqref{eq:betaUV}.
At scales where $F_\rk^l,\gr^l \sim \ak^{\efd}$ these equations are approximately the same as for $\textrm{O}(N)$ models in $\efd$ dimensions.

As a direct consequence, in the limit $k\rightarrow 0$ the flow equations are in the LPA the same as for the $\text{O}(\ccu)$-symmetric scalar field theory, i.e.~the $\textrm{O}(N)$ model with $N=\ccu$, with dimension $\efd=0$.
Only at this order in $\ak$ does the first term of the FRG equation~\eqref{eq:Wetterichresult} with $\pot{}''$ in the denominator contribute which leads to the beta function part $\bcb{n}$.
The small-$k$ leading order of the flow equation \eqref{eq:etaflow} of the anomalous dimension $\eta_k$ is linear in $\ak$ such that, considering this equation separately, $\eta_k$ would vanish in the limit.
However, in the flow equations for the couplings, $\eta_k$ occurs together with $\gr(\ak)$ which is sensitive to the summation scheme according to which it may or may not lead to a $1/\ak$ contribution (see Eq.~\eqref{eq:I2versions}).
Thus it is necessary to take the limit together such that the flow of the anomalous dimension is described by the equation
\begin{align}\label{eq:etaflowIR}
\tilde{\eta}_k := \lim_{k\rightarrow0} \left(\eta_k\gr^l(\ak)\right)
&= \frac{-2(\rk-1)\cnr{2}}{\frac{2\rk}{\ccu}(1+ \mr)^2 - \rk\cnr{2}}\, \lim_{k\rightarrow0}\ak \gr^l(\ak)\\
&= \frac{2(\rk-1)\cnr{2}}{\frac{2}{\ccu}(1+ \mr)^2 - \cnr{2}}\, \lim_{k\rightarrow0} \left(\ak^{1-\ks}\kw{1}(\ak)  \right)
\end{align}
which is the same for any order $l$. 
The limit in the last line is for example $1/3$ both for the box and the simplex sum for $\zeta=1$. It vanishes for all other cases considered here.
The flow equations of the couplings are then
\begin{eqnarray}
\frac{1}{\zeta}k\partial_k \cnr{n} &=&   - 2 n \cnr{n} 
+ \bcb{n}(\cnr{i}) + (\cct)\left(1-\frac{\tilde{\eta}_k}{\ks}\right)\bco{n}(\cnr{i})
\end{eqnarray}
which can be viewed as the Taylor expansion around $\rho=0$ of 
\[\label{eq:IRequation} \boxd{
\frac{1}{\zeta} k\partial_k\tilde{U}_k(\rho) + 2 \rho\,\tilde{U}'_k(\rho) =
\frac{1}{1 + \tilde{U}_k'(\rho) + 2\rho\,\tilde{U}_k''(\rho)}
+ (\cct)\frac{1-\frac{\tilde{\eta}_k}{\ks}}{1 + \tilde{U}_k'(\rho)}
},\]
which is in fact the FRG equation of the $\text{O}(\ccu)$ model on Euclidean space for $d=0$ dimensions~\cite{Berges:2002ga,Kopietz,Delamotte,Dupuis:2020fhh} (with $\zeta=1$ for quadratic kinetic term) except for the distinguished dependence on $\tilde{\eta}_k$. Remarkably, the same observation can be made for a scalar field on the sphere~\cite{Benedetti,Serreau:2013eoa,Guilleux:2015pma,Guilleux:2017ig,Ratra:1984yq,Serreau:2011fu,Mazzitelli:1988ib}. One may thus expect the phase structure of these theories to be similar as a consequence of the dimensional reduction. 

The physical relevance of the small-$k$ limit is not completely clear.
As the non-autonomy in the FRG equations is always in the combination $\ak=\cs k$ and not in $k$ independently of the volume scale $\cs$, strictly speaking the limit applies to scales $k\ll1/\cs$, i.e.~to modes $k$ much smaller than the size $\cs$ of the compact space.
These modes correspond to wave lengths which are much larger than $\cs$. 
It makes sense to consider such modes as corresponding to diffusion scales (``times'') or to winding modes.
But because of the compactness of the space such lengths larger than $\cs$ do not correspond to distances between points. 
In particular, correlation functions on compact space are only meaningful for distances up to $\cs$.
It is therefore not obvious whether one should consider the renormalization group flow down to arbitrary small scales $k$.

If the small-$k$ limit is meaningful, an immediate consequence of the effectively zero-dimensional equations is that there can be no phase transition between a phase of spontaneously broken and unbroken global $\textrm{U}(1)$ (or $\mathbb{Z}_2$) symmetry since the symmetry is always radiatively restored in this regime. We elaborate on this important point in particular below in Section~\ref{Section:symmetryrestoration}.

Another interpretation could be that physically the question of phase transition is only meaningful in the thermodynamic limit which corresponds to the $\cs\rightarrow\infty$ limit. Then the large-$\ak$ results of the last section would in fact apply at all scales.
A third possibility is that indeed the equations are only meaningful for scales up to small $k>1/\cs$. Then the crucial question is what happens at such intermediate scales $k\approx1$ which we will investigate in the following.

\subsubsection{Continuous rescaling and dimensional flow}

\begin{figure}
\includegraphics[width=.47\linewidth]{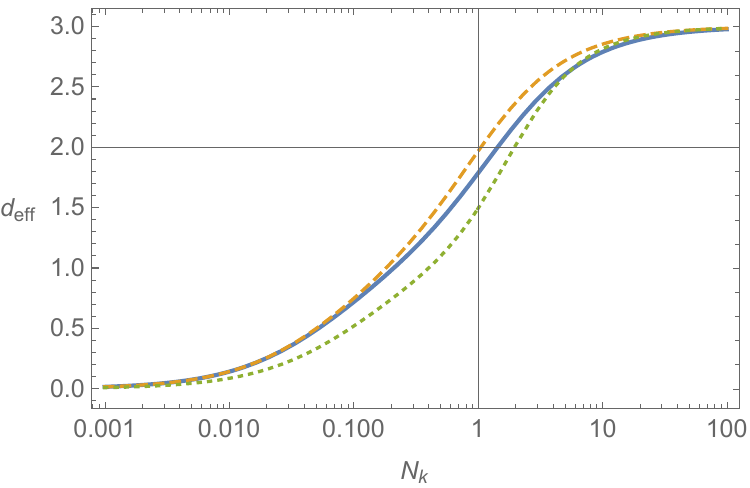}
\includegraphics[width=.46\linewidth]{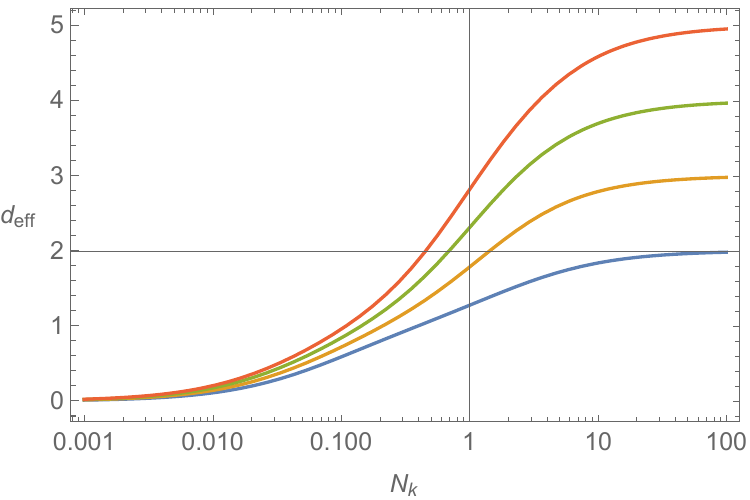}
\centering
\caption{Left: Flow of the effective dimension $\efd$ for the rank $\rk=4$ complex field ($\ccu=2$) 
comparing the different threshold functions, Eq.~\eqref{eq:I0versions}:
$\zeta=1$ integral approximation (thick line), box approximation equivalent to $\zeta\rightarrow\infty$ (dashed) and exact simplex sum $\zeta=1/2$ (dotted).
Right: Comparison of rank $\rk=3,4,5,6$ (bottom up) for complex field with $\zeta=1$ integral approximation.
}\label{fig:dimensionflow}
\end{figure}

\begin{figure}
\includegraphics[width=7cm]{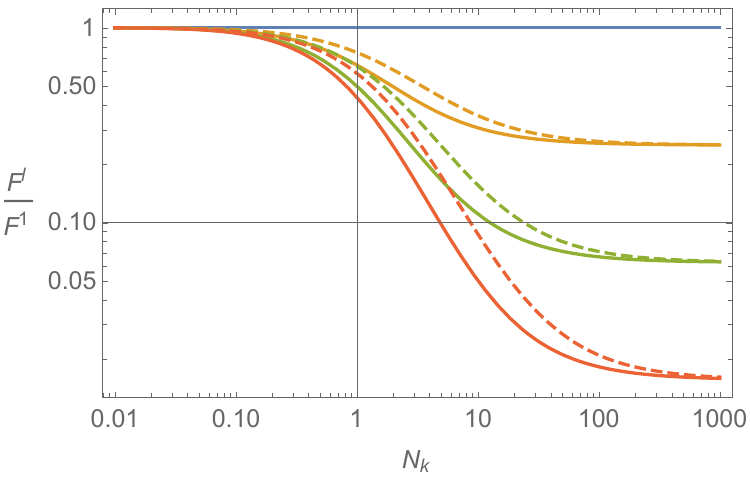}
\includegraphics[width=7cm]{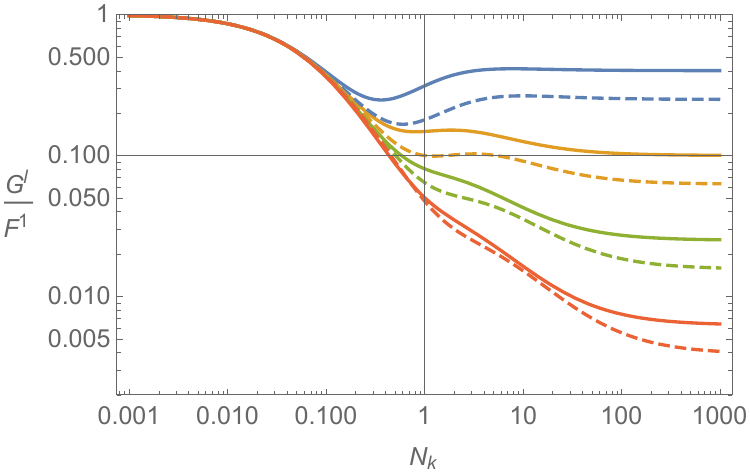}
\centering
\caption{Flow of the ratio $F_4^l/F_4^1$ (left) and $G_4^l/F_4^1$ (right) for $l=1,2,3,4$ (from top to bottom) with $\ccu=2$ in the integral approximation for $\zeta=1$ and $\zeta=1/2$ (dashed) for comparison.
}\label{fig:FGflows}
\end{figure}

For the full non-autonomous flow equations one can generalize the above rescaling to one continuously interpolating between large and small $\ak$.
The idea to invent such an interpolation to analyse non-autonomous flow equations has been used already in the case of a real $\rk=3$ field with linear propagator $\zeta=1/2$ \cite{Benedetti:2015et}.
Here, for a complex field%
\footnote{For a real field, this rescaling does not work in the $k\rightarrow0$ limit because in this regime only the $\bcb{}$ term stemming from the term with second derivative $\pot''$ in the full equation \eqref{eq:Wetterichresult} survives. Accordingly, the functions $F_\rk^l$ vanish in the limit and $F_\rk^1$ cannot be used for a rescaling. 
The physical picture of dimension flow is nevertheless the same; there is merely the technical obstacle that there is no function in $\ak$ factorizing from both the $\bcb{}$ and $\bco{}$ terms at the same time.
}, 
we have a natural $k$-dependent rescaling suggested by the flow equations themselves,
\[
\cn{n} = \wfr^{n} k^{\ks n}F_\rk^1(\cs k)^{1-n} \cnr{n} .
\] 
This rescaling leads to a more involved logarithmic derivative in the flow equations which is neatly captured by a scale-dependent generalization of the effective dimension
\[
\efd(k) := \frac{ \partial \log F_\rk^1(\cs k)}{\partial \log k} = \frac{ \partial \log F_\rk^1(\ak)}{\partial \log \ak} \, .
\]
In terms of this flowing dimension the full FRG equations are 
\begin{align}
k\partial_k \cnr{n} = & -\efd(k)\cnr{n} + n(\efd(k) - \ks  + \eta_k )\cnr{n} \\
&+\left(\zeta-\frac{\eta_k}{2}\right)\frac{\bcb{n}(\cnr{i})}{F_\rk^1(\ak)}  
+\ccu \sum_{l=1}^n \left(\zeta\frac{F_\rk^l(\ak)}{F_\rk^1(\ak)} - \frac{\eta_k}{2}\frac{\gr^l(\ak)}{F_\rk^1(\ak)}\right) \bco{n}_l(\cnr{i}) \; . \nonumber
\end{align}
Again, these equations are similar to the FRG equations of $\textrm{O}(N)$ models but modify them in four ways:
\begin{itemize}
\item First, the $\bcb{n}$ part is modified by $1/F_\rk^1$ which means that it becomes continuously switched off when going from small to large $\ak$. In this way, the equations interpolate between $\text{O}(\ccu)$-model equations at small $\ak$ and large-$N$ $\textrm{O}(N)$ model equations at large $\ak$.
\item Second, the contribution of coefficients $\bco{n}_l$ of higher order $l$ in the couplings $\cnr{i}$ to the gradient of the flow becomes continuously suppressed with larger $\ak$ by the factor $F_\rk^l/F_\rk^1 \sim \rk^{1-l}$, see Fig.~\ref{fig:FGflows}. This corresponds to the factor $\frac{\rk-s}{\rk}$ at scale ${\ak}^s$ associated to the couplings in the original FRG equation \eqref{eq:Wetterichresult}.
\item Third, there is a continuous change in the $\eta_k$-dependence in the second term governed by the ratio $G_\rk^l/F_\rk^1$ with large-$\ak$ asymptotics $\sim \rk^{1-l} \frac{\ks}{\rk-1+\ks}$, see Fig.~\ref{fig:FGflows}.
\item Fourth, and most significantly, the effective dimension $\efd(k)$ interpolates between zero and $\uvd$. We show the exact form of this interpolation in Fig.~\ref{fig:dimensionflow}.
\end{itemize}
Note that these results are only mildly sensitive to the summation schemes,
as can be seen for example from comparing the effective dimension, and thus the non-autonomous function $F_\rk^l$, in Fig.~\ref{fig:dimensionflow}. 
This justifies a posteriori to approximate the exact sum in terms of the integral, Eq.~\eqref{eq:definitionthresholdfunctions}.
The only case where the integral approximation might lead to a qualitatively different result is the small-$k$ behaviour of $\gr^l$ as discussed above, the difference being that with exact traces the anomalous dimension might not vanish completely for $k\rightarrow0$ controlled by Eq.~\eqref{eq:etaflowIR}. 

Summing up, the continuous rescaling provides an understanding of the renormalization group flow of the full non-autonomous equations in terms of standard FRG equations where some parameters, in particular the dimension, are flowing themselves.
In this way, we gain an understanding what happens when following a flow trajectory through phase space, even though we cannot solve the full non-autonomous equations.
At all scales $k$, the flow equation is a modified $\textrm{O}(N)$-model equation.
Flowing from large to small $k$, it changes continuously from the $N\rightarrow\infty$ case to the $N=\ccu$ case and the additional relative factor $\rk$ between $\m$ and the other couplings $\cn{i}$ becomes switched off.
Most significantly, the effective dimension $\efd$ of TGFT changes continuously from $\efd=\uvd$ to $\efd=0$%
\footnote{This dimensional flow should not be confused with the flow of the physical dimension of the ensemble of $\rk$-dimensional pseudo manifolds generated by the tensor fields. Tensors of any rank $\rk$ describe an ensemble of manifolds which form a branched-polymer phase at large $\ak$ and at criticality~\cite{Gurau:2013th}, i.e., they have spectral dimension $4/3$. Sub-leading contributions (e.g.~necklace interactions) yield also a planar phase with spectral dimension two~\cite{Bonzom:2015gt,Lionni:2017tk}. In general, for an ensemble of such discrete geometries one expects a flow of the spectral dimension from such reduced large-$\ak$ value to the dimension $\rk$ of the discrete manifolds \cite{Calcagni:2015is, Thurigen:2015uc}. 
But the effective dimension $\efd$ considered here is a conceptually different quantity: It is the dimension of the tensor fields according to their scaling behaviour.}.

This means, the phase diagram determining the gradient of the flow at a given scale $k$ undergoes important qualitative changes along the flow.
For it is the dimension $\efd$ which discriminates between phase diagrams with or without relevant (non-Gaussian) fixed points. 
The generic picture for rank $\rk>\crd+1$ is the following: 
At large $k$, the diagram has a GFP describing a phase transition with mean-field exponents; along the flow, when the effective dimension passes $\efd(k)=\crd$, this fixed point (the one with one positive exponent) becomes a NGFP continuously moving in phase space depending on $k$ through $\efd(k)$ (cf.~Figs.~\ref{fig:massbranches} and~\ref{fig:massbrancheseta}). 
In the LPA this Wilson-Fisher type NGFP is qualitatively the same as in the corresponding $\efd(k)$-dimensional $\textrm{O}(N)$ model and is thus expected to persist to the scale $k$ where $\efd(k)=2$.
In the LPA$'$, taking into account the anomalous dimension, this NGFP lies in a different orthant of phase space and diverges already for some $\efd(k)>2$.
In any case, below the scale $k$ where $\efd(k)=2$ there should be no fixed point separating a broken from an unbroken phase anymore. 
As a consequence, we expect that for any flow trajectory there is a finite $k$ of order one or larger at which the $\textrm{U}(1)$ symmetry of the complex field potential (or $\mathbb{Z}_2$ symmetry for real field) is restored.

\subsubsection{Symmetry restoration}\label{Section:symmetryrestoration}

As just pointed out, the effective dimension $\efd$ vanishes in the small-$\ak$ limit, i.e. the deep IR regime. One obtains an autonomous flow equation~\eqref{eq:IRequation} which is that of scalar theory on Euclidean space with vanishing dimension. Since by virtue of the Mermin-Wagner-Hohenberg theorem spontaneous breaking of continuous symmetries in two or less dimensions and that of discrete symmetries in less than two dimensions is forbidden~\cite{Hohenberg:1967zz,Mermin:1966fe,Coleman:1973ci}, we thus anticipate no phase transition between a broken and symmetric phase of the global $\textrm{U}(1)$ symmetry in TGFT on $\textrm{U}(1)^\rk$ at any rank $\rk$. Previous research using mean-field arguments~\cite{Pithis:2018bw} as well as FRG studies applied to TGFTs~\cite{Benedetti:2015et,BenGeloun:2016kw, BenGeloun:2015ej,TGFT1,TGFT2} also nourish this expectation. 

\begin{figure}
\centering
  \includegraphics[width=.45\linewidth]{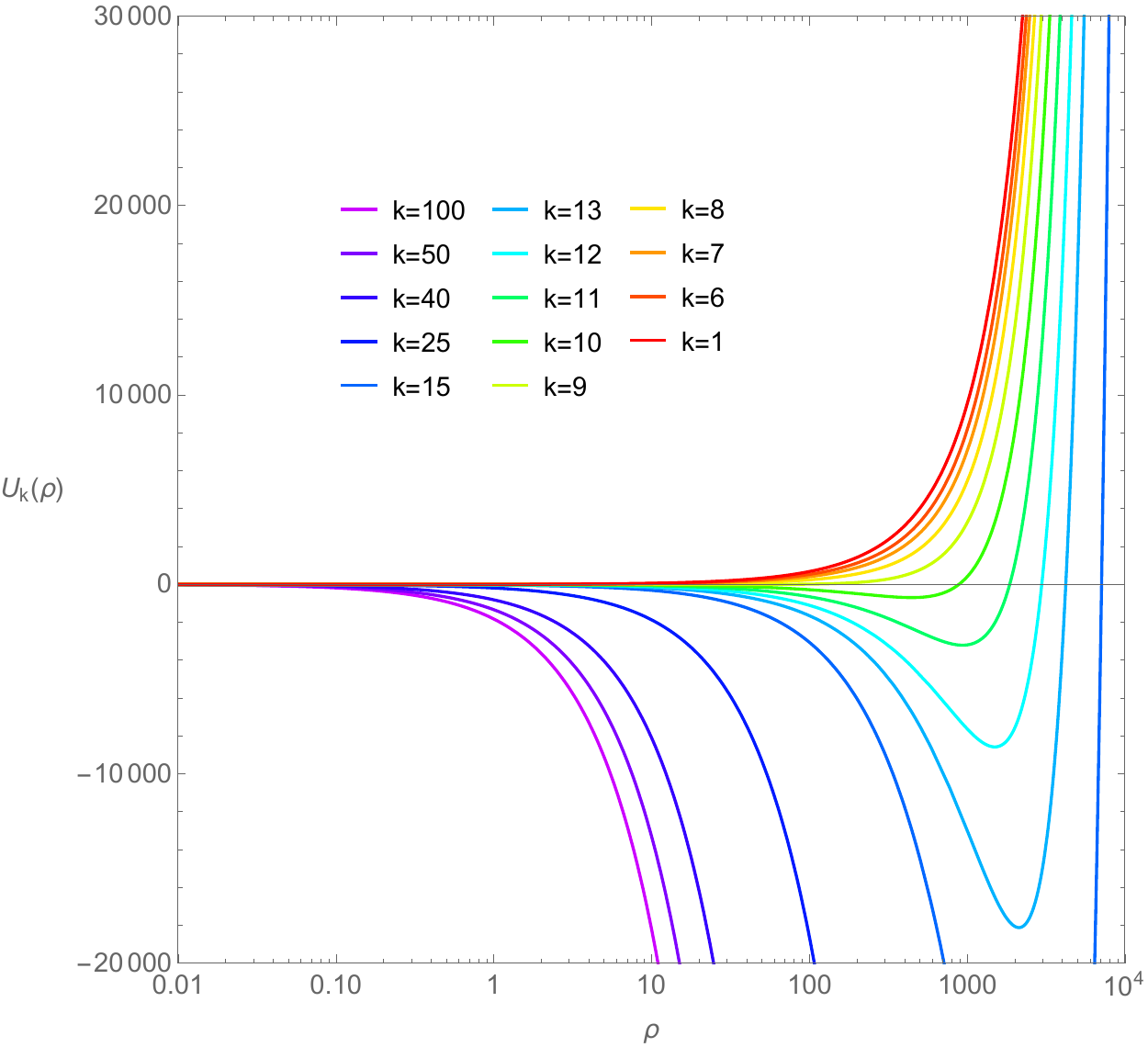}
  \includegraphics[width=.435\linewidth]{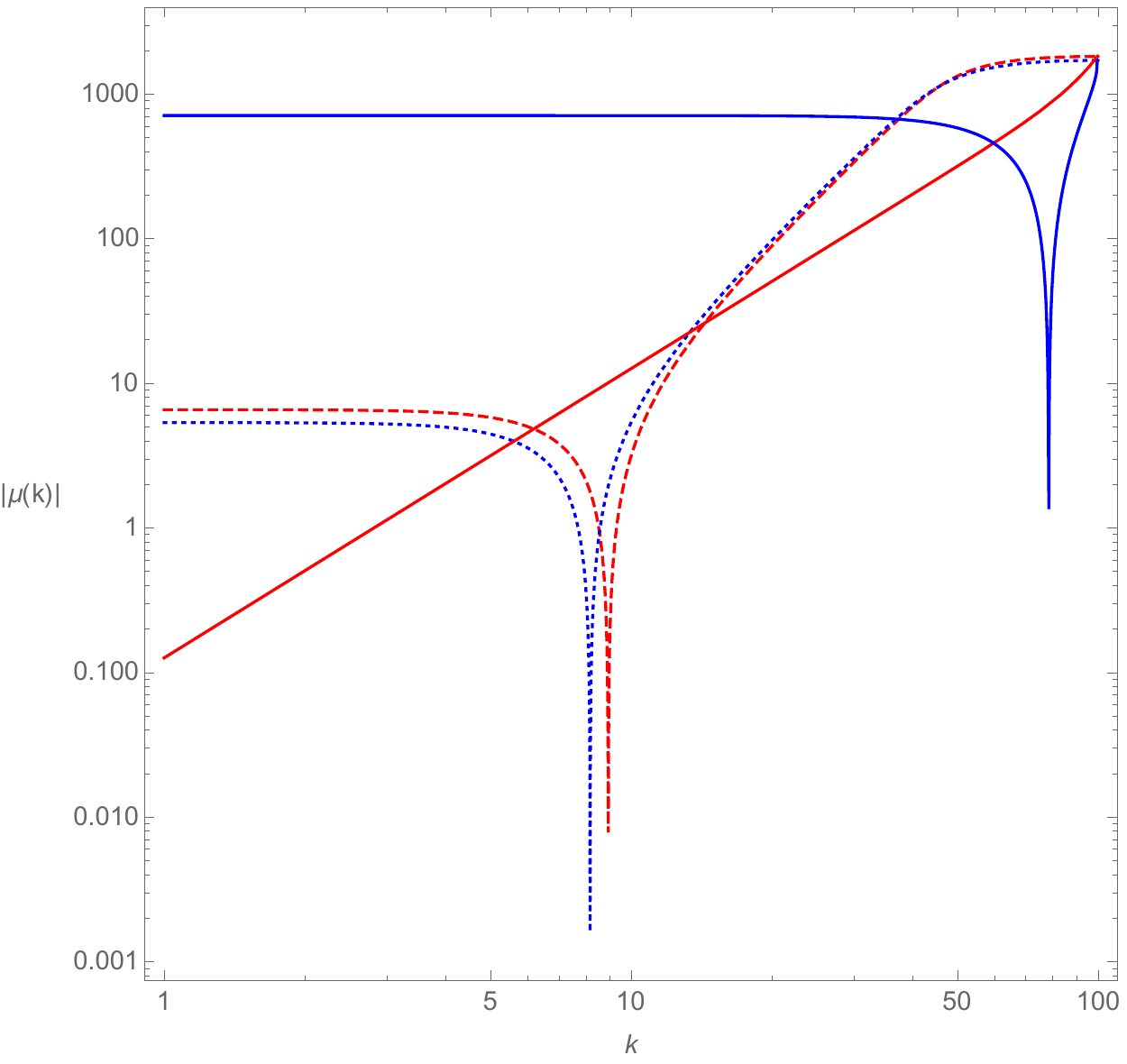}
\caption{Left panel: The flow of the dimensionful potential $\pot(\rho)$ at rank $\rk=5$ in the $\nmax=4$ truncation between $k=100$ and $k=1$ with $\cs=1$ for initial conditions at $\Lambda=100$ close to the UV non-Gaussian fixed point in that truncation: $Z(\Lambda)=1$, $\mu(\Lambda)=-0.86 \Lambda^2$, $\lambda_2(\Lambda)=0.090\Lambda^0$, $\lambda_3(\Lambda)=0.084\Lambda^{-2}$ and $\lambda_4(\Lambda)=0.075\Lambda^{-4}$. Right panel: Flow of the modulus of $\mu(k)$ in the $\nmax=4$ truncation for: (I) the system of non-autonomous $\beta$-functions~\eqref{eq:betaequations} with $\cs=1$ (dashed lines) and (II) the set of autonomous $\beta$-functions in the large-volume limit~\eqref{eq:UVequations} (continuous lines). Initial conditions are the same as in the left panel except for $\mu(\Lambda)= -0.91\Lambda^2$ (red) and $\mu(\Lambda)=-0.85 \Lambda^2$ (blue).}\label{figure:flowrank4}
\end{figure}

To assess this point, we numerically integrate%
\footnote{The present set of dimensionful flow equations is a set of coupled non-linear differential equations of $1$st order which we solve using the NDSolve routine of Mathematica employing the Runge-Kutta method at machine precision.}
the full non-autonomous equations, Eq.~\eqref{eq:betaequations}, for the dimensionful potential $\pot(\rho)=\mu\rho +  V_k(\rho)$ (cf. Eq.~\eqref{eq:isopotential}) from at $k=\Lambda$ down to to small $k$. Commencing with \textit{any} potential which explicitly exhibits spontaneous breaking of the global $\textrm{U}(1)$ symmetry in the UV, i.e. $\rho_\Lambda\ne0$, we then observe that the potential \textit{always} completely flattens out at some finite value of $k$ indicating symmetry restoration towards the IR.%
\footnote{Alternatively, the flow equation could be expanded around a non-trivial value for the field configuration $\rho=\rho_{0k}+\delta\rho_k$. Studying the flow of $\rho_{0k}$ then leads to the same qualitative observation, i.e., the system always settles into the symmetric phase at a finite value of $k$.}
This holds true for any rank $\rk$ as well as $\zeta=1/2,~1$ and is underlined here for the concrete case of the complex-valued rank-$5$ TGFT with $\zeta=1$. In Fig.~\ref{figure:flowrank4} we report the flow of the dimensionful potential and of $\mu$ in the $\nmax=4$ truncation. There, we also juxtapose the flows of $\mu$ in both phases in the large-volume limit with the flows in the compact case with same initial conditions.

We would like to draw attention to the fact that the mechanism behind the restoration of the global $\textrm{U}(1)$ (or $\mathbb{Z}_2$) symmetry is universal and thus extends well beyond the cyclic-melonic truncation considered here. That the system always settles into the symmetric phase derives its origin from the existence of the zero modes in the spectrum of the theory. Importantly, the $\rk$-fold zero mode induces the constant term in the non-autonomous part $F_r$ in the flow equations, Eq.~\eqref{eq:betaequations}. In the limit where $k\rightarrow 0$, this term dominates and prompts a scaling of the couplings with dimension $\efd=0$. It is clear that this mechanism does not depend on the combinatorial structure of the interactions and can be solely accredited to the compactness of the domain of the field. This complies with the general tenet that in systems of finite size there are no true phase transitions~\cite{zinn2002quantum,strocchi2005symmetry}. In spite of having exemplified this phenomenon by means of a cyclic-melonic potential approximation at arbitrary order here, we anticipate symmetry restoration for the full potential of any TGFT with discrete spectrum which contains isolated zero modes. In particular, our results suggest that for a phase transition of the above type to occur for TGFTs, the non-compactness of the domain is a critical prerequisite.

\section{Conclusion and discussion}\label{Section:conclusion}

The main purpose of this article was to investigate the phase structure of rank-$\rk$ TGFT on $\textrm{U}(1)$ with cyclic-melonic interactions using the FRG method. To this aim, we have for the first time derived the FRG equation in a local-potential approximation at any order and at any scale in 
TGFT and analysed its fixed point solutions. In particular, we wanted to scrutinize 
if there are phase transitions between a broken and unbroken phase of the global $\textrm{U}(1)$ (or $\mathbb{Z}_2$) symmetry.
Our main results are:

$(1)$ There are no such phase transitions on compact group configuration space of fixed volume size.
The symmetry is always radiatively restored at small enough momentum scale $k$ since the effective dimension $\efd(k)$ of the field theory flows to zero in the limit $k\to0$. Thus, one has $\efd<2$ below some $k$ and there can be no phase transitions below two dimensions.
This phenomenon is essentially due to the isolated zero modes in the spectrum on a compact space. 
Thus, we expect the same to hold true for TGFT taking into account any tensor-invariant interactions, on any compact group $G$.

$(2)$ In TGFT on a compact group $G$,  upon appropriate rescaling the FRG equation turns out to be a differential equation not just in the renormalization group scale $k$ but equally in its combination $\ak=\cs k$ with the compactness (volume) size $\cs$.
This provides a clear relation of TGFT and tensor models. Integrating out high modes up to the scale $k$, the effective group fields are tensors of size $\ak$ in momentum (representation) space. As a consequence, also the regulator is effectively a function in $\ak$. To meet regulator conditions, the power $\ak^\ks$ in this dependence is determined by the power $\ks$ of the theory's propagator.

$(3)$ In the large-$\ak$ limit, above the critical rank $\crr = \crd+1=4\zeta + 1$ the Gaussian fixed point describes a phase transition with mean-field exponents. 
The reason is simply that the scaling dimensions of couplings necessary to rescale the flow equations in this limit are the same as in standard local scalar field theory on $d$-dimensional Euclidean space when setting $d$ to $\uvd=\rk-1$.
The large-$\ak$ limit can be interpreted as a large-volume limit which corresponds to the TGFT on $\mathbb{R}^\rk$ \cite{BenGeloun:2015ej,BenGeloun:2016kw}. 
Thus, in this interpretation of the theory there can be phase transitions.

$(4)$ In the large-$\ak$ limit, below $\crr$ this phase transition persists in the LPA and is captured by a Wilson-Fisher type non-Gaussian fixed point which is qualitatively the same as in $\uvd$-dimensional $\textrm{O}(N)$ models but differs quantitatively with respect to its critical exponents. In particular, for a theory with quadratic propagator ($\zeta=1$) this non-Gaussian fixed point exists in the LPA for rank $\rk=4$.

$(5)$ In the LPA$'$, when taking into account the tensor-specific dynamics of the anomalous dimension, this non-Gaussian fixed point is modified and diverges already for some rank $\rk$ in between $3\zeta<\uvd=\rk-1<4\zeta$. 
However, at finite truncation there is a second non-Gaussian fixed point which has the same signs as the Wilson-Fisher fixed point both for the couplings and critical exponents. 
This fixed point candidate persists up to some rank $\noser$ and seems to converge for ranks close to $\noser$. The exact domain of convergence remains an open question. 
For $\zeta=1$ we find $\noser\approx5.55$ and for $\rk=5$ there are indications that this non-Gaussian fixed point exists. If true, there is a phase transition described by a non-Gaussian fixed point also for the critical rank $\crr=5$.

\

An important approximation in our calculations has been the projection onto uniform field configurations.
In the context of TGFT we have applied this for the first time.
To understand the scope of our results it is crucial to understand its implications both from a technical and conceptual point of view.

Technically, the constant field projection is a common tool in local field theories such as $\textrm{O}(N)$ models \cite{Berges:2002ga} but is more subtle for theories with tensorial interactions. 
The projection is ignorant to the distinction between combinatorially different interactions at a given order $(\gfb\gf)^n$. Also a cyclic-melonic potential generates any possible tensorial interactions under the renormalization group flow. This distinction is washed out by the projection.
However, crucial tensorial information is retained in the operator $\mathcal{O}^\ell$, Eq.~\eqref{eq:deltaoperatorg}. In particular, it covers all leading-order contributions and the approximation should thus be trustworthy in the large-$\ak$ limit.
Furthermore, up to quartic order in the truncation of the potential the projection is in good agreement with the case without projection (Sec.~\ref{Subsection:beta-functions}).
Taking into account how important it is to consider truncations with infinitely many couplings to assess whether a fixed point candidate persists for larger and larger truncations, our approximations might be viewed as a first step towards understanding the phase diagram of the full TGFT.  
In this spirit, a research strategy to this aim is to weaken the approximations, step by step including the impact of disconnected interactions, other melonic interactions, and finally non-melonic interactions \cite{Carrozza:2017dl,BenGeloun:2018ekd}.

From a conceptual point of view, it is not
obvious to understand the physical meaning of the projection on fields which are constant in configuration space. 
Given that the configuration space is related to parallel transports and thus local curvature of the gra\-vi\-tational field, the projection onto constant field configurations implies that all equally contribute to the effective dynamics. 
A potential extension could be to project onto non-trivial (non-uniform) global minima similar to those obtained for the dynamical Boulatov model~\cite{BenGeloun:2018eoe}.
In particular, to check the robustness of our results, it would be interesting to see if for such configurations the TGFT systems become effectively zero-dimensional, too.
Another possibility worth to be explored is to project onto constant field configurations in momentum space instead, which would in turn imply a peaking on vanishing curvature modes in configuration space. While this perspective has been a basic assumption to study phase transitions in spin foam models~\cite{Bahr:2017kr,Steinhaus:2018aav}, its implementation in the context of TGFT together with an adapted FRG analysis has yet to be explored. 

\

On a compact group with fixed volume we have established the very restrictive result that the U$(1)$ symmetry of the potential is always restored.
Thus, there can be no phase transition with respect to this symmetry.
But from a quantum-gravity perspective, a transition from the discrete geometries in the perturbative regime of TGFT to a phase of continuum geometries is of high interest and of particular relevance to the GFT condensate cosmology approach~\cite{Gielen:2013cr,Gielen:2014gv,Gielen_2016,Oriti:2016qtz,deCesare:2016rsf,Gielen:2017eco,Oriti_2017,Pithis_2019}.
It still has to be checked explicitly if our result of a universal symmetry restoration also applies to TGFT with other groups such as $\text{SU}(2)$ or with additional structure such as a closure constraint~\cite{Freidel:2005jy,Oriti:2012wt,Krajewski:2012wm}; but 
the zero-mode should have the same effect for such models, too.
Furthermore, it would also be desirable to rigorously establish the pendant of the Mermin-Wagner theorem of local scalar field theories~\cite{Hohenberg:1967zz,Mermin:1966fe,Coleman:1973ci} for combinatorially non-local theories such as TGFT to independently ensure the result of symmetry restoration for an effective dimension below two. 
This might be possible using operator-algebraic methods in the GFT context~\cite{Oriti:2013vv,Kegeles:2017ems,Kegeles:2018tyo}. 
But even without any more general such mathematical proof, the zero-mode effect should universally apply for compact configuration spaces.

A potential loophole out of this cul-de-sac is to consider models with an additional gravitational (Holst-Plebanski) constraint \cite{Perez:2012wv} which might provide a mechanism to remove the zero-modes from the spectrum. 
Another possibile way out could be to extend the configuration space adding degrees of freedom encoding a matter reference frame \cite{Oriti:2016qtz,Gielen:2017eco,Gielen:2018fqv}.
This adds non-compact directions to the domain of the fields which leads to theories reminiscent of SYK models~\cite{Rosenhaus:2019mfr}. 
Clearly, the zero-mode effect is absent if the tensorial degrees of freedom are not dynamic as in SYK-type models~\cite{Delporte:2018iyf}. 
But even if they are dynamic and effectively vanish at small $k$ the additional non-compact directions could persist.
Thus, it could very well be that phase transitions of the above type are actually realized in models with dynamical tensorial degrees of freedom including a matter reference frame. 

Staying in the standard TGFT setting, the obvious way to facilitate a phase transition to a broken or condensate phase with a tentative interpretation as continuum spacetime is to consider a non-compact group. This could simply be obtained in the spirit of the large-volume limit which is basically equivalent to 
TGFT on $\mathbb{R}^\rk$~\cite{BenGeloun:2016kw, BenGeloun:2015ej}. 
To this end, it would be important not only to go beyond the cyclic-melonic LPA$'$ but also to improve on our results in this LPA$'$, in particular on the domain of convergence of the second non-Gaussian fixed point to settle the question of existence of phase transitions at the critical rank $\crr=5$ and below. This might be possible generalizing analytic methods used for $\textrm{O}(N)$ models \cite{Tetradis_1996,D_Attanasio_1997}.

On the other hand, for a physical theory of quantum gravity it might in any case be necessary to choose a non-compact group since holonomies in gravity are captured by the Lorentz group.
In particular, the causal structure of spacetime is encoded therein. It has already been demonstrated for a GFT toy-model on $\textrm{SL}(2,\mathbb{R})$~\cite{Pithis:2018bw} that for such a configuration space mean-field theory is sufficient to describe a phase transition between a broken and unbroken phase. This result could serve as a motivation to study the phase structure of full-blown GFT models for Lorentzian quantum gravity in the future.

\subsection*{Acknowledgments}
The authors thank D. Benedetti, J. Ben Geloun, A. Duarte Pereira, A. Eichhorn, D. Oriti and R. Percacci for discussions and critical remarks. The authors are particularly greatful to S. Carrozza for the encouragement to study a local-potential approximation using  cyclic-melonic interactions.

The work of AGAP leading to this publication was supported by the PRIME programme of the German Academic Exchange Service (DAAD) with funds from the German Federal Ministry of Education and Research (BMBF).
The work of JT was funded by the Deutsche Forschungsgemeinschaft (DFG, German Research Foundation) in two ways,
primarily under the author's project number 418838388 and
furthermore under Germany's Excellence Strategy EXC 2044 –390685587, Mathematics M\"unster: Dynamics–Geometry–Structure.

\appendix


\section{Traces and threshold functions}\label{sec:Traces}
\renewcommand{\aa}{x}
\renewcommand{\bb}{y}
\renewcommand{\ak}{N}

Here we prove the resummation of the trace on the right-hand side in the FRG equation~\eqref{eq:WetterichLitim}.
The characteristic of the Hessian of the effective average action $\gfk$ in combinatorial non-local theories are terms with various combinations of zero modes in momentum (representation) space. 
To keep track of the relevance of the different terms, we introduce parameters $\aa, \bb, \cc, \dd$ and consider for a function $f:\mathbb{R}\rightarrow\mathbb{R}$ the sum
\[
\suma\rk{f}(\ak;\aa, \bb, \cc, \dd):=  \sum_{\vrep\in\set\rk\ak}
\frac{\sum_{\ell=1}^\rk f(\rep_\ell)}{\aa +\bb\sum_{\ell=1}^\rk \delj\ell + \cc\sum_{\ell=1}^\rk \prod_{b\ne\ell}\delj{b} +\dd \prod_{\ell=1}^\rk \delj\ell}
\]
where $\set\rk\ak$ denotes some discrete symmetric set of $\rk$-tuples $\vrep$ having some size $\ak$.
Here symmetric means that if $\vrep\in\set\rk\ak$ then also any permutation of the entries of $\vrep$ is in $\set\rk\ak$.

\subsubsection*{Volumes}

Three cases are of interest here.
The simplest case is a box, that is $\set\rk\ak$ 
is a hypercube  with zero-excluding volume 
\[\label{eq:boxvolume}
\ku\rk(\ak) := \sum_{\vrep\in\set\rk\ak} \prod_{\ell=1}^\rk (1 - \delj\ell) 
= \prod_{\ell=1}^\rk \sum_{\rep_c=-\ak}^\ak (1-\delj\ell)
= (2\ak)^{\rk} \, .
\]
We have defined $\ku\rk$ excluding all zeros $\rep_\ell=0$ since this will be the relevant sum occurring in $\suma{\rk}{f}$ due to the Kronecker symbols in the denominator.
For the quadratic cutoff we actually need the sum over a discrete $\rk$-dimensional ball $\ball\rk\ak$ of radius $\ak$ (again without zeros) which we approximate by 
\begin{align}
\ku\rk(\ak) = \sum_{\vrep\in \ball\rk\ak} \prod_{\ell=1}^\rk (1 - \delj\ell) 
&=  \sum_{\rep_1=-\ak}^\ak \sum_{\rep_2=-\sqrt{\ak^{2}-\rep_{1}^{2}} }^{\sqrt{\ak^{2}-\rep_{1}^{2}}} ... \sum_{\rep_\rk=-\sqrt{\ak^{2}-\sum_{\ell=1}^{\rk-1}\rep_{\ell}^{2}}}^{\sqrt{\ak^{2}-\sum_{\ell=1}^{\rk-1}\rep_{\ell}^{2}}}  \prod_{\ell=1}^\rk (1 - \delj\ell) \nonumber\\
&\approx \int_{\ball\rk\ak} [\extd p]^{\rk} 
= \rk \vol\rk \int_{0}^{\ak} p^{\rk-1} \extd p
= \vol\rk \ak^{\rk} 
\label{eq:ballvolume}
\end{align}
where 
\[
\vol\rk = \frac{\pi^{\rk/2}}{\Gamma(\frac{\rk}{2}+1)}
= \Bigg\{\begin{array}{cl}
        \frac{\pi^s}{s!}, & \quad \rk = 2s \\
        \frac{2(s!)(4\pi)^s}{(2s+1)!}, & \quad \rk=2s+1
        \end{array}
\]
is the volume of the continuous $\rk$-ball of radius one.
Finally, in the case of a linear regulator one might also be interested in the sum over a ``ball'' in $l_{1}$-norm $\set\rk\ak = \simp{\rk}$, that is a simplex without boundary in each quadrant, that is a generalized octahedron 
\[\label{eq:simplexvolume}
\ku\rk(\ak) = \sum_{\vrep\in\simp\rk} \prod_{\ell=1}^\rk (1 - \delj\ell) = 
2^\rk \sum_{\rep_1=1}^\ak \sum_{\rep_2=1 }^{\ak-\rep_{1}} ... \sum_{\rep_\ell=1}^{\ak-\sum_{\ell=1}^{\rk-1}\rep_{\ell}} 1
= 2^\rk \poch\rk(\ak)
\]
where
\[
\poch{\rk}(\ak) := \frac{1}{\rk!}\prod_{i=0}^{\rk-1}(\ak-i)
= \frac{1}{\rk!} (\ak-\rk+1)_\rk 
= \frac{1}{\rk!} \frac{\Gamma(\ak+1)}{\Gamma(\ak+1-\rk)} 
\]
is expressed in terms of Pochhammer symbols $(\cdot)_\rk$ or, respectively, the Gamma function $\Gamma$.

\subsubsection*{The general trace formula}

In the following, we will prove by induction that for $\rk\ge2$
\[\label{eq:trace}\boxd{
\suma\rk{f}(\ak;\aa, \bb, \cc, \dd) = 
\frac{\rk f(0)}{\aa+\rk(\bb+\cc)+\dd} + 
\sum_{s=1}^\rk \binom{\rk}{s}\frac{s\kf{s}(\ak) + (\rk-s)f(0)\ku{s}(\ak)}{\aa +(\rk-s)\bb + \delta_{s,1}\cc}
}
\]
where 
\[
\kf{s}(\ak) := 
\sum_{\rep_1,..\rep_s \ne 0} 
f(\rep_\ell) 
\]
is the sum over all $(\rep_1,...,\rep_s)\in\set{s}\ak$ excluding any $\rep_\ell=0$,
and $\rep_\ell$ is any of them (which is meaningful because $\set{s}\ak$ is symmetric).
For constant functions $f(\rep)=c$ we have $\kf{s} = c\ku{s}$ such that, in particular,
\[\label{eq:trace1}
\suma\rk{{1/\rk}}(\ak;\aa, \bb, \cc, \dd) = 
\frac{1}{\aa+\rk(\bb+\cc)+\dd} + 
\sum_{s=1}^\rk \binom{\rk}{s}\frac{\ku{s}(\ak)}{\aa +(\rk-s)\bb + \delta_{s,1}\cc}.
\]
On the other hand, for functions with $f(0)=0$ the sum simplifies to
\[\label{eq:trace2}
\suma\rk{f}(\ak;\aa, \bb, \cc, \dd) 
= \sum_{s=1}^\rk \binom{\rk}{s}\frac{s\kf{s}(\ak)}{\aa +(\rk-s)\bb + \delta_{s,1}\cc}.
\]
These two cases are used to evaluate the trace in the FRG equation, Eq.~\eqref{eq:WetterichLitim}, with 
$x=\wfr k^\ks+\m$, $y=z= V_k^\ell{}'(\rho) \equiv 
V'_k(\rho)/\rk$ and $u = \epsilon 2\rho V''_k(\rho) - V'_k(\rho)$ to find Eq.~\eqref{eq:Wetterichresult}.

\subsubsection*{The proof}

To start the proof, the $\rk=2$ case is straightforward:
\begin{align}
\suma2{f}(\ak;\aa, \bb, \cc, \dd) &= 
\sum_{\rep_1} \sum_{\rep_2}
\frac{f(\rep_1)+f(\rep_2)}{\aa +\bb(\delj 1 + \delj 2) + \cc(\delj{2}+\delj{1}) +\dd \delj{1}\delj{2}} \\
&=\sum_{\rep_1}\left[
\frac{f(\rep_1)+f(0)}{\aa +\bb(\delj 1 + 1) + \cc(1+\delj{1}) +\dd \delj{1}}
+\sum_{\rep_2\ne 0} \frac{f(\rep_1)+ f(\rep_2)}{\aa +\bb\delj 1  + \cc\delj{1}} \right]\nonumber \\
&= \frac{\sum_\ell f(0)}{\aa +2(\bb+ \cc) +\dd}
+ 2\frac{\sum_{\rep\ne0}f(\rep)+f(0)\ku1(\ak)}{\aa+\bb+\cc} 
+ \sum_{\rep_1,\rep_2\ne0}\frac{f(\rep_1)+f(\rep_2)}{\aa} \, . \nonumber
\end{align}
Then, for any $\rk+1>2$ we evaluate first the sum over $\rep_{\rk+1}$, 
\begin{align}
\suma{\rk+1}{f}(\ak;\aa, \bb, \cc, \dd) =&  \sum_{\rep_1,...,\rep_{\rk}} \Bigg(
\sum_{\rep_{\rk+1}\ne 0} \frac{\sum\limits_{\ell=1}^{\rk}f(\rep_\ell)+f(\rep_{\rk+1})}{\aa +\bb \sum\limits_{\ell=1}^{\rk} \delj\ell 
+ \cc \prod\limits_{\ell=1}^{\rk} \delj\ell} \\
&+\frac{\sum\limits_{\ell=1}^{\rk}f(\rep_\ell)+f(0)}{\aa +\bb \left(\sum\limits_{\ell=1}^{\rk} \delj\ell +1 \right)
+ \cc \left(\prod\limits_{\ell=1}^{\rk} \delj\ell + \sum\limits_{\ell=1}^{\rk} \prod\limits_{b\ne\ell}\delj{b} \right)
+\dd \prod\limits_{\ell=1}^{\rk} \delj\ell}   \Bigg) \nonumber\\
=& \,\suma{\rk}{f}(\ak;\aa+\bb, \bb, \cc,\cc+ \dd) 
+ f(0) \suma{\rk}{{1/\rk}}(\ak;\aa+\bb, \bb, \cc,\cc+ \dd) \\
&+\sum_{\rep_{\rk+1}\ne 0} \left(\suma{\rk}{f}(\ak;\aa, \bb, 0,\cc)
+ f(\rep_{\rk+1}) \suma{\rk}{{1/\rk}}(\ak;\aa, \bb, 0,\cc) \right) \, , \nonumber
\end{align}
and use then the induction hypothesis Eqs.~\eqref{eq:trace},~\eqref{eq:trace1} for the four resulting terms individually 
\begin{align}
\suma{\rk+1}{f}(\cdot;\aa, \bb, \cc, \dd)
=&\frac{\rk f(0) + f(0)}{\aa+(\rk+1)(\bb+\cc)+\dd} + 
\sum_{s=1}^\rk \binom{\rk}{s}\frac{s\kf{s} + (\rk-s)f(0)\ku{s}+f(0)\ku{s}}{\aa +(\rk+1-s)\bb + \delta_{s,1}\cc} \nonumber\\
&+\frac{\rk f(0)\ku1 + \kf{1}}{\aa+\rk\bb +\cc} + 
\sum_{s=1}^\rk \binom{\rk}{s}\frac{ s\kf{s+1} + (\rk-s)f(0)\ku{s+1} + \kf{s+1}}{\aa +(\rk-s)\bb} \nonumber\\
=& \frac{(\rk+1) f(0)}{\aa+(\rk+1)(\bb+\cc)+\dd} + 
\sum_{s=1}^{\rk} \binom{\rk}{s}\frac{s\kf{s} + (\rk+1-s)f(0)\ku{s}}{\aa +(\rk+1-s)\bb + \delta_{s,1}\cc} \nonumber\\
&+ \sum_{s=1}^{\rk} \binom{\rk}{s-1}\frac{s\kf{s} + (\rk+1-s)f(0)\ku{s}}{\aa +(\rk+1-s)\bb +\delta_{s,1}\cc} 
+\frac{(\rk+1)\kf{\rk+1}}{x}\nonumber\\
=& \frac{(\rk+1) f(0)}{\aa+(\rk+1)(\bb+\cc)+\dd} + 
\sum_{s=1}^{\rk+1} \binom{\rk+1}{s}\frac{s\kf{s} + (\rk+1-s)f(0)\ku{s}}{\aa +(\rk+1-s)\bb + \delta_{s,1}\cc}
\end{align}
where in the first step one shifts in the second sum $s\mapsto s-1$ and uses 
$\binom{\rk}{s-1}+\binom{\rk}{s}=\binom{\rk+1}{s}$ to then combine the two series in the next step.
This proves Equation \eqref{eq:trace}.

\subsubsection*{Threshold functions}
\renewcommand{\ks}{\gamma}
\renewcommand{\vks}[1]{v_{#1}^{(\ks)}}

For $\eta_k\ne 0$ there is a part in the FRG equation~\eqref{eq:WetterichLitim}  with the Casimir in the numerator.
Thus, we need threshold functions $\kf{s}$ for $f(\rep)=|\rep|^{\ks}$,
\[
\kw{s}(\ak):=\sum_{\vrep\in\set{s}\ak} |\rep_\ell|^{\ks} \, .
\]
For the hypercube case they are given by $H_n^{(m)}$, the $n$'th generalized harmonic number of order $m$, as
\[
\kw{s}(\ak)= 2H_\ak^{(-\ks)} \ku{s-1}(\ak)
 = 2H_\ak^{(-\ks)} (2\ak)^{s-1} 
 = (2\ak)^{s}\cdot \biggl\{\begin{array}{lr}
        \frac{1}{2} (\ak+1) & \text{for } \ks = 1, \\
        \frac{1}{6} (\ak+1)(2\ak+1) & \text{     } \ks = 2,\\
        ... &
        \end{array} 
\,~.
\]
For the simplex case we are aware of closed expressions of the general $\ks$-dependent sum only up to $s=3$,
\begin{eqnarray}
\kw{1}(\ak) &=& 2 H_\ak^{(-\ks)},\\
\kw{2}(\ak) &=& 4\left(-H_\ak^{(-\ks-1)} +(\ak+1)H_\ak^{(-\ks)} \right),\\
\kw{3}(\ak) &=& 8\cdot\frac{1}{2}\left(H_\ak^{(-\ks-2)} - (2\ak+3)H_\ak^{(-\ks-1)} + (\ak+1)(\ak+2)H_\ak^{(-\ks)}\right) \,.
\end{eqnarray}
Alternatively, in the special cases $\ks=1,2$ we find for arbitrary dimension $s$,
\begin{eqnarray}
I_1^{(s)}(\ak) &=& \frac{2^s}{(s+1)!}\prod_{i=0}^s(\ak+i), \\ 
I_2^{(s)}(\ak) &=& \frac{2^s}{(s+2)!}(2\ak+s)\prod_{i=0}^s(\ak+i) \, .
\end{eqnarray}
For the integral approximation over the ball $\ball{s}\ak$ it is more convenient to consider the function as a weighted sum over the $l_{\ks}$-norm of the s-tuple $\vrep$,
\[
\kw{s}(\ak)
= \frac{1}{s} \sum_{\vrep\in\set{s}\ak} s |\rep_\ell|^{\ks} 
= \frac{1}{s} \sum_{\vrep\in\set{s}\ak} \sum_{\ell=1}^s |\rep_\ell|^{\ks} 
= \frac{1}{s} \sum_{\vrep\in\set{s}\ak} {(||\vrep||_{\ks})}^\ks \, .
\]
This expression has now a straightforward integral approximation
\begin{align}\label{eq:thresholdzeta}
\kw{s}(\ak) 
&\approx \frac{1}{s}\int [\extd p]^{s} p^{\ks}
= \vks{s}\int_{0}^{\ak} p^{s +\ks -1} \extd p 
= \frac{\vks{s}}{s+\ks} \ak^{s+\ks} 
\end{align}
where, consequently, we have now integrated over the ball in $L^{\ks}$ norm with unit volume
\[
\vks{s} = 2^s  \frac{\Gamma(\frac1{\ks}+1)^s}{\Gamma(\frac{s}{\ks}+1)} \, .
\]
In this way, one also has a generalization of the integral-approximated volume Eq.~\eqref{eq:ballvolume}
\[\label{eq:integralvolume}
\ku{s}(\ak) = \vks{s}\ak^s 
\]
used above.

\section{Traces at quadratic order in the field expansion}\label{sec:etaequation}
\renewcommand{\ks}{2\zeta}
\renewcommand{\m}{\mu_k}
\renewcommand{\ak}{N_k}
\renewcommand{\aa}{\alpha}
\renewcommand{\bb}{\beta}

We present here the calculations for the FRG equation~\eqref{eq:Wetterich} at quadratic order without projection onto constant field, that is we calculate the $\gfb\gf$ terms.
On the left of the equation one has
\[\label{eq:lhsquadratic}
k\partial_k \gfk |_{\gfb\gf} 
= \sum_\vrep \left(\frac{1}{\cs^{\ks}} \cas_\vrep k\partial_k\wfr  + k\partial_k \m \right)\gfb_\vrep \gf_\vrep \,.
\]
Thus, on the right-hand side of the equation only the terms of order $\gfb_\vrep \gf_\vrep$ and $C_\vrep \gfb_\vrep \gf_\vrep$ in the fields and in the momenta are of interest.
At quadratic order in the fields we have 
\[
\frac{1}{2} \overline{\Tr} \left[ \frac{k \partial_k \reg}{\Gamma_k^{(2)} + \reg }  \right]_{\gfb\gf} 
= - \frac{\ccu}{2} \sum_{\vrep\in\set\rk\ak} \frac{1}{\cs^{\ks}}\left(\aa {\cas_\vrep} + \bb \ak^{\ks} \right) \sum_{\ell=1}^\rk \frac{\cn{2}^\ell}{2} 2 \left(\mop_{\rep_\ell} + \nop_{\crep}\right)
\]
where we abbreviate the contributions in Eq.~\eqref{eq:2x2Wetterich} stemming from $P_\textsc{r}^2 \, k \partial_k \reg $ with
\[
\aa = - \frac{ k\partial_k\wfr}{(\wfr k^{\ks} + \m)^2}
\quad , \quad 
\bb =   \frac{\ks\wfr + k\partial_k\wfr}{(\wfr k^{\ks} + \m)^2} .
\]
To compare the two sides, one has to perform in the sum over $\vrep\in\set\rk\ak$ only the partial sums over the momenta $\rep_\ell$ on which the $\gfb\gf$ terms are \emph{not} depending. 

At this stage the specific choice of the summation set $\set\rk\ak$ makes a huge difference. 
The $\rep_\ell$ dependence of the right-hand side enters solely through these partial sums, that is through the dependence of their bounds on the momenta \emph{not} summed over.
Thus, for the hypercube case where these bounds never depend on momenta one always obtains trivial flow equations for the wave function renormalization (yielding $\eta_k=0$).
On the other hand, if the bound depends on the momenta linearly as in the simplex case, one obtains terms linear in $\rep_\ell$ which have a comparison with the left-hand side Eq.~\eqref{eq:lhsquadratic} only if the kinetic term is linear. 
In this sense, the flow equation of the wave function renormalization is very sensitive to choosing the appropriate regulator $\reg$.

For these reasons we perform the partial sum over the four relevant terms $\mop$, $\nop$, $\cas_\vrep\mop$ and $\cas_\vrep\nop$ now for the case of the quadratic regulator, that is the sum over the discrete ball $\ball\rk\ak$.
For $\mop$ and $\nop$ we have to calculate sums which, in contrast to the threshold function $\ku{s}$, Eq.~\eqref{eq:generalizedvolume}, include the zeros,
\[\label{eq:generalizedvolume}
\kv{s}(\ak) := \sum_{\vrep\in\ball\rk\ak} 1 \approx 1 + \sum_{s=1}^\rk \binom{\rk}{s} \vks{s}\ak^s \, .
\]
Using this we calculate
\begin{eqnarray}
\frac{1}{(\gf,\gf)}\sum_{\vrep\in\ball\rk\ak} \nop_{\crep} &= & 
\kv{1}\left( (\ak^{\ks}-\cas_\crep )^{\frac{1}{\ks}} \right)
= 1 + \vks1 (\ak^{\ks}-\cas_\crep)^{\frac{1}{\ks}} \nonumber\\
&=& 1 + \vks1\ak - \frac{\vks1}{\ks} \ak^{1-\ks} \cas_\crep + \mathcal{O}(\cas_\crep{}^2) \, ,
\end{eqnarray}
\begin{eqnarray}
\frac{1}{(\gf,\gf)}\sum_{\vrep\in\ball\rk\ak} \mop_{\rep_\ell} 
&=& \kv{\rk-1}\left( ({\ak^{\ks} - |\rep_\ell|^{\ks} })^\frac{1}{\ks} \right)
= 1 + \sum_{s=1}^{\rk-1} \binom{\rk-1}{s} \vks{s}(\ak^{\ks}-|\rep_\ell|^{\ks})^{\frac{s}{\ks}} \nonumber\\
&=& \red{1 + \sum_{s=1}^{\rk-1} \binom{\rk-1}{s} \vks{s}
\left(\ak^{s} - \frac{s}{\ks}\ak^{s-\ks}|\rep_\ell|^{\ks} \right)} +\mathcal{O}(\rep_\ell^4).
\end{eqnarray}
For the two sums over squared momenta we find using the threshold functions Eq.~\eqref{eq:thresholdzeta}
\begin{eqnarray}
\sum_{\vrep\in\ball\rk\ak} C_\vrep & \nop_{\crep} &
= \sum_\crep \nop_\crep \sum_{|\rep_\ell|^{\ks}\le {\ak^{\ks}-\cas_\crep}}\left(\cas_\crep + |\rep_\ell|^{\ks} \right) \nonumber\\
= \sum_\crep & \nop_\crep & \left(\cas_\crep  
\kv{1}\left( (\ak^{\ks}-\cas_\crep )^{\frac{1}{\ks}} \right) + \kw{1}\left( (\ak^{\ks}-\cas_\crep )^{\frac{1}{\ks}} \right) \right) \\
= \sum_\crep &\nop_\crep& \left(
\green{ \cas_\crep (1+\vks{1}\ak) + \frac{\vks1}{1+\ks}\left(\ak^{1+\ks} - \frac{1+\ks}{\ks}\ak \cas_\crep\right)} + \mathcal{O}(\cas_\crep{}^2) \right) \, , \nonumber
\end{eqnarray}
\begin{eqnarray}
\sum_{\vrep\in\ball\rk\ak}  C_\vrep &\mop_{\rep_\ell}& = 
\sum_{\rep_\ell} \mop_{\rep_\ell} \sum_{\crep\in \ball{\rk-1}{\sqrt{\ak^2-|\rep_\ell|^{\ks}}}}\left(|\rep_\ell|^{\ks} + \cas_\crep  \right) \nonumber\\
= \sum_{\rep_\ell} &\mop_{\rep_\ell}& 
\left[ |\rep_\ell|^{\ks}\kv{\rk-1}\left( ({\ak^{\ks} - |\rep_\ell|^{\ks} })^\frac{1}{\ks} \right) 
+ (\rk-1)\kw{\rk-1}\left( ({\ak^{\ks} - |\rep_\ell|^{\ks} })^\frac{1}{\ks} \right) \right] \nonumber\\
= \sum_{\rep_\ell} &\mop_{\rep_\ell}& \Bigg[ 
\blue{|\rep_\ell|^{\ks} \left( 1 + \sum_{s=1}^{\rk-1} \binom{\rk-1}{s} \vks{s}\ak^{s} \right)} \\
&& + \blue{ \frac{\rk-1}{\rk-1+\ks}\vks{\rk-1} \left(\ak^{\rk-1+\ks} - \frac{\rk-1+\ks}{\ks}\ak^{\rk-1}|\rep_\ell|^{\ks} \right) +  \mathcal{O}(\rep_\ell^4) } \Bigg] \, . \nonumber
\end{eqnarray}
Taking all terms together we have up to second order in momenta
\begin{align}
\frac{1}{2} \overline{\Tr} \left[\frac{k \partial_k \reg}{\Gamma_k^{(2)} + \reg}  \right]_{\gfb\gf}
= &- \frac{\ccu}{2 \cs^2}\frac{\cn{2}}{\rk} \sum_\vrep \gfb_\vrep \gf_\vrep 
\bigg\{
\aa\rk\left(
\green{\frac{\vks1}{1+\ks}\ak^{1+\ks}}
\blue{+\frac{\rk-1}{\rk-1+\ks}\vks{\rk-1}\ak^{\rk-1+\ks}}\right) \nonumber\\
&+\bb\rk\left(\ak^{\ks} + \vks1\ak^{1+\ks} + 
\red{\ak^{\ks} + \sum_{s=1}^{\rk-1} \binom{\rk-1}{s} \vks{s} \ak^{s+\ks}} \right) \nonumber\\
&+\aa\cas_\vrep \Bigg( 
\green{(\rk-1) (1+ (1-\frac{1}{\ks})\vks1 \ak) } \nonumber\\ 
& \quad \quad \blue{+ 1 + \sum_{s=1}^{\rk-2} \binom{\rk-1}{s} \vks{s}\ak^{s} -\frac{\rk-1-\ks}{\ks}\vks{\rk-1}\ak^{\rk-1} }\Bigg) \nonumber\\
&-\bb\cas_\vrep \left((\rk-1)\frac{\vks1}{\ks}\ak + 
\red{\sum_{s=1}^{\rk-1} \binom{\rk-1}{s} \frac{s}{\ks}\vks{s}\ak^{s}} \right) \biggl\} \, .
\end{align}
Note that this step, which is summing partial Casimirs to the full $\cas_\vrep$, is only possible upon identifying the quartic couplings of different colour. 
It is not clear how to obtain the full Casimir, i.e.~the sum over all colours $\ell$, on the right-hand side if couplings $\cn{2}^\ell$ are distinguished for different $\ell$.

Comparing with the left-hand side Eq.\,\eqref{eq:lhsquadratic} we find the two flow equations for the anomalous dimension $\eta_k$ and $\m$.
We have 
\[
\eta_k \equiv -\frac{1}{\wfr}k\partial_k \wfr = - \frac{\ks-\eta_k}{\ks}\frac{\ccu}{2\rk}\frac{\cn{2}}{(\wfr k^{\ks}+ \m)^2} F_\rk^\bb(\ak)
+ \frac{\eta_k}{\ks}\frac{\ccu}{2\rk} \frac{\cn{2}}{(\wfr k^{\ks}+ \m)^2} 
F_\rk^\aa(\ak)
\]
where
\begin{eqnarray}
F_\rk^\bb(\ak) &= 2(\rk-1)&\ak + \red{\sum_{s=1}^{\rk-1} \binom{\rk-1}{s} s\,\vks{s}\ak^{s}},\\
F_\rk^\aa(\ak) &= \green{(\rk-1)}&\green{(\ks+ (\ks-1)\vks1 \ak)} \\
& &+ \blue{\ks + \ks\sum_{s=1}^{\rk-2} \binom{\rk-1}{s} \vks{s}\ak^{s}  -(\rk-1-\ks)\vks{\rk-1}\ak^{\rk-1} }
\, . \nonumber
\end{eqnarray}
Solving for $\eta_k$ the equation is
\begin{align}
\eta_k &= \frac{ \ks\cn{2} \ccu  F_\rk^\bb(\ak)}{- 2\rk\cdot\ks(\wfr k^{\ks}+ \m)^2 +\cn{2} \ccu \left(F_\rk^\bb(\ak) + F_\rk^\aa(\ak) \right) } \\
&= -\cn{2}\frac{2(\rk-1)\ak + \red{\sum_{s=1}^{\rk-1} \binom{\rk-1}{s} s\,\vks{s}\ak^{s}}}
{\frac{2\rk}{\ccu}(\wfr k^{\ks}+ \m)^2 
- \cn{2} \left( \rk +
2(\rk-1)\ak  + \red{\sum_{s=1}^{\rk-2} \binom{\rk-1}{s} \frac{s + \blue{\ks} }{\ks}\vks{s}\ak^{s}}
+\vks{\rk-1}\ak^{\rk-1}
\right)} \, . \nonumber
\end{align}
For the mass term we have 
\begin{align}
k\partial_k \m = -\wfr k^{\ks} \cn{2} \ccu &\Bigg(\frac{\ks-\eta_k}{2}
\frac{1 + \vks1\ak + 
\red{1 + \sum_{s=1}^{\rk-1} \binom{\rk-1}{s} \vks{s} \ak^{s}}
}{(\wfr k^{\ks}+ \m)^2} \nonumber\\
&+ \frac{\eta_k}{2} \frac{
\green{\frac{1}{1+\ks}2 \ak}
\blue{+\frac{\rk-1}{\rk-1+\ks}\vks{\rk-1}\ak^{\rk-1}}
}{(\wfr k^{\ks}+ \m)^2}  
\Bigg) 
\end{align}
which is in good agreement with the result from the expansion of the full FRG equation for a constant field, Eq.~\eqref{eq:massflow}.

\section{Scaling dimensions \label{sec:Dimensions}}

The scaling dimensions necessary to rescale the FRG equation in the large-$\ak$ limit can be taken from results on the renormalizability of TGFTs~\cite{BenGeloun:2014gp,Carrozza:2013uq}.
A peculiarity of tensorial group field theories is that canonical dimension and scaling dimension differ. 
The following discussion follows and slightly generalizes the arguments in the Appendix of Ref.~\cite{BenGeloun:2016kw}.

From the kinetic part of the action \eqref{eq:regularizedkinetic} one derives the canonical dimension of the group field. The measure has canonical dimension $[\extd\vg] = -\gd\rk$ in terms of rank $\rk$ and group dimension $\gd$ and for the kinetic term we assume the general case $[\kin] = \ks$~\cite{BenGeloun:2014gp}. Thus, the canonical dimension of the field is
\[
[\gf] = [\gfb] = \frac{\gd\rk-\ks}{2}.
\]
Using $[\gf]$ one derives the canonical dimension of a coupling for an interaction with combinatorics captured by the coloured graph $b$ from the action as 
\[
[\cb] = -[(\extd\vg)^{\nb\rk}] - \nb[\gf\gfb] = \gd{\nb  \rk} - 2\nb\frac{\gd\rk-\ks}{2} = \ks \nb
\]
where $\nb$ is its order, that is half the number of vertices of the graph $b$.
In contrast to standard (combinatorially local) QFT, the canonical dimension of the couplings does not depend on any configuration space dimension but only on the order of the interaction $\nb$ (and the scaling $\zeta$ of the kinetic term).

The scaling dimension is the asymptotic scaling exponent of amplitudes for given external structure.
At cutoff $\Lambda$ the amplitude associated with a diagram $\Gamma$ scales asymptotically 
\[
|A^\Lambda_\Gamma| \propto |\prod_{v}\lambda_v|\Lambda^{\sdd_\Gamma},
\]
where $\sdd_\Gamma$ is the superficial (power counting) divergence degree of $A^\Lambda_\Gamma$.
If one now demands that all diagrams $\Gamma$ with given boundary $b=\partial\Gamma$ have the same scaling $|A^\Lambda_\Gamma| \propto \Lambda^{\scd{b}}$, one has the system of equations for the scaling dimensions $\scd{b}$
\[\label{eq:scaldimequation}
\scd{\partial\Gamma} = \sum_b V^b_\Gamma \scd{b} + \sdd_\Gamma
\]
for all diagrams $\Gamma$ wherein $V^b_\Gamma$ is the number of vertices with boundary $b$. 
The equations can be decoupled and solved expressing the superficial divergence degree in terms of the vertex numbers $V^b_\Gamma$.
From renormalization analysis~\cite{BenGeloun:2014gp} it is known that
\begin{eqnarray}
\sdd &=& \gd F - \ks E \\
&=& - \gd \gdeg + \frac{\gd(\rk-s)-\ks}{2}2\left(\sum_b V^b \nb - \next\right) - \gd(\rk-s)(V-1) 
\end{eqnarray}
for any diagram $\Gamma$ ($\Gamma$-subscripts dropped here) where $\next$ is the number of black (or white) vertices of $\partial\Gamma$, $V$ is the total number of internal vertices and $\gdeg$ is the Gurau degree~\cite{GurauBook}.
Furthermore, the degree is different for TGFT without gauge constraint where $s=1$ and with gauge constraint where $s=2$.
Except for the Gurau degree $\gdeg$, this is the same result as for standard scalar field theory with an effective dimension
\[
\uvd := \biggl\{\begin{array}{lr}
        \gd(\rk-2), & \text{with gauge constraint} \\
        \gd(\rk-1), & \text{without gauge constraint}
        \end{array}
\]
that is 
\[
\sdd + (\uvd-\ks)\next-\uvd = -\gd\gdeg + \sum_b \left[(\uvd-\ks)\nb-\uvd\right]V^b.
\]

In this work we are only interested in melonic interactions, thus also all the diagrams $\Gamma$ of the theory are melonic such that $\gdeg=0$.
Inserting into Eq.~\eqref{eq:scaldimequation} one finds then
\begin{eqnarray}
\scd{b} &=& \uvd - ({\uvd-\ks})\nb\\
&=& [\cb] + \uvd\left(1-\nb\right) .
\end{eqnarray}

Thus, the scaling dimension differs from the canonical dimension for all but the quadratic ($\nb=1$) term.

\bibliographystyle{JHEP}
\bibliography{main}

\end{document}